\documentclass[fleqn,usenatbib]{mnras}

\usepackage[T1]{fontenc}

\DeclareRobustCommand{\VAN}[3]{#2}
\let\VANthebibliography\thebibliography
\def\thebibliography{\DeclareRobustCommand{\VAN}[3]{##3}\VANthebibliography}

\usepackage{graphicx}	
\usepackage{amsmath}	
\usepackage{amssymb}	
\usepackage{subcaption} 
\usepackage{upgreek}    
\usepackage{newtxtext,newtxmath}  

\newcommand{\sfrunit}{\ensuremath{\rm M_\odot\,yr^{-1}}}
\newcommand{\massunit}{\ensuremath{\rm M_\odot}}
\newcommand{\Mpc}{\ensuremath{\,\text{Mpc}}}
\newcommand{\mass}{\ensuremath{M_{\star}}}
\newcommand{\kms}{\ensuremath{\,\text{km}\,\text{s}^{-1}}}
\newcommand{\microns}{\ensuremath{\rm \umu m}}
\newcommand{\asec}{\ensuremath{\,\rm arcsec}}
\newcommand{\hec}{\textit{HECATE}}
\newcommand{\hyper}{\textit{HyperLEDA}}
\newcommand{\ned}{\textit{NED}}
\newcommand{\nedd}{\textit{NED-D}}
\newcommand{\twomass}{\textit{2MASS}}
\newcommand{\sdss}{\textit{SDSS}}
\newcommand{\iras}{\textit{IRAS}}
\newcommand{\wise}{\textit{WISE}}
\newcommand{\akari}{\textit{AKARI}}
\newcommand{\planck}{\textit{Planck}}
\newcommand{\gsw}{\textit{GSWLC-2}}
\newcommand{\rifz}{\textit{RIFSCz}}
\newcommand{\lga}{\textit{2MASS-LGA}}
\newcommand{\masx}{\textit{2MASS-XSC}}
\newcommand{\masp}{\textit{2MASS-PSC}}
\newcommand{\galex}{\textit{GALEX}}
\newcommand{\rbgs}{\iras{}-\textit{RBGS}}
\newcommand{\vvir}{\ensuremath{v_{\rm vir}}}
\newcommand{\ngc}{NGC\,4993}
\newcommand{\code}[1]{{\lq{}{#1}\rq{}}}

\newcommand{\portal}{\url{http://hecate.ia.forth.gr}}

\title[The Heraklion Extragalactic Catalogue]
{
    The Heraklion Extragalactic Catalogue (HECATE):
    a value-added galaxy catalogue for multi-messenger astrophysics
}

\author[K. Kovlakas, A. Zezas, J. Andrews et al.]{%
K. Kovlakas,$^{1,2}$\thanks{e-mail: kkovlakas@physics.uoc.gr}
A. Zezas,$^{1,2,3}$
J. J. Andrews,$^{4}$
A. Basu-Zych,$^{5,6}$
T. Fragos,$^7$
\newauthor
A. Hornschemeier,$^{5,8}$
K. Kouroumpatzakis,$^{1,2}$
B. Lehmer,$^{9}$ and
A. Ptak$^{5}$
\\
$^1$Physics Department, University of Crete, GR 71003, Heraklion, Greece\\
$^2$Institute of Astrophysics, Foundation for Research and Technology-Hellas, GR 71110 Heraklion, Greece\\
$^3$Harvard-Smithsonian Center for Astrophysics, 60 Garden Street, Cambridge, MA 02138, USA\\
$^4$Center for Interdisciplinary Exploration and Research in Astrophysics (CIERA), Northwestern University, 1800 Sherman Ave, Evanston, IL 60201, USA\\
$^5$NASA Goddard Space Flight Center, Laboratory for X-ray Astrophysics, Greenbelt, MD 20771, USA\\
$^6$Department of Physics, University of Maryland Baltimore County, Baltimore, MD 21250, USA\\
$^7$Geneva Observatory, University of Geneva, Chemin des Maillettes 51, 1290 Sauverny, Switzerland\\
$^8$The Johns Hopkins University, Homewood Campus, Baltimore, MD 21218, USA\\
$^{9}$Department of Physics, University of Arkansas, 825 West Dickson Street, Fayetteville, AR 72701, USA
}

\date{Accepted XXX. Received YYY; in original form ZZZ}

\pubyear{2020}

\begin{document}
\label{firstpage}
\pagerange{\pageref{firstpage}--\pageref{lastpage}}
\maketitle

\begin{abstract}
    We present the {\it Heraklion Extragalactic Catalogue}, or \hec{}, an all-sky value-added galaxy catalogue, aiming to facilitate present and future multi-wavelength and multi-messenger studies in the local Universe.
    It contains 204,733 galaxies up to a redshift of 0.047 (D${\lesssim}200\Mpc{}$), and it is ${>}$50\% complete in terms of the $B$-band luminosity density at distances in the 0--170\,Mpc range.
    By incorporating and homogenising data from astronomical databases and multi-wavelength surveys, the catalogue offers positions, sizes, distances, morphological classifications, star-formation rates, stellar masses, metallicities, and nuclear activity classifications. 
    This wealth of information can enable a wide-range of applications, such as: (i) demographic studies of extragalactic sources, (ii) initial characterisation of transient events, and (iii) searches for electromagnetic counterparts of gravitational-wave events.
    The catalogue is publicly available to the community at a dedicated portal, which will also host future extensions in terms of the covered volume, and data products.
\end{abstract}


\section{Introduction}
\label{txt:intro}

With the availability of all-sky surveys across the electromagnetic spectrum (e.g., {\it LSST}, {\it ZTF}, {\it eROSITA}) and the advent of the era of multi-messenger observations (e.g., gravitational-wave, neutrino, cosmic-ray observatories) there is an increasing need for homogenised extragalactic catalogues that can be used for the characterisation of individual sources.

Astronomical databases like \ned{} \citep{Helou91}, \textit{SIMBAD} \citep{Wenger00} and \hyper{} \citep{Makarov14} have significantly boosted extragalactic research via the collection and organization of data such as positions, distances, photometric fluxes, and morphological classifications. However, due to the diversity of the different sources of these data, they cannot be readily used for studies requiring derived galaxy properties such as star-formation rate (SFR), stellar mass (\mass{}), metallicity, and nuclear activity, for large samples of objects. Although detailed catalogues based on focused surveys provide such information (e.g., {\it MPA-JHU}; \citealt{Kauffmann03,Brinchmann04,Tremonti04}), the lack of all-sky coverage limits their usefulness for many astrophysical applications, such as characterisation of sources in multi-wavelength all-sky or serendipitous surveys \citep[e.g., X-ray surveys;][]{Kim07,Saxton08}.
 
The rapid identification of counterparts of transient sources such as gamma-ray bursts (GRBs) or rare events (e.g., high-energy cosmic rays), and the strategic planning of follow-up observations, are possible with the aid of all-sky galaxy catalogues. Furthermore, the use of astrophysical information has been used to increase the effectiveness of identifying the hosts of gravitational-wave (GW) sources \citep[e.g.,][]{LigoVirgo17}. To this extent there is a growing effort to build galaxy catalogues providing information on \mass{} or SFR (or their proxies) aiming to aid GW follow-up observations \citep[e.g.,][]{Kopparapu08,White11,Gehrels16,Cook19,Dalya18,Ducoin20}. However, these catalogues do not provide metallicity, which can be a key factor for the identification of GW hosts \citep[e.g.,][]{Artale19}. Since the aforementioned galaxy catalogues were designed for applications focusing on distant galaxies (e.g., GWs, GRBs), the provided data may not be very accurate for nearby galaxies (e.g., $D{<}40\Mpc{}$), which often require special treatment (e.g., extended vs. point-source photometry, and distance measurements vs. application of the Hubble-Lema\^itre law). Therefore, studies involving nearby galaxy samples often invest in compiling the necessary galaxy data from scratch.

In order to enable large-scale studies of transient events such as those described earlier, or multi-wavelength properties of galaxies \citep[e.g., X-ray or $\gamma$-ray scaling relations;][]{Ackermann12,Komis19,Kovlakas20}, we require an all-sky catalogue that gives accurate locations, galaxy dimensions, distances, multi-band photometry, and most importantly derived stellar population parameters. For this reason, we compiled an all-sky value-added catalogue of 204,733 nearby galaxies within a distance of 200\Mpc{}: the {\it Heraklion Extragalactic Catalogue} (\hec{}\footnote{
    \textit{Hek\'at\=e}
    (greek, $\rm E\upkappa\acute{\upalpha}\uptau\upeta$),
    goddess of crossroads and witchcraft in ancient Greek mythology. Pronunciation: {\it hek-UH-tee}.}%
).
This catalogue provides all the aforementioned quantities based on a variety of sources. Special care is taken to develop procedures that consolidate the available data, maximize the coverage of the parameters, and address possible biases and offsets between different parent catalogues. The derivation of homogenised stellar population parameters, including metallicity, and nuclear activity classifications, highlight the usefulness of the \hec{} as a reference sample for the characterisation of sources in multi-wavelength and/or multi-messenger observations. The catalogue is publicly available at the \hec{} portal: \portal{}.

In \S\ref{txt:selection} we describe the selection of galaxies from the \hyper{} database, and the incorporation of redshift ($z$) and size information. The assembly and combination of distance measurements, as well as the derivation of $z$-dependent distances for galaxies without distance measurements is described in \S\ref{txt:distances}. The compilation of multi-wavelength data and the derivation of stellar population parameters is presented in \S\ref{txt:multisection}. In \S\ref{txt:discussion} we compare the \hec{} with other galaxy catalogues, discuss its limitations, and present various applications. Finally, in \S\ref{txt:summary} we present future extensions of the catalogue. Throughout the paper, unless stated otherwise, uncertainties correspond to 68\% confidence intervals.

\section{Sample selection}
\label{txt:selection}

\begin{figure*}
    \centering
    \includegraphics[width=\textwidth]{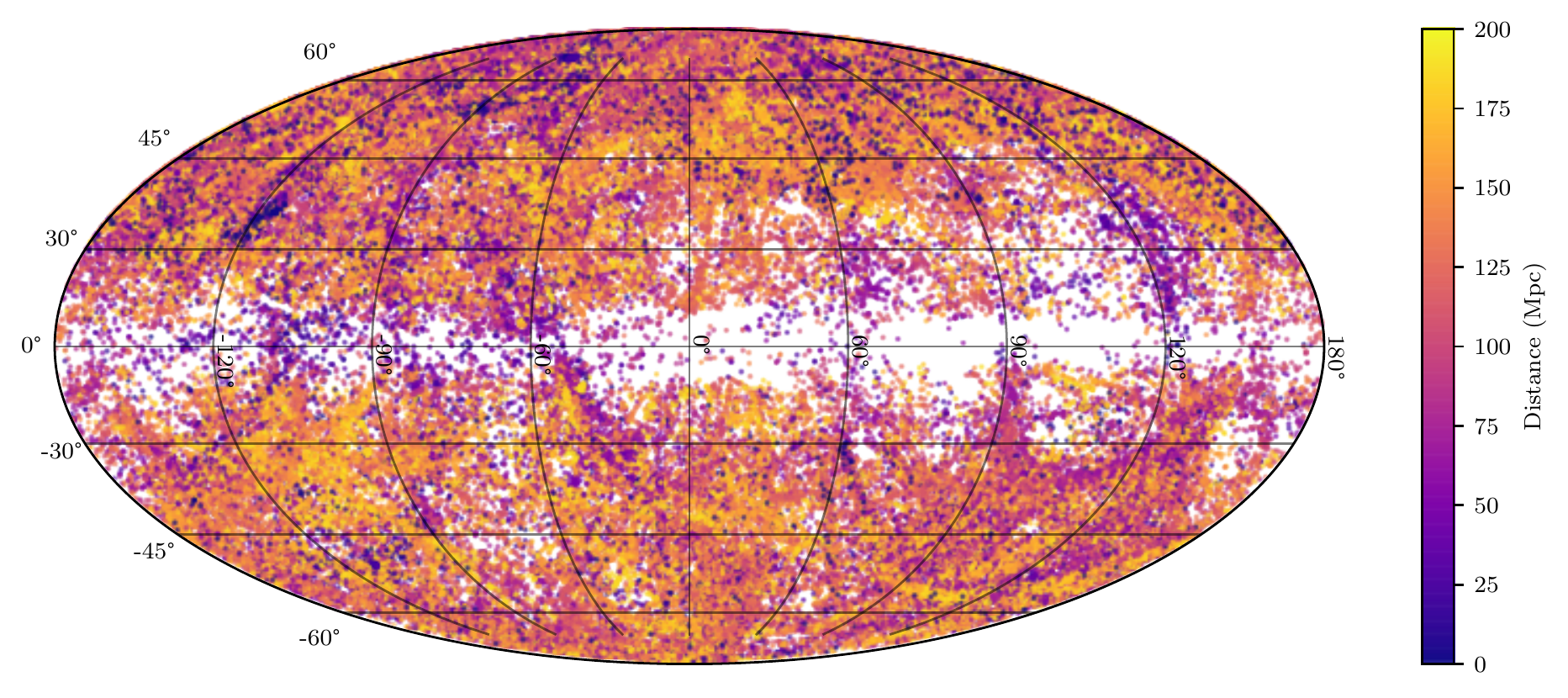}
    \caption{Sky map of the galaxies in the \hec{} in Galactic coordinates, colour-coded according to their distance.}
    \label{fig:allsky}
\end{figure*}

As the basis of our catalogue, we use the \hyper{} database
\citep{Makarov14}, which includes, combines and homogenises extragalactic data in the literature, without explicit flux or volume limits. Furthermore, common problems such as misprints, duplication, poor astrometry and wrong associations that can be found in legacy catalogues (e.g., \textit{UGC}: \citealt{Nilson73}, \textit{RC3}: \citealt{RC91}) are generally identified and rectified by the \hyper{} pipeline. 

Out of the 5,377,544 objects in the \hyper{} (as of October 2019), we select 3,446,810 (64\%) that are characterised as individual galaxies (\code{objtype=G}), excluding multiple systems (but not their members), groups, clusters, parts of galaxies, stars, nebul\ae{} etc.

Since the distances for the majority of the galaxies have not been measured, we perform the selection of local Universe galaxies based on a recession velocity limit.  We note that reported heliocentric radial velocity measurements typically contain the components of the peculiar motions of the Sun and the Milky Way. The peculiar motions of the galaxies are generally not known. We correct for those of the Sun with respect to the local Universe by computing the Virgo-infall corrected radial velocity, \vvir{}, which corrects for all motions of the Sun, and Milky Way up to the level of the infall of the Local Group to the Virgo Cluster. We select all galaxies with $\vvir{}{<}14,000\kms{}$ (corresponding to $z{<}0.047$ and $D{\lesssim{}}200\Mpc{}$ assuming Hubble parameter $h{=}0.678$; \citealt{PLANCK15}). 
The Virgo-infall corrected radial velocity in \hyper{} is outdated (D. Makarov, private communication). Therefore, we compute it for all galaxies (see Appendix~\ref{app:vvir} for details on the computation).

204,467 objects are found in \hyper{} with $\vvir{<}14,000\kms{}$ while 2,560,816 exceeded the velocity limit. However, for the 681,527 galaxies in \hyper{} without radial velocity measurements, we attempted to get measurements from \ned{}. The association to \ned{} is described in Appendix~\ref{app:ned}. In total, we recover the radial velocities for 1,494 additional objects with $\vvir{<}14,000\kms{}$.

Note that in the above procedures,
we performed various manual inspections to exclude duplicates in \hyper{} or misclassified objects (stars, artefacts from diffraction light, \lq{}parts of galaxies\rq{}, etc.) In total 1,228 objects were rejected in this process. The final sample consists of 204,733 galaxies. \autoref{fig:allsky} shows a sky map of the \hec{}.

Out of the 204,733 galaxies in our sample, there are 39,251 objects without size information, restricting the cross-matching capabilities of our sample. For the majority of them, the semi-major axis is complemented via cross-linking of our sample with other databases and surveys, resulting into 199,895 galaxies with size information (97.6\%). The procedure is described in detail in Appendix~\ref{app:diameters}.

Finally, in the \hec{} we include additional information from \hyper{} such as: astrometric precision, object name, morphological classification, optical photometry, inclination, and Galactic absorption. A full list of the information provided in the \hec{} is given in Appendix~\ref{app:columns}.

\section{Distance estimates}
\label{txt:distances}

Robust distance estimates for the galaxies in the \hec{} are essential for the purposes of this catalogue, and required for estimating the stellar population parameters of the galaxies.

While redshift-derived distances can be calculated using the Hubble-Lema\^itre law for the majority of the galaxies in the \hec{} (positive $z$), this approach is not accurate in the case of nearby galaxies for which recessional velocities are dominated by their peculiar motions. In addition, this method cannot be used for blue-shifted galaxies\footnote{In fact, the most blue-shifted galaxy in our sample with a reliable distance measurement is a Virgo Cluster member, VCC~0815, at distance of $19.8\Mpc{}$, which corresponds to a recession velocity ${\sim}1400 \kms{}$ but its heliocentric radial velocity is $-700{\pm}50 \kms{}$.}. Furthermore, at the distance range of the \hec{}, the unknown peculiar motion of a given galaxy adds to the uncertainty on its distance, equally or more than the propagation of the uncertainties of the galaxy's $z$ and the Hubble parameter.
For this reason, we use $z$-independent distance measurements from \nedd{} where available (for ${\approx}10\%$ of the galaxy sample), and combine them with the method described in \S\ref{txt:nedd}. For the remainder of the galaxies (${\approx}90\%$), we estimate the distance of the galaxies using a regression method, described in \S\ref{txt:zdependent}, based on the sample of galaxies with known distances.

\subsection{Redshift-independent distances}
\label{txt:nedd}

The largest resource of $z$-independent distances is the \nedd{} compilation, containing 326,850 measurements (as of March 2020) for 183,062 objects, based on 96 different distance indicators \citep{Steer17}. However, for objects with multiple measurements, \nedd{} does not readily provide a summary of these distance estimates. On the other hand, {\it CosmicFlows 3.0} ({\it CF3}; \citealt{Tully16}), reports distance estimates for 17,669 galaxies at $z{\lesssim}0.05$, calculated as uncertainty-weighted averages of individual measurements. Aiming at an as-large-as-possible sample of galaxies with distance determinations, we obtain distance measurements from \nedd{} in order to combine them into unique estimates for each galaxy, and use the {\it CF3} for consistency checks.

We reject measurements that are not based on peer-reviewed sources, and those using outdated distance moduli for the Large Magellanic Cloud (i.e. outside the $18.3{-}18.7\rm\,mag$ range; \citealt{LMCmodulus}) or distance scales calibrated for Hubble constants outside the range $60{-}80\kms{}\Mpc{}^{-1}$. Many of the 93 distance indicators reported in \nedd{} are appropriate for objects at distances greater than the volume limit of the \hec{} (e.g., SNIa) and therefore we do not consider them. We also avoid methods applied in less than three publications, as their systematic uncertainties or validity may be insufficiently understood. To be conservative, we select $10$ commonly used indicators that are considered relatively reliable at distances ${<}200\Mpc$ \citep[e.g.,][]{Steer17}, listed in \autoref{tbl:distindicators}.
\begin{table}
 \caption{List of distance indicators used in \hec{} (see \citealt{Steer17} and references therein), the number of galaxies ($N_{\rm gal}$) for which the measurements (number $N_{\rm meas}$) based on each indicator were considered in the final distance estimates, and the corresponding typical uncertainty of the distance moduli, $\langle\sigma_\mu\rangle$, in mag. 
 }
 \label{tbl:distindicators}
 \begin{tabular}{lrrr}
  \hline
  Distance indicator                & $N_{\rm gal}$ & $N_{\rm meas}$\textsuperscript{*} & $\langle\sigma_\mu\rangle$ \\
  \hline
  Cepheids                          &    75 &  1416 & 0.11 \\
  Eclipsing binary                  &     4 &    45 & 0.09 \\
  Fundamental plane                 & 10697 & 26975 & 0.35 \\
  Horizontal branch                 &    29 &    65 & 0.10 \\
  Red clump                         &    14 &   102 & 0.09 \\
  RR Lyr\ae                         &    38 &   282 & 0.09 \\
  Sosies                            &   280 &   280 & 0.29 \\
  Surface brightness fluctuations   &   482 &  1650 & 0.18 \\
  Tip of the red giant branch       &   358 &  1361 & 0.13 \\
  Tully-Fisher                      & 10780 & 11309 & 0.40 \\
  \hline
 \end{tabular}
 
\begin{minipage}{\columnwidth}
\textsuperscript{*}~We note that $N_{\rm meas}{\geq}N_{\rm gal}$ because for many galaxies there are multiple distance measurements based on the same indicator.
\end{minipage}

\end{table}
For publications reporting for the same galaxy multiple individual measurements based on the same indicator (e.g., Cepheid distances for different stars within a galaxy), we adopt the concluding measurement in each publication. Reported zero distance uncertainties (10 cases) were treated as undefined. Preference is given to measurements with reported uncertainties over those without uncertainties. In total, we associate 43,511 distance measurements with 21,174 galaxies in the \hec{}.

For the 13,247 galaxies with single distance measurements, we adopt them as they are, 97\% of which have reported uncertainties.

For the 7,336 galaxies with multiple distance measurements and uncertainties, we calculate the final distances and corresponding uncertainties using a weighted Gaussian Mixture (GM) model.
The weights depend on the year of publication (penalising old measurements) to reduce historical biases (e.g., older calibrations, unknown biases) and measurement uncertainties. The weight for the $i$-th measurement is:
\begin{equation}
    w_i = {\delta^{y_i - y_{\text{ref}}}}  {\sigma_i^{-2}},
    \label{eq:yearerrorweight}
\end{equation}
where $\delta$ is the penalty per year -- we set $\delta{=}2^{0.1}$ so that the weight is halved for every decade passed\footnote{We chose this value because: (i) we found systematic offsets ($0.05$--$0.2\rm\,mag$) in the distance moduli measured at times with differences ${>}20\rm yr$, and (ii) for small values of $\delta$ (corresponding to $^1$/$_2$-folding time-scales of less than 5 years), we found an increased scatter (${>}0.1\rm\,mag$) because, effectively only few newer measurements contribute to the distance.}, $y_i$ is the year of measurement, and $y_{\text{ref}}$ is an arbitrary reference year.
The GM distribution of the distance modulus $\mu$ of a galaxy is derived by combining $M$ individual measurements:
\begin{equation}
    f_{\rm GM}(\mu) = \sum\limits_{i=1}^{M} w_i f_i(\mu) \Big/ \sum\limits_{i=1}^{M}{w_i},
    \label{eq:gm}
\end{equation}
where $w_i$ are the weights calculated in \autoref{eq:yearerrorweight}, and $f_i$ is the PDF of the distribution of each measurement. We consider each measurement to be Gaussian-distributed, with mean and standard deviation equal to the distance modulus and its uncertainty reported in \nedd{}. We note that the mean of the distribution resulting from \autoref{eq:gm} is mathematically equivalent to the weighted average of the individual means (and therefore consistent with the methods for galaxies with single measurement, or without uncertainties [see below]), while its spread accounts for both the dispersion of the measurements and their uncertainties.

For the 591 galaxies with multiple measurements but no uncertainties, we use their weighted mean as the final estimate, and their weighted standard deviation as the uncertainty. In these cases, the weights are:
\begin{equation}
 w_i = \delta^{y_i - y_{\text{ref}}},
 \label{eq:yearweight}
\end{equation}
where the parameters are the same as in \autoref{eq:yearerrorweight}.

We note that for seven galaxies out of these 591, the standard deviation was 0 (possibly duplicate measurements), and therefore we do not report the uncertainty of the final distance estimate.

\subsection{Redshift-dependent distances}
\label{txt:zdependent}

\begin{figure}
    \centering
    \includegraphics[width=\columnwidth]{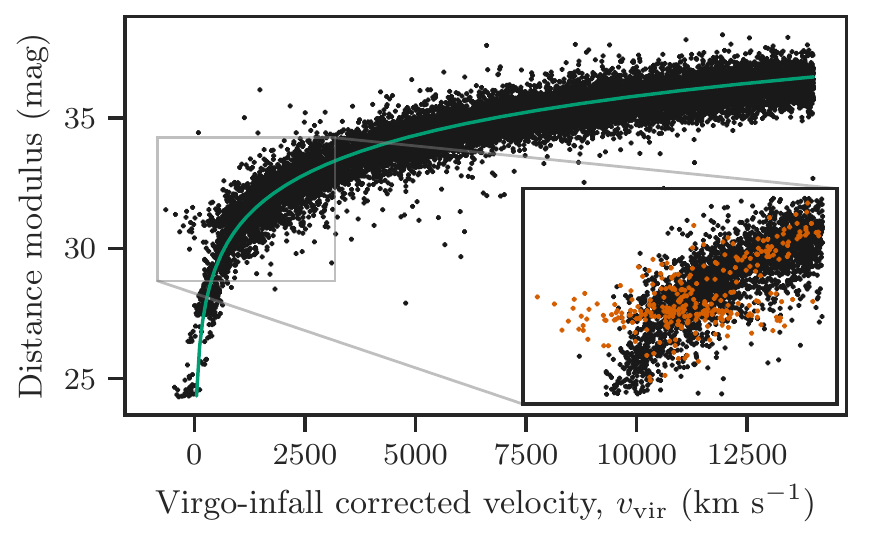}
    \caption{
        Hubble diagram of the galaxies in the \hec{} with $z$-independent distances. For reference the green} line shows the Hubble-Lema\^itre law with $H_0{=}(67.8{\pm}0.1)\kms \Mpc^{-1}$ \citep{PLANCK15}.
        We note that the majority of the points are following the law, albeit with significant dispersion at low values of $\vvir$, and the existence of a \lq{}branch\rq{} at distance modulus ${\sim}31$ (see inset) caused by Virgo Cluster galaxies (orange; see \S\ref{txt:virgomemebrship}) which present significant velocity dispersion (cf. fig. 10 in \citealt{Tully16}).
    \label{fig:hubblediagram}
\end{figure}

For galaxies without distance measurements (${\simeq}90\%$), we rely on the spectroscopic redshift information. While we could simply use the Hubble-Lema\^itre law and the redshift of each object in order to calculate their distances, the proximity of the galaxies in the \hec{} sample makes them very sensitive to peculiar velocities and local deviations from the Hubble flow. For this reason we adopt a data-driven approach where the galaxies with $z$-independent distances (10\% of the full sample) are used as the training data set in a regression model that infers the distances (and uncertainties) at similar recession velocities for the rest of the sample. The uncertainties of the radial velocities were not accounted for in the regression since, in the case of spectroscopic redshifts, they are negligible compared to the uncertainties of the distance measurements in the training sample.

\autoref{fig:hubblediagram} shows the distance modulus as a function of the radial velocity for the galaxies with $z$-independent distances in our sample (calculated as described in \S\ref{txt:nedd}). We observe 
    (i) that -- unsurprisingly -- the distance correlates with radial velocity even for nearby and blueshifted galaxies, albeit with higher dispersion, and
    (ii) a horizontal branch at distance modulus $\approx 31\,\rm mag$ that is caused by Virgo Cluster galaxies (see inset of \autoref{fig:hubblediagram}).
In order to account for such local deviations from the Hubble flow we employ a data-driven approach for robust distance estimates as follows:
\begin{enumerate}
\item  
    the galaxy sample is separated into two subsamples: galaxies in the Virgo Cluster and the rest (\S\ref{txt:virgomemebrship});
\item 
    for each subsample, a regressor is trained using the galaxies with redshifts and $z$-independent distances, so that the distance and its uncertainty are predicted from the recession velocity;
\item
    the distance and its uncertainty for the galaxies without $z$-independent measurements, is predicted using the two regressors.
\end{enumerate}

\subsubsection{Virgo-cluster membership}
\label{txt:virgomemebrship}

As we discussed in the previous paragraphs, and shown in \autoref{fig:hubblediagram}, special treatment of Virgo Cluster members is necessary for estimating their distance from the recession velocity. The most up-to-date catalogue of galaxies of the Virgo Cluster, the \textit{Extended Virgo Cluster Catalog (EVCC)}, was produced by \citet{Kim2014} using the radial velocities and a cluster infall model, as well as morphological and spectroscopic classification schemes. The EVCC is cross-matched with our sample to identify the galaxies associated with the Virgo cluster.

\subsubsection{Local average and standard deviation of Hubble diagram}
\label{txt:localaverage}

We use the Kernel Regression technique \citep{Nadaraya64} to compute the intrinsic distance modulus $\mu_{\text{int}}(\vvir)$ at a given Virgo-infall corrected radial velocity, $u{\equiv}\vvir$. The sample is split into Virgo members (VC) and non-Virgo members (nVC). For each subsample, we compute the local (at $u$) distance modulus, $\mu_{\text{int}}(u)$ as the weighted average of the distance moduli $\mu_i$ of the $N$ galaxies it contains, with weights ($q_i(u)$) given by the Guassian kernel with bandwidth $h$ (or \lq{}averaging length\rq{}). Similarly, for each subsample (VC and nVC) we calculate the \lq{}local standard deviation\rq{} in terms of the bias-corrected weighted standard deviation:
\begin{equation}
 \sigma_i(v) = \sqrt{\frac{V_1}{V_1^2 - V_2} \sum\limits_{i=1}^{N} q_i(u)\left[\mu_i - \mu_{\rm int}(u)\right]^2},
 \label{eq:weightedstd}
\end{equation}
where $V_1=\sum\limits_{i=1}^{N} q_i(u)$ and $V_2 = \sum\limits_{i=1}^{N} q_i^2(u)$.

The choice of the bandwidth $h$ effectively sets the resolution, in radial velocity, of the derived statistical quantities. Due to the significant curvature of the radial velocity vs. distance modulus diagram (\autoref{fig:hubblediagram}) for $u{\lesssim}2000\kms$, the resolution should be increased in this region in order to capture the shape and prevent mixing of data from regions of significantly different slopes. On the other hand, at greater radial velocities (or distances) a relatively large bandwidth would allow more data points to contribute, and hence provide an estimate that is less influenced by outliers. For these reasons, we set the bandwidth $h$ for the Gaussian Kernel to be a function of the radial velocity, increasing with radial velocity but also kept constant in the steep part of the diagram by enforcing a minimum value, $h_{\min}$:
\begin{equation}
    h(u) = h_{\min} \times \max\left\{1, u / 2000\kms\right\},
\end{equation}
where $h(u)$ is the bandwidth of the Gaussian Kernel for points evaluated at Virgo-infall corrected radial velocity $u$. For the Virgo Cluster model, we keep the bandwidth fixed as the distance modulus is expected to be roughly the same, rendering such considerations irrelevant.

The baseline of the bandwidth, $h_{\min}$, must be chosen carefully as it can easily result into \lq{}overfitting\rq{} if too small (only a few of the data points are considered for the fit in each bin), or, \lq{}underfitting\rq{} if too large (introducing \emph{lack-of-fit} variance). We find the optimal bandwidth, by minimising the total regression error $\mathcal{S}$, i.e. the quadratic sum of the regression errors, $\mathcal{S}_i$, corresponding to each galaxy. $\mathcal{S}_i$ is evaluated by employing the leave-one-out cross-validation technique: the $i$-th galaxy is removed from the sample, and its distance is estimated using the kernel regression. The residual between the true distance modulus, and the regressed one is $\mathcal{S}_i$. Additionally, when the optimal bandwidth has been found, we remove outliers based on the true distance modulus and the predicted one (and its uncertainty), by performing sigma clipping at the $3\sigma$-level, and re-optimise for $h_{\min}$ iteratively until no outlier is found. We applied the above procedure and found optimal minimum bandwidth $h_{\min}=68.2\kms$ for non-Virgo galaxies, after removing 149 outliers (${<}1\%$ of the nVC subsample). For the Virgo galaxies, the optimal bandwidth was $h_{\rm VC}=294.5\kms$, while only one outlier was found (${<}1\%$ of the VC subsample).

\subsubsection{Local intrinsic dispersion}
\label{txt:intrinsicdisp}

The local standard deviation we compute in \autoref{eq:weightedstd} encompasses the uncertainties of the distance moduli due to measurement uncertainties and the \emph{intrinsic scatter} of the true distance modulus. The latter is attributed to the peculiar velocities of the galaxies and the systematic uncertainties due to the distance ladder calibration. Given the model $\mu_{\text{int}}(v)$ and following \citet{Kelly07} we formulate the error model
\begin{equation}
    \mu_i = \mu_{\text{int}}(v_i) + \epsilon_i + \epsilon_{\text{int}}(v_i),
    \label{eq:errormodel}
\end{equation}
where $\epsilon_i$ is a Gaussian-distributed random variate with mean equal to $0$ and standard deviation equal to the distance modulus uncertainty of the $i$-th galaxy, $\mu_{\rm int}$ is the local average and $\epsilon_{\text{int}}(v_i)$ is a Gaussian-distributed variate with mean equal to $0$ and standard deviation $\sigma_{\text{int}}(v)$, which is a \emph{local intrinsic scatter} model. We apply a Maximum Likelihood Estimator (MLE) to calculate the local intrinsic scatter, $\epsilon_{i}(v)$. We note that the uncertainties on radial velocities have not been considered in our analysis as they are typically one order of magnitude smaller (${\sim}10\kms$) than the optimal kernel bandwidth for both VC and nVC models (${\sim}100\kms$) and the typical peculiar velocities of galaxies (${\sim}100\kms$; e.g., \citealt{Hawkins03}).

We apply the above Kernel Regression model to 617 Virgo galaxies and 182,326 nVC galaxies to derive their distances and uncertainties. Also, for 37 Virgo galaxies and 317 nVC galaxies with $z$-independent distances but no uncertainties, we apply the local intrinsic scatter model to estimate their uncertainty. We ensure that the two models are applied only to galaxies with radial velocities covered by the training data sets: $\vvir{\in}\left[-792\kms, 2764\kms\right]$ for VC and $\vvir{\in}\left[-481\kms, 14,033\kms\right]$ for nVC in order to avoid extrapolation (note that the ranges are expanded by half optimal bandwidth, $h_{\min}$), leaving only 12 objects in the \hec{} without distance estimates.

For quick reference, in Appendix~\ref{txt:app:empirical} we provide empirical formul\ae{} for the distance modulus of a galaxy $\mu_{\rm int}$, and its uncertainty $\epsilon_{\rm int}$, given its Virgo-infall corrected velocity, based on the results of the aforementioned methods.

\subsubsection{Validation of the regression technique}

\begin{figure}
    \centering
    \includegraphics[width=\columnwidth]{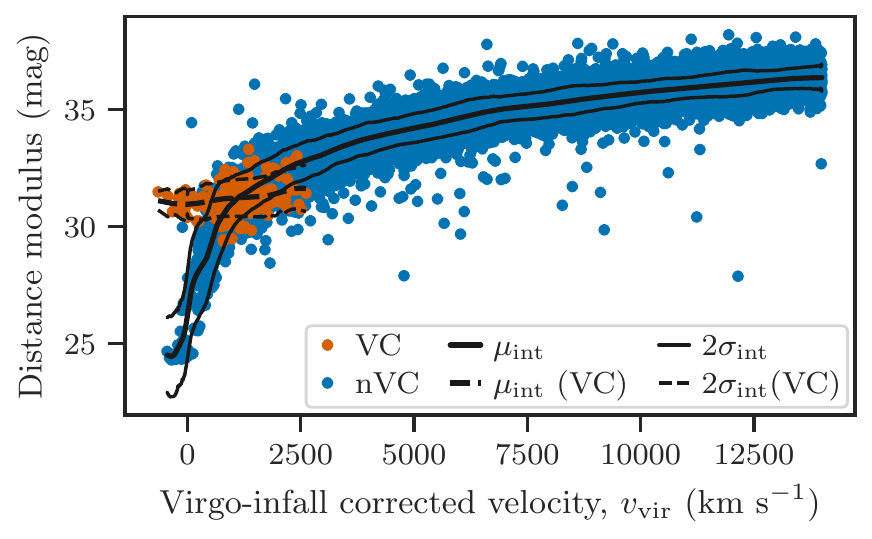}

    \includegraphics[width=\columnwidth]{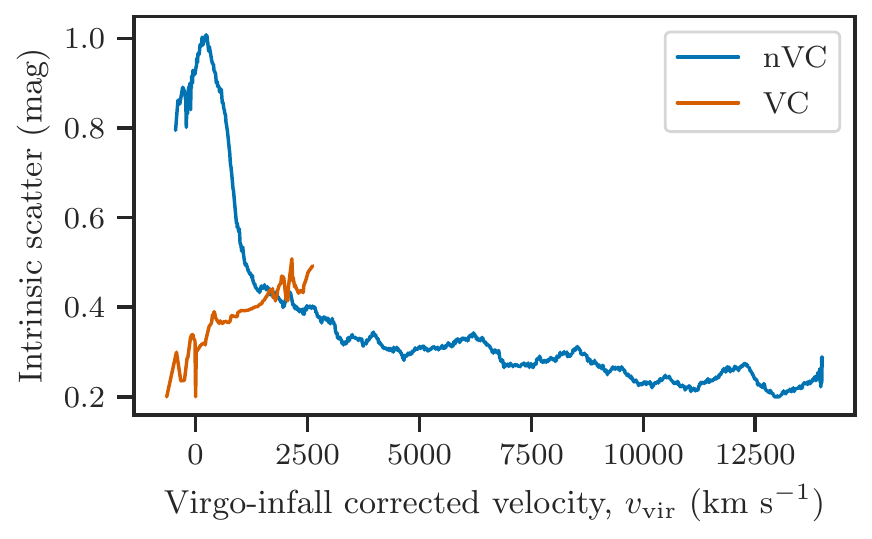}

    \includegraphics[width=\columnwidth]{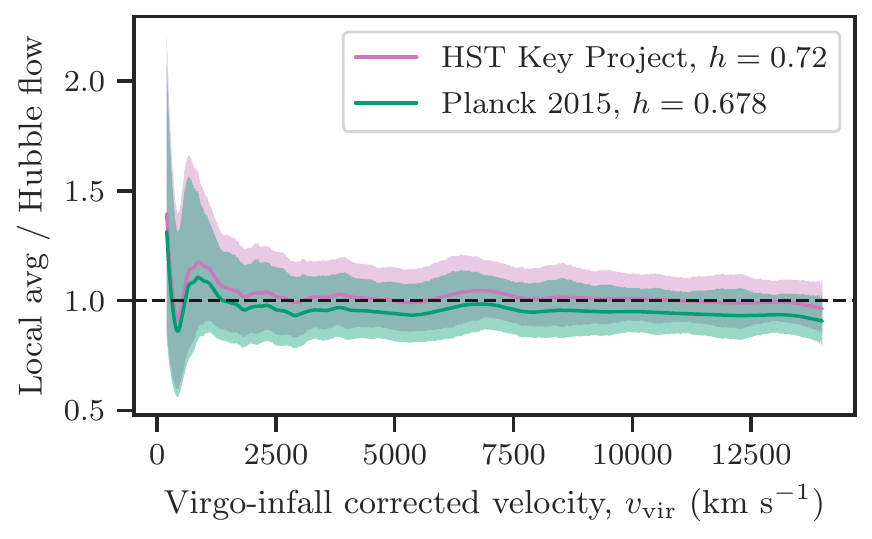}

    \caption{
        Assessment of the accuracy of the Kernel Regression models. The models capture the trends in the $D$-$z$ in the local Universe, and provide accurate distances ($0.2$-$0.4\rm\,mag$ or $10$-$20\%$) particularly for $\vvir{}{\gtrsim}{2500}$ ($D{\gtrsim{}}35\Mpc{}$).
        \textbf{Top:}
            $z$-independent distances in our sample (points), separated to Virgo Cluster members (orange) and nVC galaxies (blue). The black lines depict the local mean and the $2$-sigma regions (using the local standard deviation) according to the two regressors (dashed for VC, and continuous for nVC). 
        \textbf{Middle:}
            The local standard deviation of the distance modulus of the two models.
        \textbf{Bottom:}
            The ratio of the local average distance (\S\ref{txt:localaverage}) and the distance inferred from Hubble-Lema\^itre law for two values of Hubble parameter: $0.72$ (HST Key Project; magenta line) and $0.678$ (\planck{} 2015; green line). For each ratio, we plot with the same colours, the $68\%$ confidence region that reflects the local intrinsic scatter (\S\ref{txt:intrinsicdisp}). \label{fig:panelhecate}
    }
    \label{fig:v_model}
\end{figure}

The resulting distances from the Kernel Regression technique described above should reflect the trends of the $z$-independent distances used in the \hec{}, and converge to the Hubble-Lema\^itre law for large distances.

The top panel of \autoref{fig:v_model} shows the distance moduli as a function of the radial velocity of the two subsamples: 273 Virgo Cluster galaxies and 15,294 non-Virgo galaxies. We see the local average and the $2\sigma$ confidence intervals in terms of local intrinsic scatter. The latter is shown independently in the middle panel of \autoref{fig:v_model}, where we observe that the accuracy of the non-VC model drops significantly for $v{\lesssim}1,500\kms{}$ as expected from the domination of the peculiar velocities over the Hubble-flow component. Conversely, the VC model presents a slight increase in the distance with increasing radial velocity, which is possibly due to the contamination from background galaxies in the EVCC. For the same reason, the uncertainty of the inferred distances at high radial velocities for VC members is higher than that at the low radial velocities.

The convergence to the Hubble-Lema\^itre law is shown in the bottom panel of \autoref{fig:v_model}, where we plot the ratio of the local average model (\vvir{}; see \S\ref{txt:localaverage}) to the Hubble flow distance, for $\vvir{\in}\left[200, 14,000\right]\kms$. Two different values of $H_0$ are considered: $72\kms\Mpc^{-1}$ \citep[HST Key Project; ][]{Freedman01} and $67.8\kms\Mpc^{-1}$ \citep{PLANCK15}. We see that the $z$-independent distances converge to the Hubble flow distances, and agree to the local Universe estimate of the Hubble constant (HST Key Project).

Finally, we check the distance estimates in the \hec{} against {\it CF3}. For 1,949 galaxies with $z$-dependent distances in the \hec{}, but $z$-independent in {\it CF3}, we find agreement within the expected scatter from the regression model for 99\% of the galaxies in common. Furthermore,
the mean and median difference is $0.009\,\rm mag$, and the scatter in the difference between the distance moduli from {\it CF3} and the one calculated in the regression model is $0.25\,\rm mag$.

\section{Multi-wavelength data and stellar population parameters}
\label{txt:multisection}

One of the main objectives for the compilation of the \hec{} is to provide stellar population parameters for galaxies in the local Universe. To do so, we obtain photometric and spectroscopic data by cross-correlating the \hec{} with surveys from the infrared (IR) to the optical. In \S\ref{txt:multi} we evaluate the required data to attain the most reliable galaxy properties and discuss the selection and cross-matching criteria for each survey. In \S\ref{txt:derived} we describe the methodology we use for deriving the parameters from the associated multi-wavelength data.

\subsection{Associated photometric and spectroscopic data}
\label{txt:multi}

Star-formation rate estimates can be obtained by photometric data from IR to UV bands (or combinations of them; for a review, see \citealt{Kennicutt12}). While optical and UV-based SFR indicators are sensitive to dust absorption, IR indicators overcome this limitation by measuring the dust-reprocessed stellar emission. Although UV+IR composite SFR indicators \citep[e.g.,][]{Hao11} are now becoming more widely used (especially in the case of dwarf metal-poor galaxies) their implementation relies in the availability of integrated UV photometry. The all-sky \galex{} UV survey does not provide integrated photometry for large, nearby galaxies, hampering the use of these SFR indicators. Therefore, we rely on mid- and far-IR indicators using \iras{} and \wise{} photometry (see \S\S\ref{txt:iras} and \ref{txt:wise}), aiming at a homogeneous and as-complete-as-possible compilation of SFR estimates.

For the computation of the galaxy stellar masses, one of the most reliable photometric indicators is the $K_{s}$-band luminosity \citep[e.g.,][]{Gardner97}. In order to account for the stellar-population age dependence of the  mass-to-light ratio ($M/L$) we use calibrations that incorporate optical colours \citep{Bell03}. For this reason, we obtain \twomass{} and \sdss{} photometry, as described in \S\S\ref{txt:twomass} and \ref{txt:sdss}.

Spectroscopic data can be used to estimate the metallicity of the galaxies, as well as characterise them on the basis of their nuclear activity. In \S\ref{txt:sdss} we describe the acquisition of spectroscopic data from \sdss{}.

\subsubsection{Far-infrared: \iras{}}
\label{txt:iras}

We cross-link the \hec{} galaxies to \iras{} objects. For the cross-correlation with the \iras{} catalogue we adopt the following approach. When a galaxy is included in the \textit{IRAS Revised Bright Galaxy Sample} (\rbgs{}), we adopt this photometric information, which is more reliable for extended galaxies \citep{Sanders03}. In total, we associate 589 galaxies with the \rbgs{} catalogue (Appendix~\ref{app:rbgs}). For the remaining galaxies, we use the \textit{Revised \iras{}-FSC Redshift Catalogue} (\rifz{}; \citealt{Wang14}) which provides a clean (excluding poor quality and cirrus sources) sample of \iras{} galaxies at $60\umu\mathrm{m}$. This also gives more reliable positions than its parent \iras{} {\it Faint Source Catalog} (\iras{}{\it-FSC}; \citealt{Moshir90}) via the association to more recent surveys. In total, we associate 19,082 galaxies in our sample with \rifz{} (Appendix~\ref{app:rifz}).

\subsubsection{Mid-infrared: \wise{}}
\label{txt:wise}

The previous cross-matches with \rbgs{} and \rifz{} objects incorporate \iras{} photometry for 19,671 objects in the \hec{} (9.6\%). To obtain a more complete census of the IR emission of the galaxies in the \hec{} sample we could also use the deeper all-sky, surveys (e.g., \wise{}, \akari{}). However, at the time of compilation of the \hec{}, there are no extended source catalogues of \wise{} and \akari{} that can provide reliable flux measurements for nearby galaxies.
For this reason, we use the \lq{}forced photometry\rq{} catalogue by \citet{Lang16} who extracted fluxes from {\it unWISE} images \citep{unwise14} for \sdss{}-DR10 photometric objects using the \sdss{} apertures. We cross-correlate this catalogue with the \hec{} by matching the \sdss{} ID, which is already specified in the \hec{} (\S\ref{txt:sdss}). As the \wise{} forced photometry catalogue is organised in {\it unWISE} tiles, there are galaxies in overlapping regions. For these cases, we select the data from the tile in which the galaxy is closer to the tile's centre.
123,699 \hec{} galaxies to objects are linked to \sdss{} objects with \wise{} forced photometry. We note, however, that the use of this catalogue restricts our \wise{} photometric data to the \sdss{} footprint. \wise{} photometry is available for the wider \hec{} sample, but as mentioned earlier it is not reliable for the resolved galaxies.

\subsubsection{Near-infrared: \twomass{}}
\label{txt:twomass}

To incorporate \twomass{} photometry in our sample, we cross-match the \hec{} with three \twomass{} catalogues in the following order of priority: (i) {\it Large Galaxy Atlas} \citep[\lga{};][]{Jarrett03}, (ii) {\it Extended Source Catalog} \citep[\masx{};][]{Skrutskie06}, and (iii) {\it Point Source Catalog} \citep[\masp{};][]{Cutri12}. This order ensures that for the resolved galaxies we use the most reliable measurements of their flux. Specifically, from \lga{} and \masx{} we obtain the $JHK$ \lq{}total\rq{} magnitudes from the extrapolated surface brightness profiles (see \lga{} home page\footnote{\url{https://irsa.ipac.caltech.edu/applications/2MASS/LGA/intro.html}} and \S4.5.a.iv in the Explanatory Supplement\footnote{\url{https://old.ipac.caltech.edu/2mass/releases/allsky/doc/explsup.html}}). From the \masp{} we obtain the \lq{}default\rq{} magnitudes. We note that when no uncertainty is provided, the listed magnitudes are upper limits.
We link \hec{} galaxies to 609 objects in the \lga{}, 117,713 in the \masx{}, and 25,224 in the \masp{}, overall providing \twomass{} photometry for 143,546 galaxies. More details about the cross-matching procedure can be found in Appendix~\ref{app:twomass}.

\subsubsection{Optical: \sdss{}}
\label{txt:sdss}

For the cross-matching of the \hec{} and the \sdss{}, we use the DR12 photometric catalogue, and select only primary\footnote{\url{www.sdss.org/dr12/help/glossary/\#surveyprimary}} objects. We use a match radius of 3\asec{} around the \hec{} coordinates, and we select the closest match (typical separation of the matched objects is ${\sim}0.2\asec{}$), resulting in 123,711 matches.

We opt to use spectroscopic data from the {\it MPA-JHU DR8} catalogue \citep{Kauffmann03,Brinchmann04,Tremonti04}, which are based on the emission-line component of the spectrum after subtracting the underlying stellar component, to estimate the metallicities and classify the galaxies on the basis of their nuclear activity. By cross-matching the catalogue with the \hec{}, we obtain measurements for 93,714 out of the 123,711 \sdss{} objects in the \hec{}.

The {\it GALEX-SDSS-WISE Legacy Catalog 2} (\gsw{}) of \citet{Salim16,Salim18}
provides SFR and \mass{} estimates through optical-UV spectral energy distribution (SED) fits to galaxies within the \sdss{} footprint and distance ${>}50\Mpc$. By matching 75,672 \hec{} galaxies to \gsw{} objects on the basis of their object IDs in \sdss{}, we obtain additional SFR and \mass{} estimates.

\subsection{Derived parameters}
\label{txt:derived}

The following paragraphs describe the methods employed for the estimation of parameters from the acquired multi-wavelength data (\S\ref{txt:multi}). An overview of the provided data is listed in \autoref{tbl:columns}.

\subsubsection{Stellar masses}
\label{txt:stellarmass}

The stellar masses are estimated by combining the $K_{s}$-band luminosities of the galaxies with the appropriate mass-to-light ratios. The integrated $K_{s}$-band luminosities of the galaxies are calculated from their \twomass{} photometry and distances (we adopted $3.29\,\rm mag$ for the absolute magnitude of the Sun; \citealt{Blanton07}). We exclude objects without uncertainties in their photometry, or uncertainty higher than 0.3 mag, resulting in $L_K$ measurements for 133,017 (65\%) galaxies in the \hec{}. The $K_{s}$-band $M/L$ ratio (${\equiv}\mass{}/L_{K_s}$) is computed using the calibration of \citet{Bell03} which accounts for differences in the stellar populations by means of the $g-r$ colour of the galaxies:
\begin{equation}
    \log\left(M/L\right) = -0.209 + 0.197\left(g-r\right).
\end{equation}
$g-r$ colours are available for 53,171 (26\%) galaxies with reliable photometry (\sdss{} flags q\_mode=\lq{}+\rq{} and Q=3, and uncertainties $<0.1\,\rm mag$ on $g$ and $r$). The mean $M/L$ ratio of the galaxies in the \hec{} is 0.822, while the scatter is 0.091. This mean value is used for the 79,846 (39\%) galaxies without \sdss{} photometry. The scatter gives us an estimation of the $M/L$ ratio variations due to the different $g-r$ colours of the galaxies, and it can be used to assess the uncertainty on the \mass{} of galaxies without \sdss{} photometry. For the remainder (35\%) of the \hec{} sample that does not have $K_{s}$-band measurements in \twomass{} we do not estimate \mass{}.

\begin{figure}
    \centering
    \includegraphics[width=\columnwidth]{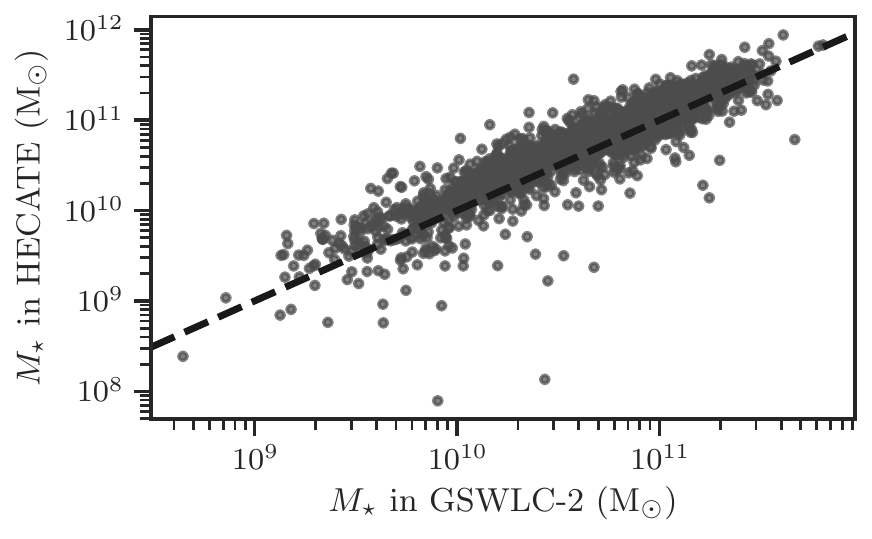}
    \caption{Comparison of the stellar mass estimates in the \hec{} and the \gsw{} for the common galaxies. The galaxies closely follow the 1:1 line, indicated by the black, dashed line, with an intrinsic scatter of 0.21\,dex.}
    \label{fig:compM}
\end{figure}

The \gsw{} provides \mass{} derived using a different method (SED-fitting using UV to IR data \citep{Salim16,Salim18}). In \autoref{fig:compM} we compare these estimates with our derived \mass{} using near-IR photometry. We find very good agreement (scatter of 0.21\,dex), although SED-based \mass{} are slightly lower on average (factor of $-0.11\,\rm dex$), possibly due to assumptions of stellar population models, or star-formation histories.

\subsubsection{Star-formation rates}
\label{txt:sfr}

The SFR estimates of the \hec{} galaxies are based on measurements of IR luminosity from the \iras{} or \wise{} surveys. These surveys provide the optimal combination of reliable, well-calibrated, SFR indicators \citep{Kennicutt12}, sensitivity, and sky coverage. Since the sensitivity of the two surveys varies depending on the band, we use a combination of SFR indicators depending on the  availability of reliable measurements. For \iras{}, we use the total-IR (TIR), far-IR (FIR) and 60\microns{} luminosities, depending on the bands with reliable \iras{} fluxes (\lq{}FQUAL\rq{}$\geq$2, i.e, excluding upper limits). For \wise{} we use the monochromatic Band-3 (W3; $12\microns{}$) and Band-4 (W4; $22\microns{}$) fluxes from the \wise{} forced photometry catalogue as discussed in \S\ref{txt:wise}. \wise{} SFR estimates are not provided for objects with uncertainties greater than $0.3\rm\,mag$, or those that were considered as point sources in the analysis of \citet{Lang16}\footnote{Using the galaxies with \wise{} and \iras{} photometry we found that the \wise{} SFRs are significantly lower for ${\sim}$2500 sources with the flag \code{treated\_as\_pointsource} set in the catalogue of \citet{Lang16}, 20\% of which have SFR estimates from \iras{} data.}.

\begin{figure*}
    \centering
    \includegraphics[width=\textwidth]{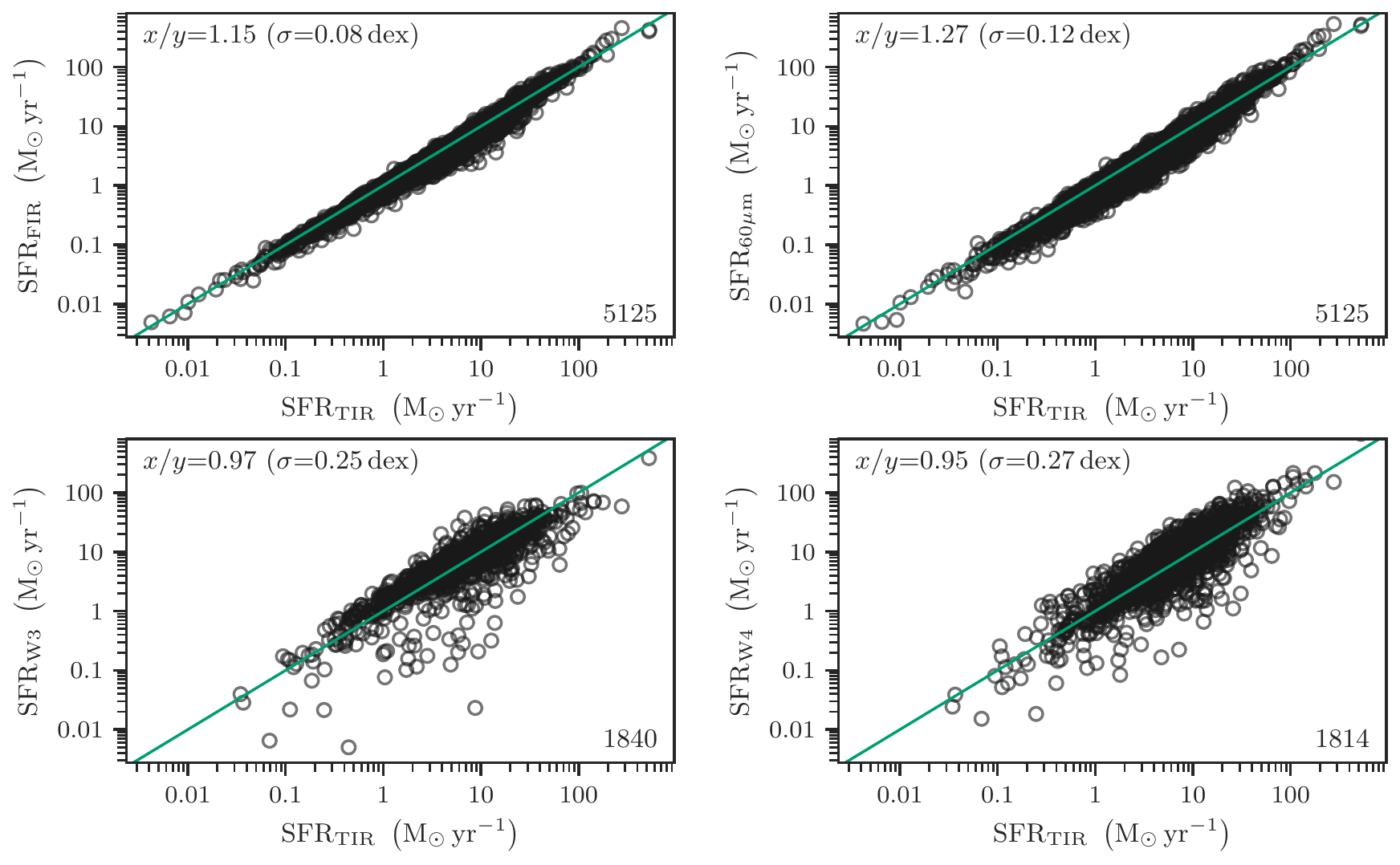}
    \caption{Comparison between the TIR-SFR indicator and the other four indicators used in the \hec{} (not rescaled as in the computation of $\mathrm{SFR_{HEC}}$); the 1:1 line is shown as a green line. For each SFR indicator, the linear scaling factor and the scatter are reported in the top left corner, while the number of overlapping galaxies used for the scaling is reported in the bottom right corner.
    The four SFR indicators scale with the reference indicator (TIR) well, and present intrinsic scatter upto 0.27\,dex (typical of photometric SFR estimates; \citealt{Kennicutt12}). The scaling factors are used for the computation of the homogenised SFR column in the \hec{}.
    }
    \label{fig:rescaleSFR}
\end{figure*}

For completeness, we provide in our catalogue SFR measurements based on all indicators (\autoref{tbl:sfrindicators}) available for each galaxy (including the SED-based SFR from the \gsw{} for convenience to users focusing in the \sdss{} footprint). This is particularly important since the various surveys used to derive the SFRs, cover different subsets of the \hec{}. In order to have consistent SFRs for the largest possible set of objects, and given the fact that different indicators often result in systematic offsets in the derived SFRs, we also provide a homogenised SFR ($\mathrm{SFR_{HEC}}$), calculated as follows.

First, we account for offsets between the different SFR indicators (\autoref{tbl:sfrindicators}) by calculating their ratio with respect to the TIR-based SFR which we consider as a reference since it probes star-formation regimes at different timescales \citep{Kennicutt12}. The mean ratio for each indicator is adopted as the correction factor. \autoref{fig:rescaleSFR} shows comparisons between the SFR indicators, also giving the scaling factor, standard deviation, and the number of galaxies used in each comparison.
The homogenised SFR of an object in the \hec{} is the TIR-SFR if available, otherwise we use, in order of preference, the rescaled FIR-SFR, $60\microns{}$, $12\microns{}$ and $22\microns{}$ based SFR. Although the $22\microns{}$-band (probing hot dust associated with young star-forming regions) is a better-calibrated SFR than the $12\microns{}$-based one (probing emission from polycyclic aromatic hydrocarbons; e.g., \citealt{Parkash18}), preference is given to the latter due to the higher quality of the W3 \wise{} data \citep[e.g.,][]{Cluver17}. We note that no rescaling is performed in the individual SFR indicator columns.

The TIR luminosity includes emission in the 100\microns{} band, which in the case of galaxies with low specific SFR may have a non-negligible contribution from stochastically heated dust from older stellar populations \citep[e.g.,][]{Galliano18}. Although this may overestimate the SFR in early-type galaxies, it is a widely used and well understood SFR indicator that gives reliable SFR for actively star-forming galaxies, which are the majority of the \hec{} (fig. 2 in \citealt{Kovlakas20}). We note that the catalogue provides all SFR indicators (Appendix~\ref{app:columns}) before rescaling, and the homogenised SFR, where indicators were rescaled according to the procedure described above. A flag is provided, denoting which indicator was used in the homogenised SFR (cf. \autoref{tbl:sfrindicators}). Therefore, based on the scaling factors reported in the table, one can translate the provided SFR to the reference indicator of their choice. Due to the selection of the TIR as a reference, the homogenised SFRs are consistent with the \citet{Kroupa01} IMF, constant SFH, and \citet{Leitherer99} stellar population models.

\begin{figure*}
    \centering
    \includegraphics[width=\textwidth]{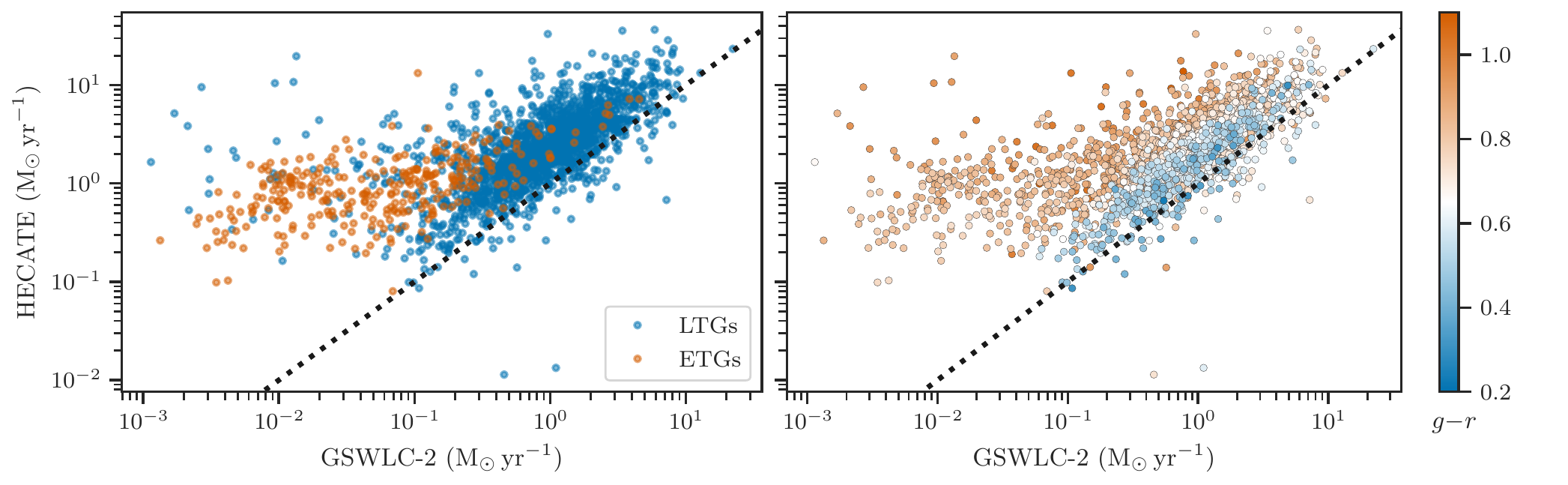}
    \caption{(a) Comparison between the homogenised SFRs and the SED-based SFRs from the \gsw{} for early (orange) and late-type (blue) galaxies. In the bulk of the sample, mainly consisting of late-type galaxies, the \gsw{} underestimates the SFR due to the lack of the dust component associated with the star-forming activity, whereas in early-type galaxies, the \hec{} overestimates the SFR due to the dust emission caused by the stochastic heating from old stellar populations, rather than star-formation. (b) Same as panel (a), but now, the galaxies are colour-coded according to their \sdss{} $g{-}r$ colour (in mag), which is a more reliable indicator of the stellar population age of the galaxies than the morphological classifications.}
    \label{fig:SFRcomp}
\end{figure*}

In \autoref{fig:SFRcomp} we compare the homogenised SFRs against the SED-based SFRs from the \gsw{}. We see that at ${\rm SFR }{\gtrsim{}}0.1\,\sfrunit{}$ (typical for star-forming galaxies), the \hec{} provides SFRs that scale with, but are a factor of ${\simeq}2.3$ larger than those of \gsw{}. This could be because the SEDs used in the \gsw{} do not include IR emission above $22\microns{}$, therefore missing the dominant, relatively cold, dust component associated with star-forming activity (probed in the ${\sim}$60\microns{} band). Furthermore, differences in the IMFs (only 0.02\,dex in this comparison), stellar population models, and SFHs might produce additional offsets (see discussion in \citealt{Salim16}). As discussed above, the IR-based SFRs may overestimate the SFR in low specific-SFR galaxies, which can explain the flattening observed at low SFRs, and the difference between early-type and late-type galaxies. This is demonstrated in the right-hand panel of \autoref{fig:SFRcomp} where the $g-r$ colour is used to indicate the contribution of the older stellar populations. Galaxies with redder colours (and hence higher $g-r$ tend to have an excess of IR-based SFRs with respect to SED-based SFRs. We see qualitatively that for $g-r > 0.65$, the discrepancy between the two becomes fairly significant.

\begin{table*}
\caption{
    The five different SFR indicators used in the \hec{}. The TIR and WISE indicators assume a constant star-formation history (SFH), {\it Starburst99} stellar population models \citep{Leitherer99}, and the \citet{Kroupa01} initial mass function (IMF). On the other hand, the FIR indicator assumes starbursts with timescales of 10--100 Myr, \citet{Leitherer95} stellar models, and the \citep{Salpeter55} IMF, whereas the 60\microns{} indicator relies on a universal SFH \citep{Rowan99}, the \citet{Bruzual93} stellar models, and the \citet{Salpeter55} IMF. The offsets between the TIR and the other {\it IRAS} indicators can be explained on the basis of the different assumptions.
    \label{tbl:sfrindicators}
}
\begin{tabular}{lrccccr}
    \hline
    Survey / bands & $N_{\rm gal}$ & $L$ calibration & SFR scaling relation & Scaling in SFR$_{\rm HEC}$ & Flag; number in SFR$_{\rm HEC}$ \\
    (1) & (2) & (3) & (4) & (5) & (6) \\\hline
    IRAS 25, 60, 100\microns{} (TIR) & 5,125 & \citet{Dale02} & \citet{Kennicutt12} & (reference) & RT/FT\textsuperscript{*}; 5,125 \\
    IRAS 60, 100\microns{} (FIR) & 5,721 & \citet{Helou88} & \citet{Kennicutt98} & 1.15 \phantom{-}(0.08\,dex) & FF; 596 \\
    IRAS 60\microns{} & 19,671 & \multicolumn{2}{c}{\citet{Rowan99}} & 1.27 \phantom{-}(0.10\,dex) & R3/F3\textsuperscript{*}; 13,950 \\
    WISE 12\microns{} & 81,948 & \multicolumn{2}{c}{\citet{Cluver17}} & 0.97 (-0.01\,dex) & W3; 72,726 \\
    WISE 22\microns{} & 46,078 & \multicolumn{2}{c}{\citet{Cluver17}} & 0.95 (-0.02\,dex) & W4; 1,872\\\hline
\end{tabular}
\begin{minipage}{\linewidth}
Description of columns:
(1) the survey (\iras{} or \wise{}), and the photometric bands used for the computation of the flux; 
(2) the number of galaxies for which the SFR indicator is computed; 
(3), (4) references to the definition of the composite band and SFR scaling calibration; 
(5) The scaling factor used only for the \lq{}homogenisation\rq{} of the SFR indicator with respect to the TIR indicator. In parenthesis we give the scatter of the \lq{}homogenisation\rq{} relation; 
(6) the flag (in the column \code{logSFR\_flag}; see \autoref{tbl:columns}) and the number of galaxies for which the homogenised SFR is based on this indicator.
\textbf{Note}: \textsuperscript{*} the first letter indicates whether the photometry was taken from the \rbgs{} (R) or \rifz{} (F).
\end{minipage}
\end{table*}

\subsubsection{Metallicity estimates}
\label{txt:metallicity}

To measure the gas-phase metallicities for our sample, we use the optical emission-line fluxes provided in the {\it MPA-JHU DR8} value-added \lq{}galSpecLine\rq{} catalogue \citep[see details for methods in][]{Brinchmann04} based on the \sdss{}-DR8 data. 
Especially relevant for measuring accurate nebular emission lines, this catalogue applies stellar-population synthesis models to accurately fit and subtract the stellar continuum, including stellar absorption features.
We calculate the gas-phase metallicities, $12+\log(O/H)$, using the \citet{PP04} O3N2 (henceforth, PP04 O3N2) prescription, which has been shown by \citet{Kewley08} to be robust (i.e. it can trace a wide range of metallicities, it has relatively low scatter, and most importantly it is less sensitive to extinction effects than other indicators).
Our metallicity analysis is subject to the quality of the [\ion{O}{III}], [\ion{N}{II}], $\rm H\beta$, or $\rm H\alpha$ emission lines and the PP04 O3N2 relation limitations. Therefore, we set the following flags (see column \lq{}flag\_metal\rq{} in Appendix~\ref{app:columns}) to mark uncertain results: sources with \code{1} have O3N2 ${>}2$ ratios (670 sources), where the PP04 O3N2 relationship is invalid, and therefore the extrapolated metallicities are highly uncertain; \code{2} marks emission lines with low signal-to-noise ($\sigma{<}3$; 882 sources). Therefore, only sources with flags set to \code{0} have reliable metallicity measurements (62,728 sources). Objects without metallicity estimates are flagged with \code{-1} (140,453 sources).

\subsubsection{Nuclear activity}
\label{txt:activity}

Using the \sdss{}-DR8 emission-line data (see \S\ref{txt:sdss} and \S\ref{txt:metallicity}), we identify AGN based on the location of the galaxies in the emission-line ratio diagnostic of \citet[][which has been trained on the same dataset]{Stampoulis19}. This diagnostic takes into account all available line ratios in order to provide a single robust activity classification that avoids the contradictory classifications that can be obtained from the use of the traditional two-dimensional line-ratio diagrams. We consider the ([\ion{S}{II} $\lambda\lambda6717,6731$\AA{}]/H$\alpha$, [\ion{N}{II} $\lambda6584$\AA{}]/H$\alpha$, [\ion{O}{III} $\lambda5007$\AA{}]/H$\beta$) three-dimensional diagram and when we have reliable measurements for the [\ion{O}{I}] $\lambda6300$\AA{} line we use the four-dimensional ([\ion{S}{II}]/H$\alpha$, [\ion{N}{II}]/H$\alpha$, [\ion{O}{I}]/H$\alpha$, [\ion{O}{III}]/H$\beta$) diagram. In this way, we provide nuclear activity classification for 64,280 (31\%) galaxies with signal-to-noise ratio greater than 2 in the emission lines used. Out of these 64,280 galaxies, 9,987 (15\%) are characterised as AGN, leaving a non-AGN sample of 54,293 galaxies.

One of the motivations for the compilation of this galaxy catalogue, was the study of X-ray source populations in \lq{}normal\rq{} (i.e., non AGN-hosting) nearby galaxies. Therefore, we also include the AGN classifications from \citet{She17} who studied galaxies that had been observed with {\it Chandra} at distances less then 50\,Mpc. In total, we obtain classifications for 716 galaxies.

Finally, we combine the classifications in a single estimate. For galaxies with classifications from only one of the two sources, we adopt them as they are. For galaxies both in the \sdss{} and \citet{She17} sample, they are characterised as AGN if they are classified as such by either of the two sources, otherwise as non-AGN. The \sdss{}, \citet{She17}, and the combined classifications are all provided in the catalogue (for 64,280, 716, and 64,910 objects respectively, leaving 139,823 galaxies without classification).

\section{Discussion}
\label{txt:discussion}

\subsection{Comparison with other catalogues}
\label{txt:comparisons}

Out of the available all-sky galaxy catalogues only the {\it Galaxy List for the Advanced Detector Era} ({\it GLADE}; \citealt{Dalya18}), the {\it Mangrove} \citep{Ducoin20}, and the {\it Census of the Local Universe} ({\it CLU}; \citealt{Gehrels16,Cook19}) are similar in scope (i.e., offer multi-wavelength photometry and galaxy characterisation) as the \hec{}.

The {\it GLADE} galaxy catalogue provides coordinates, distances, and photometry in the $B$, $J$, $H$, and $K$-bands by cross-matching five catalogues: \hyper{}, \twomass{}-XSC, {\it GWGC}, the {\it 2MASS photometric redshift catalogue}, and \sdss{}-DR12Q. Without an explicit limit on $z$, it is an ideal tool for low-redshift cosmology, and studies of distant transient events such as long GRBs. A recent extension of the {\it GLADE} is the {\it Mangrove} catalogue, which provides \mass{} estimates via mid-IR photometry obtained by cross-matching the {\it GLADE} and the {\it AllWISE} catalogue.

Over the distance range covered by the \hec{}, the completeness of {\it GLADE} and {\it Mangrove} in terms of the $B$-band luminosity is similar to that of the \hec{} (cf., fig. 2 in \citealt{Dalya18} and \autoref{fig:completeness}).
However, the two catalogues do not include size information for the galaxies, limiting their usability for the association of host galaxies with sources from serendipitous and all-sky surveys \citep[e.g.,][]{Webb20}. Another important difference between the \hec{} and the {\it GLADE} or {\it Mangrove} is that the \hec{} provides robust distances for local Universe galaxies\footnote{While the use of photometric redshifts in {\it GLADE} and {\it Mangrove} provides distances estimates for distant galaxies without spectroscopic measurements, their typical uncertainty of $\Delta z{=}0.015$ \citep{Dalya18} is prohibitive for galaxies in the local Universe ($z<0.047$). In addition, redshift-independent distances are provided through the {\it GWGC} catalogue which is limited to 100\,Mpc.}, SFRs based on a wide suite of indicators, as well as, homogenised SFRs that bridge the systematic differences between the individual indicators, and integrated \twomass{} and \wise{} photometry for nearby galaxies.

The {\it CLU} catalogue has been progressively constructed since 2016 to aid the identification of GW hosts \citep{Gehrels16}, and provide a census of emission-line galaxies with $D{<}200\Mpc$ using new observations \citep{Cook19}. Including information from the \ned{}, \hyper{}, Extragalactic Distance Database, \sdss{}-DR12, 2dF Galaxy Redshift Survey, the Arecibo Legacy Fast ALFA, \galex{}, and \wise{}, it provides multi-wavelength data, SFRs and \mass{} based on \wise{} photometry.
However, for studies of nearby galaxies, the {\it CLU} has the same limitations as in the case of {\it GLADE}: it does not provide size information on the sample galaxies, and the \wise{}-based photometry is problematic for nearby, extended objects (\S\ref{txt:wise}).

Concluding, the \hec{} provides robust distances (an important parameter for nearby galaxies; see \S\ref{txt:distances}), and additional data that are not readily-available in the other catalogues: reliable homogenised SFRs, metallicities, as well as morphological and AGN classifications.

\subsection{Completeness}
\label{txt:completeness}

\begin{figure*}
    \centering
    \includegraphics[width=\textwidth]{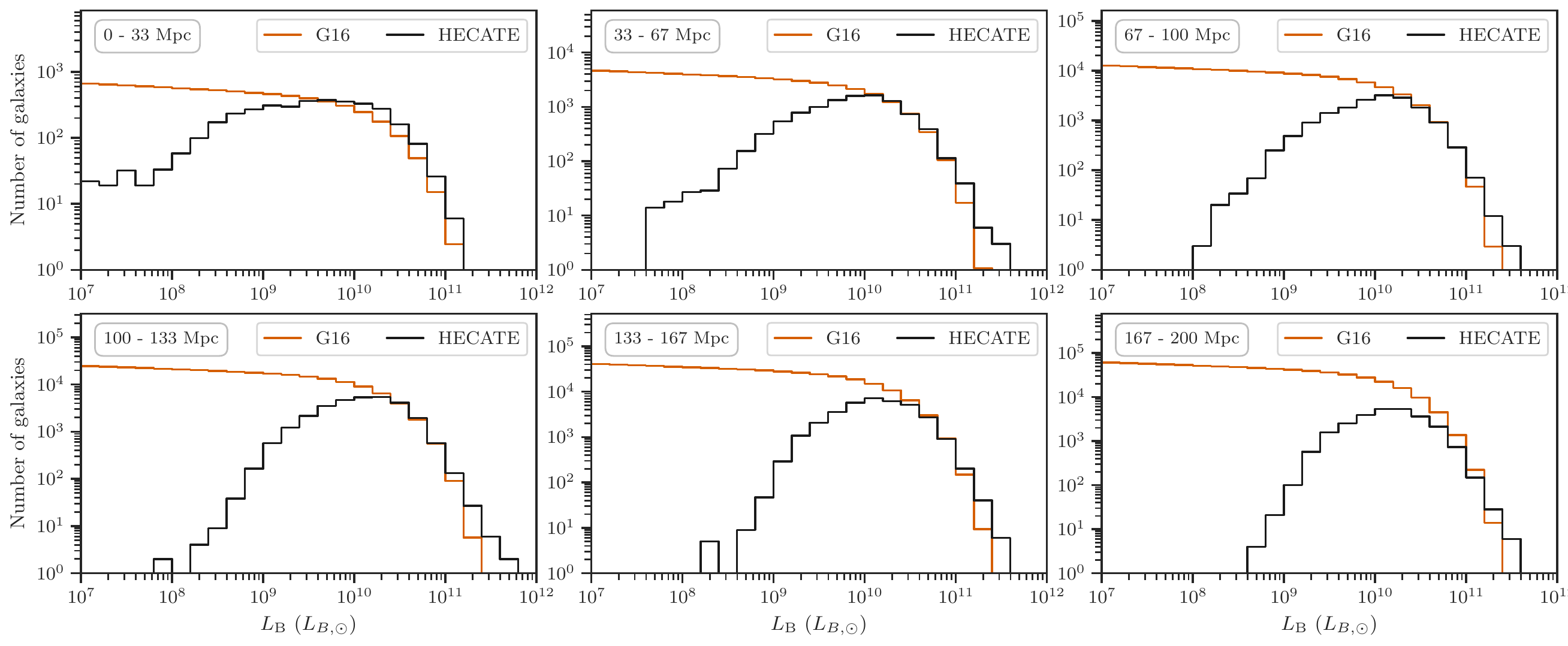}
    \caption{The distribution of the $B$-band luminosities of the galaxies in the \hec{} at different distance ranges (top left boxes), and comparison against the expectation from the $L_{\rm B}$ LF in \citet{Gehrels16}.}
    \label{fig:schechter}
\end{figure*}

\begin{figure}
    \centering
    \includegraphics[width=\columnwidth]{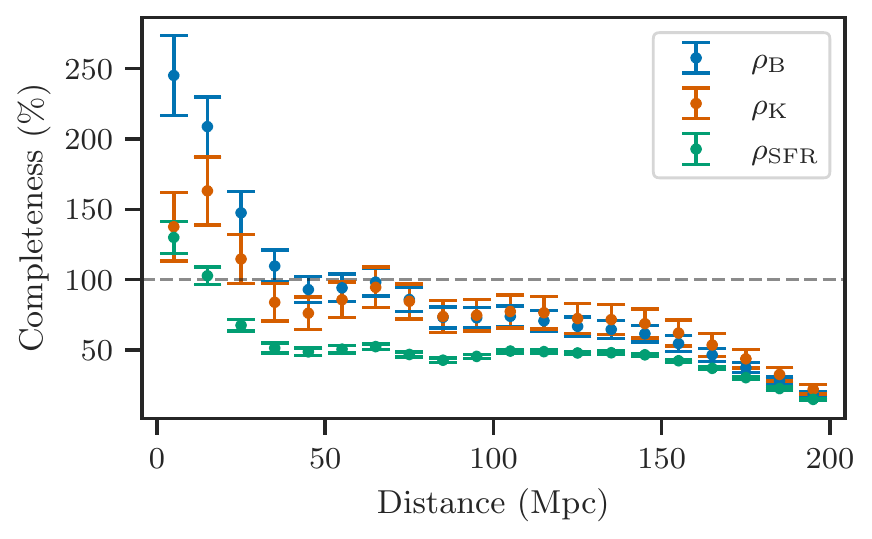}
    \caption{The completeness of the \hec{} in terms of the included $B$-band, $K_s$-band, and SFR density with respect to the expectation from observational estimates. The completeness is between 50\% and 100\% within 150\Mpc{}. At small distances, the completeness exceeds 100 per cent because of the overdensity in the neighbourhood of the Milky Way \citep[cf.][]{Gehrels16}.}
    \label{fig:completeness}
\end{figure}

The completeness of the \hec{} cannot be robustly calculated due to the unknown selection function of the \hyper{}, which is further complicated by the selection effects introduced by the other catalogues that it is cross-correlated with. However, we can obtain an estimate of the completeness by comparing the distribution of $B$-band luminosities with the expectation from the galaxies LF, following the approach of \citet{Gehrels16} and \citet{Dalya18}. Using the same Schechter LF as in the aforementioned papers\footnote{$\Phi{=}1.6{\times}10^{-2}\,h^3 \Mpc{}^{-3}$, $a{=}{-}1.07$, and $L_\star{=}1.2{\times} 10^{10}\,{\rm L_{B,\odot}}$ \citep[cf.][]{Gehrels16}. We adopt $h{=}0.7$ as an intermediate value between the published $H_0$ calibrations.}, we compute the expected number of galaxies in different bins of luminosities and distances, shown in orange in \autoref{fig:schechter}, which we compare with the number of galaxies in the \hec{} in the respective bins (black). We find that the \hec{} is complete down to $L_{\rm B}{\sim}10^{9.5} L_{B,\odot}$ at distances less than 33\Mpc{}, and down to $L_{\rm B}{\sim}10^{10} L_{B,\odot}$ at $67{<}D{<}100\Mpc{}$. However, at distances greater than $167\Mpc{}$ the \hec{} suffers by incompleteness even at the high-end of the LF.

Since many applications of the \hec{} are related to the stellar content of the galaxies, we can quantify its completeness in terms of the ratio of the integrated $B$-band luminosity of galaxies at distance $D$, with respect to the mean $L_{\rm B}$-density of the local Universe. This approach has been followed in several studies of nearby samples of galaxies: \citet{Kopparapu08,White11,Gehrels16,Dalya18}. We adopt the mean $L_{\rm B}$ density of $\left(1.98\pm0.16\right)\times10^{8} \Mpc{}^{-3}$ \citep{Kopparapu08} that was used by the aforementioned works. To account for the different sources of uncertainties. we sample from the distributions of the various quantities involved in the computation (i.e., the mean $L_{\rm B}$-density, and the galaxy distances), and compute the completeness in bins of $10\Mpc{}$. This is performed for 10,000 iterations to obtain the mean and standard deviation of the completeness as a function of the distance. The $L_{\rm B}$-completeness is shown by blue error bars in \autoref{fig:completeness}. We find that the \hec{} is ${>}75\%$ complete in terms of the blue light at $D{<}100\Mpc{}$, and ${\sim}50\%$ at $D{\sim}170\Mpc{}$. The completeness above 100\% at small distances ($D{<}30\Mpc{}$) is the result of the over-density in the neighbourhood of the Milky Way.

Similarly, we calculate the completeness of the \hec{} in terms of the \mass{}. For this reason we perform the same exercise with the $K_s$-band luminosity, which is a tracer of the \mass{} of the galaxies. We adopt a $K_s$-band luminosity density of $5.8\times10^{8} h {\rm L}_{K,\odot}\Mpc{}^3$ \citep{Bell03}. The result is similar to the $L_{\rm B}$-completeness as shown by orange in \autoref{fig:completeness}, exhibiting both the excess at small distances and the cut-off at large distances.

The completeness in terms of the SFR is calculated in the same way (shown by green points in \autoref{fig:completeness}), adopting a local Universe SFR density of $0.015\,\sfrunit{}\Mpc{}^{-3}$ \citep{Madau14}, and using the homogenised SFR for the \hec{} galaxies. In this case, the \hec{} is incomplete at all distances in its regime, with ${\sim}50\%$ completeness at $30{<}D{<}150\Mpc{}$. The lower completeness in SFR with respect to the other parameters ($L_{\rm B}$ and $\mass{}$) stems from the fact that the \wise{}-based SFRs in the \hec{} do not have all-sky coverage since they are based on forced photometry on \sdss{} objects. Nevertheless, due to the all-sky coverage of \iras{} and despite its shallowness, it covers more than 50\% of the star-forming activity in the Galactic neighbourhood. Finally, the completeness might be overestimated with respect to estimates from SED methods, since the homogenised SFRs are systematically higher for galaxies in the low-SFR regime (see \autoref{fig:SFRcomp}).

\subsection{Limitations}
\label{txt:limitations}

The parent sample of the \hec{}, the \hyper{} database, includes objects and related data, from hundreds of surveys with different sky coverage and sensitivity limits. Therefore, the selection function of the \hyper{}, and as a consequence, that of the \hec{}, is intractable (\S\ref{txt:completeness}). Generalisations based on the provided galaxy compilation should be treated carefully.

At low distances ($D{\lesssim}20\Mpc{}$) peculiar velocities dominate the Hubble flow (\S\ref{txt:zdependent}). This is accounted for by the regression model for estimating distances based on the recession velocities of the galaxies, however the increased scatter reduces the accuracy of the inferred distances for velocities $\vvir{}{\lesssim{}}1,500\kms{}$ (\autoref{fig:v_model}). This can be remedied by measuring $z$-independent distances for the nearby galaxies. In addition, there are a few cases where distance measurements are significantly different from the Hubble-flow distance\footnote{e.g., NGC\,5434 is reported to have a distance of 3.8\,Mpc both in \nedd{}, and subsequently in the \hec{}, but its $z$ implies $D{\approx}70\Mpc{}$.}. The causes of these discrepancies are diverse and difficult to identify in most cases (e.g., problematic distances due to biases in distance indicators, wrong redshifts because of superimposed stars, typos, etc.) In the future, the methods for estimation of distances will include special treatment for such outliers.

Furthermore, the derived stellar population parameters are based on multi-wavelength data from combinations of surveys and calibrations. The statistical treatments presented in this paper (e.g., homogenisation of SFR estimates, fixed $M/L$ ratio for galaxies without $M/L$ estimate) provide estimates of stellar populations for a large fraction of galaxies in the local Universe. While this allows for statistical studies of large galaxy samples, or quick searches for objects of interest, more accurate methods ought to be preferred when focusing on individual galaxies.

The IR-based SFR estimates are based on calibrations that assume \lq{}normal\rq{} star-forming galaxies. In the case of quenched, early-type galaxies, the SFRs may be overestimated \citep[e.g.,][]{Hayward14}. Indicators based on optical-UV SED analysis could be more reliable for these galaxies.

\begin{figure}
    \centering
    \includegraphics[width=\columnwidth]{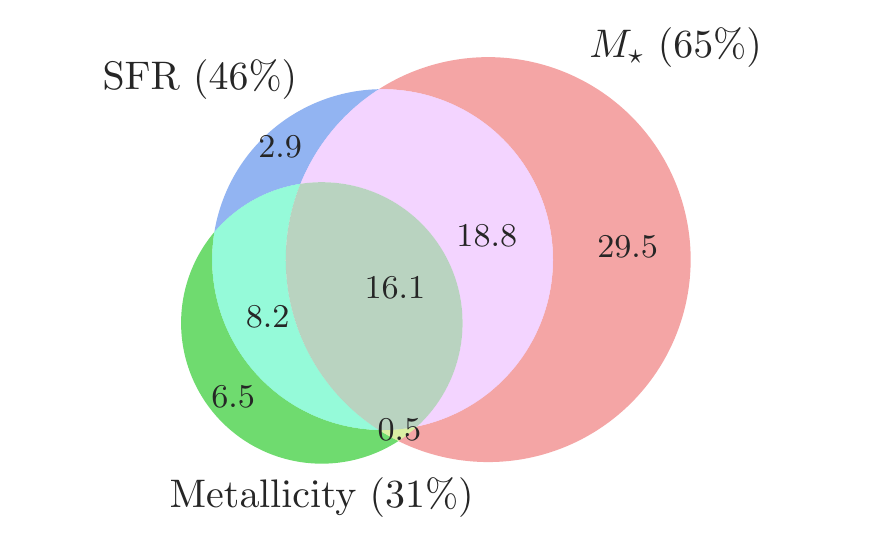}
    \caption{Venn diagram of the coverage of the stellar population parameters, SFR, \mass{}, and metallicity, in the \hec{}. The per cent coverage for each is reported next to its label, while the numbers in the coloured areas denote the percentage for the different combinations of parameters.}
    \label{fig:venn_spp}
\end{figure}

One of the most important limitations of the \hec{} is its non-uniform coverage in terms of the SFR and \mass{}. In \autoref{fig:venn_spp} we show the coverage of stellar population parameters in the \hec{}. SFR, \mass{}, and metallicity estimates are available for 46\%, 65\%, and 31\% of the galaxies, respectively.
Currently, the \wise{} photometry is obtained through the forced photometry catalogue of \citet{Lang16} which is limited to the \sdss{} footprint. While this is driven by the need for accurate photometry for the extended galaxies (which are the majority of the \hec{} galaxies), it leaves a significant fraction of the sample without sensitive IR photometry that could provide reliable and uniform SFR measurements. For specific cases, this limitation can be remedied by including in the analysis data from additional catalogues. In a future version of the \hec{} we will apply the forced photometry method to all galaxies in the \hec{}, thus providing robust stellar population parameters. In addition, incorporation of additional photometry and spectroscopy from other surveys (e.g., Pan-STARSS, LAMOST; \citealt{Chambers16}; \citealt{Deng12}), will increase the multi-wavelength, AGN classification, and metallicity coverage of the \hec{}. This will also allow the computation of SED fits, that will provide additional SFR and \mass{} estimates for the galaxies.

\subsection{Applications}
\label{txt:applications}

The motivation for creating an all-sky galaxy catalogue with positions, sizes, multi-wavelength data, and derived parameters (e.g., SFR, \mass{}, metallicity) was to enable several applications relying on the initial characterisation of sources in the context of the host galaxy, or identifying counterparts of transient events for follow-up observations. In this section we outline some specific use cases.

\subsubsection{Application to all-sky and serendipitous surveys}
\label{txt:allsky}

The distance limit of the \hec{} and the large array of the information it provides make it an ideal sample for designing wide-area multi-wavelength surveys, or characterising sources within. For example, it can form the baseline sample for realistic simulations of the data expected to be obtained with future surveys \citep[see][ for an application to the {\it eROSITA} survey]{Basu20}, but also it can be a reference sample for the initial characterisation of newly-identified sources (e.g., with Dark Energy Survey: \citealt{Flaugher05}; Large Synoptic Survey Telescope, LSST survey: \citealt{Ivezic19}).

A demonstration of the potential of the \hec{} is given in \citet{Kovlakas20}, a comprehensive study of ultraluminous X-ray sources in the local Universe based on the {\it Chandra Source Catalog 2.0} \citep{Evans10}. The positional and size information available in \hec{} allowed the association of X-ray sources with their host galaxies and the robust estimation of the fraction of interlopers. In addition the SFR, \mass{}, and metallicity information was used to derive scaling relations between the ULXs and the stellar populations in their host galaxies. The special treatment of nearby galaxies (e.g., extended photometry) in the \hec{} was essential for the science in this project since the target sample was limited in a volume out to 40\Mpc{}. Similarly, the combination of \hec{} with {\it XMM-Newton} has been is the basis for the largest study of the X-ray scaling relations of galaxies (Anastasopoulou et al., in preparation), and the largest {\it XMM-Newton} census of ULXs in nearby galaxies up to date (Bernadich et al., in preparation).

\subsubsection{Application in search of EM counterparts to GW sources}
\label{txt:gwloc}

All-sky galaxy catalogues are crucial for the timely identification of electromagnetic (EM) counterparts to GW sources \citep[e.g.,][]{Nissanke13,Gehrels16}. This is a key step for constraining their nature (e.g., \citealt{LigoVirgo17}), understanding the formation and evolution of their progenitors \citep[e.g.,][]{Kalogera2007,Abbot17}, or even using them as standard \lq{}sirens\rq{} to measure the Hubble constant \citep[e.g.,][]{Schutz86,Chen18,AbbottNature}.

The poor localisation of GW sources by the contemporary GW detectors (${\gtrsim}100\rm{\, deg^{2}}$; \citealt{AbbottLocal}), makes the search for EM counterparts a daunting task. The adopted solution is to perform targeted follow-up observations of a list of potential hosts prioritised based on properties such as their distance, or the parameters of their stellar populations \citep[e.g.,][]{Kanner08,Nuttall10,Gehrels16,Arcavi17,Kasliwal17,GCN21516,GCN21519,DelPozzo18,Yang19,Salmon20,Wyatt20}.

This approach has already led to the compilation of galaxy catalogues that provide in addition to positions and distances, photometric information \citep[as proxies to SFR; e.g.,][]{Kopparapu08,White11,Gehrels16}, or directly SFR and \mass{} determinations \citep[e.g.,][]{Dalya18,Cook19,Ducoin20}. This is driven by models which predict that GW populations scale with SFR \citep[e.g.,][]{Phinney91}, and/or \mass{} \citep[e.g.,][]{Mapelli17,Artale19,Toffano19,Adhikari20}.
However, these catalogues lack information on metallicity which can be an important factor in the GW rates \citep[e.g.,][]{OShaughnessy17,Mapelli18,Artale19,Neijssel19,Artale20a,Bavera20}.

The \hec{}, having a distance limit (${\sim}200\Mpc$) that is sufficient for searches of EM counterparts to GW sources from binary neutron stars (BNS) until the mid-2020s \citep[e.g.,][]{Sensitivity20},
and providing stellar population parameters, can be used for assigning likelihoods to putative GW hosts for observational follow-up campaigns. In this section, we use as an example the GW event GW170817, the only case of verified EM counterpart of a BNS, to
\begin{enumerate}
\item 
    illustrate the use of the \hec{} in producing priority lists of galaxies for EM counterpart searches,
\item 
    study the effect of the different pieces of information (direction, distance, and stellar population parameters) in the prioritisation of host candidate galaxies,
\item 
    assess, post facto, the ability of various schemes in giving high priority to the host galaxy of GW170817, \ngc{}.
\end{enumerate}

The priority lists are the result of ordering the galaxies based on their probability of being the hosts,
\begin{equation}
    P \propto P_{\rm 3D} \times G_{\rm intr}
\end{equation}
where $P_{\rm 3D}$ is the volume-weighted probability given the position and distance of the galaxy, and $G_{\rm intr}$ is a factor (or grade) which scales with the probability for a galaxy to host a GW event given its intrinsic properties (e.g., \mass{} or SFR proxy, or merger rate).

As a first step, we acquire the HEALPix map \citep{Gorski05} produced by BAYESTAR \citep{Singer16} which contains the 2-D localisation probability, i.e. the probability that the GW event is on a specific direction of the sky, and the corresponding distance probability distribution.
By cross-matching the \hec{} with the HEALPix map we find 2,249 candidate host galaxies in the 99.9\% region of GW170817 (based on the 2-D localisation probability). As the \lq{}directional\rq{}, namely the 2-D probability of the galaxy, $P_{\rm 2D}$, we assign the value of the HEALPix pixel which contains the centre of the galaxy. 
The 3-D probability, $P_{\rm 3D}{=}P_{\rm 2D}{\times}P_{\rm d}$, is computed by combining the $P_{\rm 2D}$ with the GW event distance probability density ($P_{\rm d}$) for the corresponding pixel in the HEALPix map, and the distance of the galaxy in the \hec{}.

Subsequently, the 3-D probabilities are multiplied by \lq{}astrophysical\rq{} terms ($G_{\rm intr}$) which are assumed to be proportional to the merger rate of BNSs, and therefore the probability of a merger. The astrophysical terms are generally parametrised in terms of the $L_{\rm B}$ (cf. \citealt{Arcavi17,Salmon20}), stellar mass (cf. \citealt{Ducoin20}), and the theoretical predictions on the merger rate of BNS, as a function of different combinations of the stellar population parameters:
(i) $n(\mass{})$, 
(ii) $n(\mass{}, \mathrm{SFR})$, and 
(iii) $n(\mass{}, \mathrm{SFR}, Z)$,
based on the results of \citet{Artale20a} for $z=0$ (cf. their table 1), where $Z$ is the metallicity of the galaxy.

Since the three stellar population parameters may not be known for all galaxies in the HEALPix map, we also use a \lq{}combined\rq{} estimate, where the appropriate merger rate is used depending on the available information:
\begin{equation}
    n_{\rm comb} = 
    \begin{cases}
        n(\mass{}, \mathrm{SFR}, Z) & \text{if \mass{}, SFR and $Z$ are defined} \\
        n(\mass{}, \mathrm{SFR}) & \text{if \mass{} and SFR are defined} \\
        n(\mass{}) & \text{if \mass{} is defined}
    \end{cases}.
\end{equation}
 
Finally, in order to include in our analysis galaxies without \mass{} estimates (for which $n$ cannot be inferred), we employ the weighting scheme of \citet{Ducoin20}:
\begin{equation}
    P_{\rm Du} \propto P_{\rm 3D} \left(1 + \alpha n_{\rm comb}\right), 
    \quad \text{where} \quad
    \alpha = \frac{\sum P_{\rm 3D}}{\sum P_{\rm 3D} n_{\rm comb}}.
\end{equation}

The quantities, $P_{\rm 2D}$, $P_{\rm 3D}$, $P_{\rm 3D}{\times}L_{\rm B}$, $P_{\rm 3D}{\times}\mass{}$, $P_{\rm 3D}{\times}n(\mass{})$, $P_{\rm 3D}{\times}n(\mass{},{\rm SFR})$, $P_{\rm 3D}{\times}n_{\rm comb}$, and $P_{\rm Du}$, are used to produce priority lists of the host galaxy candidates, to test the aforementioned schemes for prioritising candidate host galaxies (\autoref{tab:priorities}). The scheme which accounts for the metallicity dependence of the merger rate is omitted due to lack of metallicity estimates in the sky region of the GW event.

\begin{table}
    \centering
    \caption{Prioritisation lists of host galaxy candidates (first five sources), and computed probabilities based on different schemes. The true host (bold text), \ngc{}, is successfully recovered as first or second most probable host galaxy once the astrophysical information is accounted for.}
    \label{tab:priorities}
    \begin{tabular}{llll}
    \hline
    Galaxy\textsuperscript{(i)} & $P_{\rm 2D}$ &  Galaxy\textsuperscript{(ii)} & $P_{\rm 3D}$ \\\hline
    PGC4690279 & 0.002   &  ESO508-004 & 0.054\\
    PGC3799401 & 0.002   &   ESO575-055 & 0.051\\
    PGC3798804 & 0.002   &   ESO575-053 & 0.049\\
    PGC4690296 & 0.002   &   PGC4692149 & 0.045\\
    PGC4690280 & 0.002   &   PGC169673  & 0.045\\
    \hline\hline
    Galaxy & $P_{\rm 3D}{\times}L_{\rm B}$ &  Galaxy & $P_{\rm 3D}{\times}\mass{}$ \\\hline
    {\bf NGC4993}    & 0.096  &  {\bf NGC4993} & 0.163 \\
    ESO508-019 & 0.079    & NGC4830 & 0.148 \\
    IC4197     & 0.074    & IC4197  & 0.119 \\
    NGC4830    & 0.073    & NGC4970 & 0.115 \\
    NGC4970    & 0.072    & NGC4968 & 0.103 \\
    \hline\hline
    Galaxy & $P_{\rm 3D}{\times}n(\mass{})$ &  Galaxy & $P_{\rm 3D}{\times}n(\mass{},\rm SFR)$ \\\hline
    {\bf NGC4993}    & 0.164  &    NGC4968 & 0.180 \\
    NGC4830    & 0.151  &    {\bf NGC4993} & 0.135 \\
    IC4197     & 0.121  &    NGC4830    & 0.102 \\
    NGC4970    & 0.117  &    IC4187     & 0.100 \\
    NGC4968    & 0.102  &    NGC4970    & 0.087 \\
    \hline\hline
    Galaxy & $P_{\rm 3D}{\times}n_{\rm comb}$ &  Galaxy & $P_{\rm Du}$ \\\hline
    NGC4968    & 0.180  &    NGC4968    & 0.082\\
    {\bf NGC4993} & 0.135  &  {\bf NGC4993}    & 0.071\\
    NGC4830    & 0.102  &    NGC4830    & 0.050\\
    IC4197     & 0.100  &    IC4197     & 0.049\\
    NGC4970    & 0.087  &    NGC4970    & 0.044\\
    \hline
    \end{tabular}
    
    \begin{minipage}{0.45\textwidth}
    \textbf{Notes}:
    \textsuperscript{(i)} The rank of \ngc{} is 461.
    \textsuperscript{(ii)} The rank of \ngc{} is 7.
    \end{minipage}
\end{table}

\begin{figure*}
    \centering
    \begin{subfigure}{0.475\textwidth}
    \includegraphics[width=\textwidth]{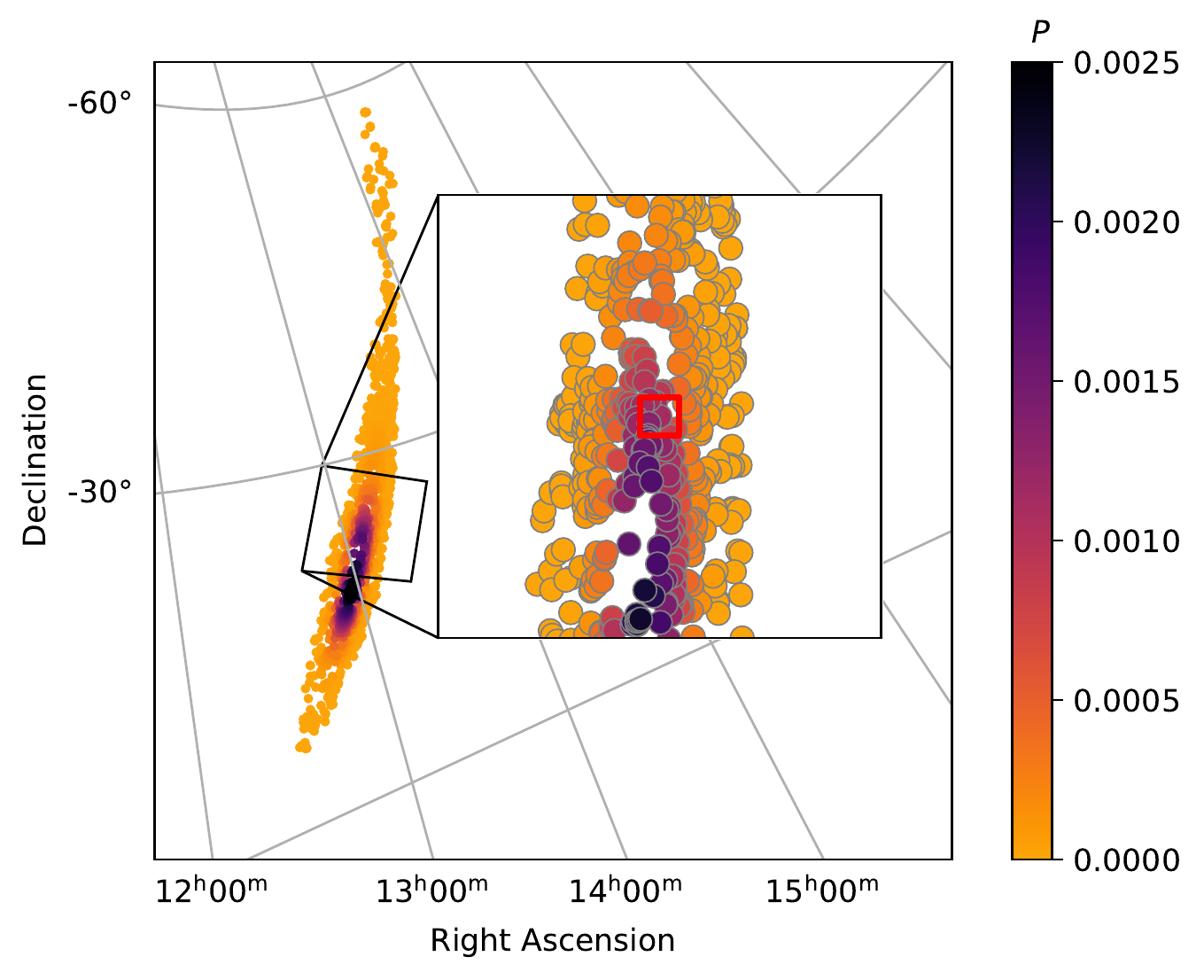}
    \caption{2-D probability}
    \end{subfigure}
    \begin{subfigure}{0.475\textwidth}
    \includegraphics[width=\textwidth]{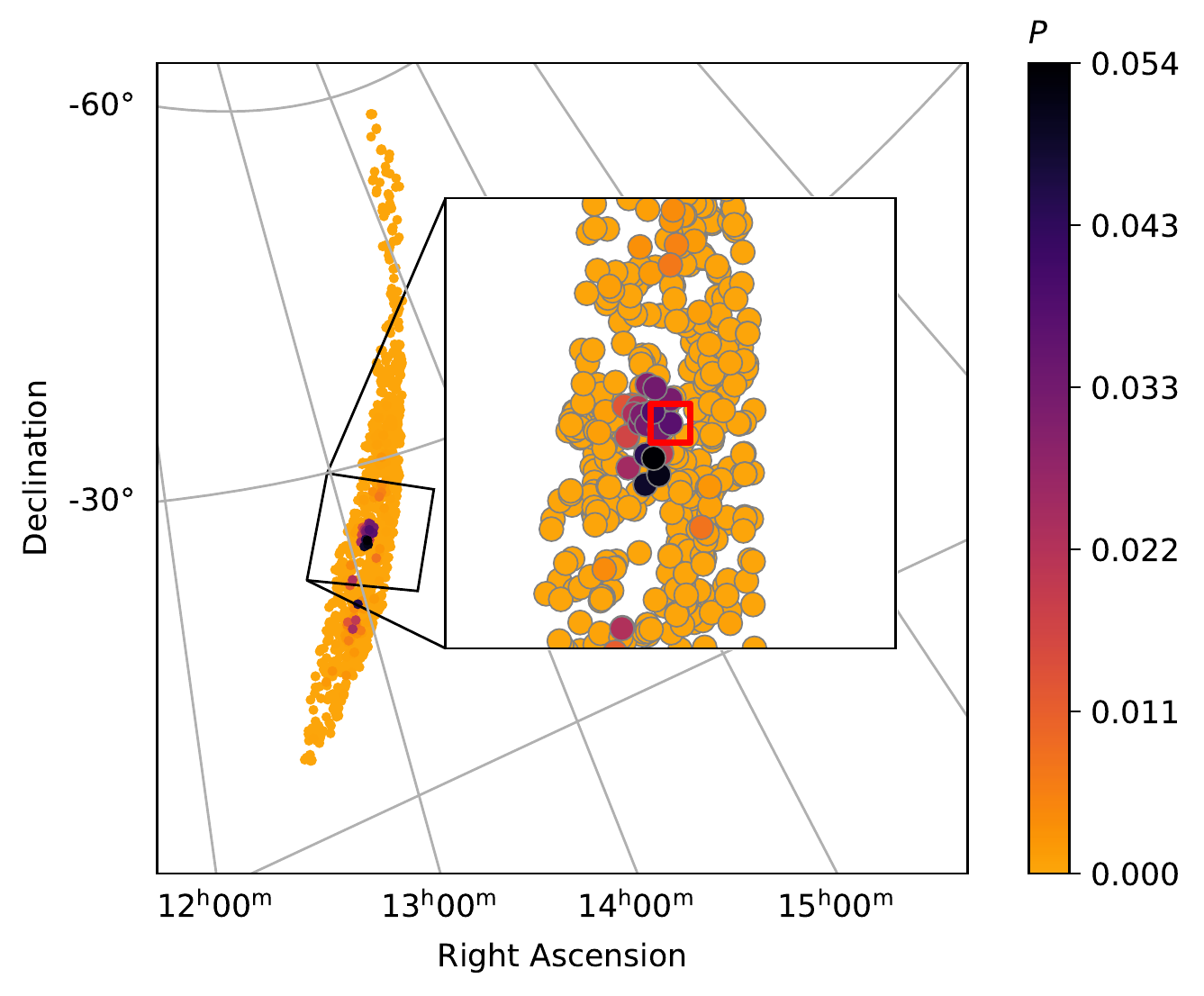}
    \caption{3-D probability}
    \end{subfigure} 
    \begin{subfigure}{0.475\textwidth}
    \includegraphics[width=\textwidth]{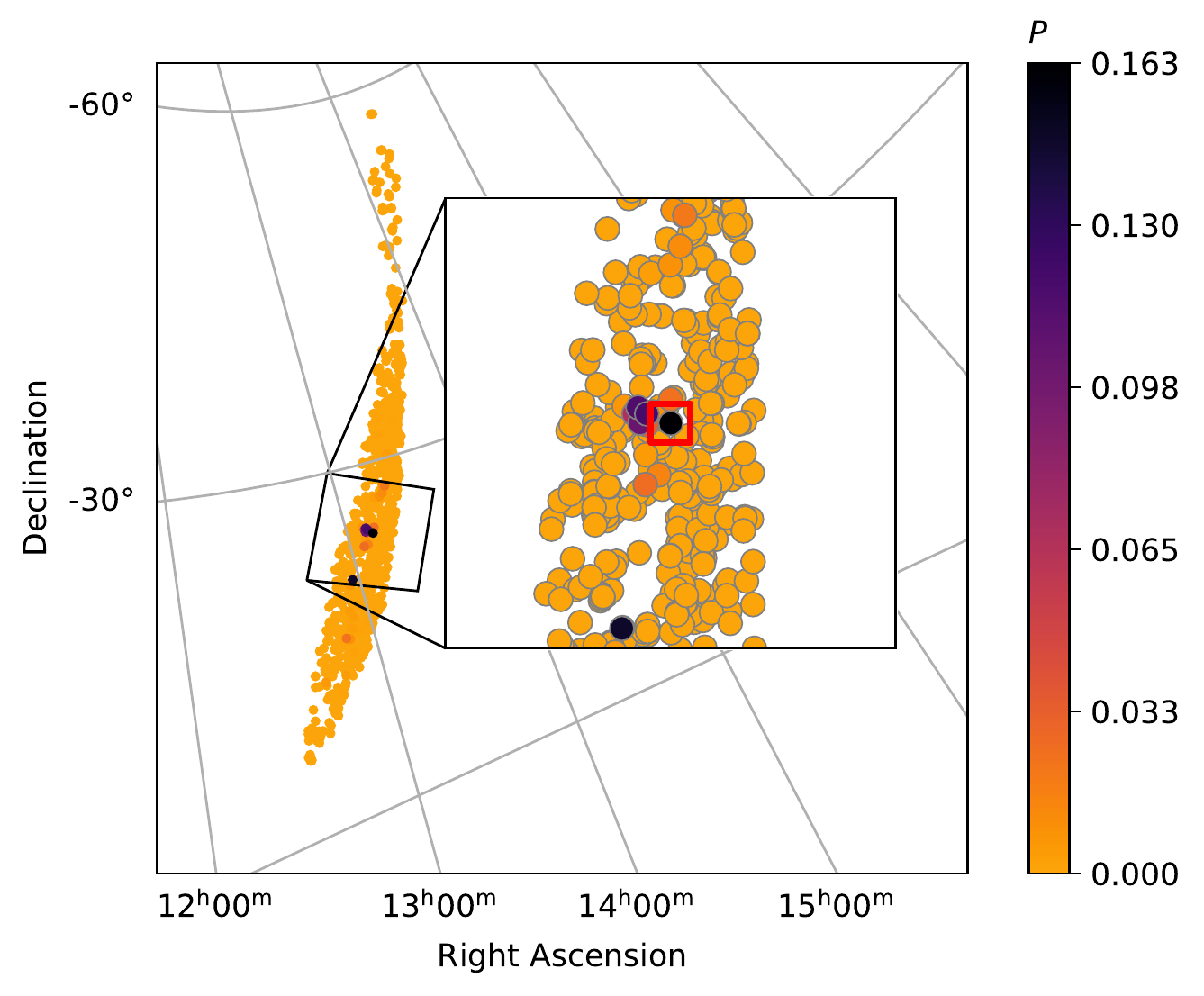}
    \caption{3-D probability combined with $\mass{}$}
    \end{subfigure}
    \begin{subfigure}{0.475\textwidth}
    \includegraphics[width=\textwidth]{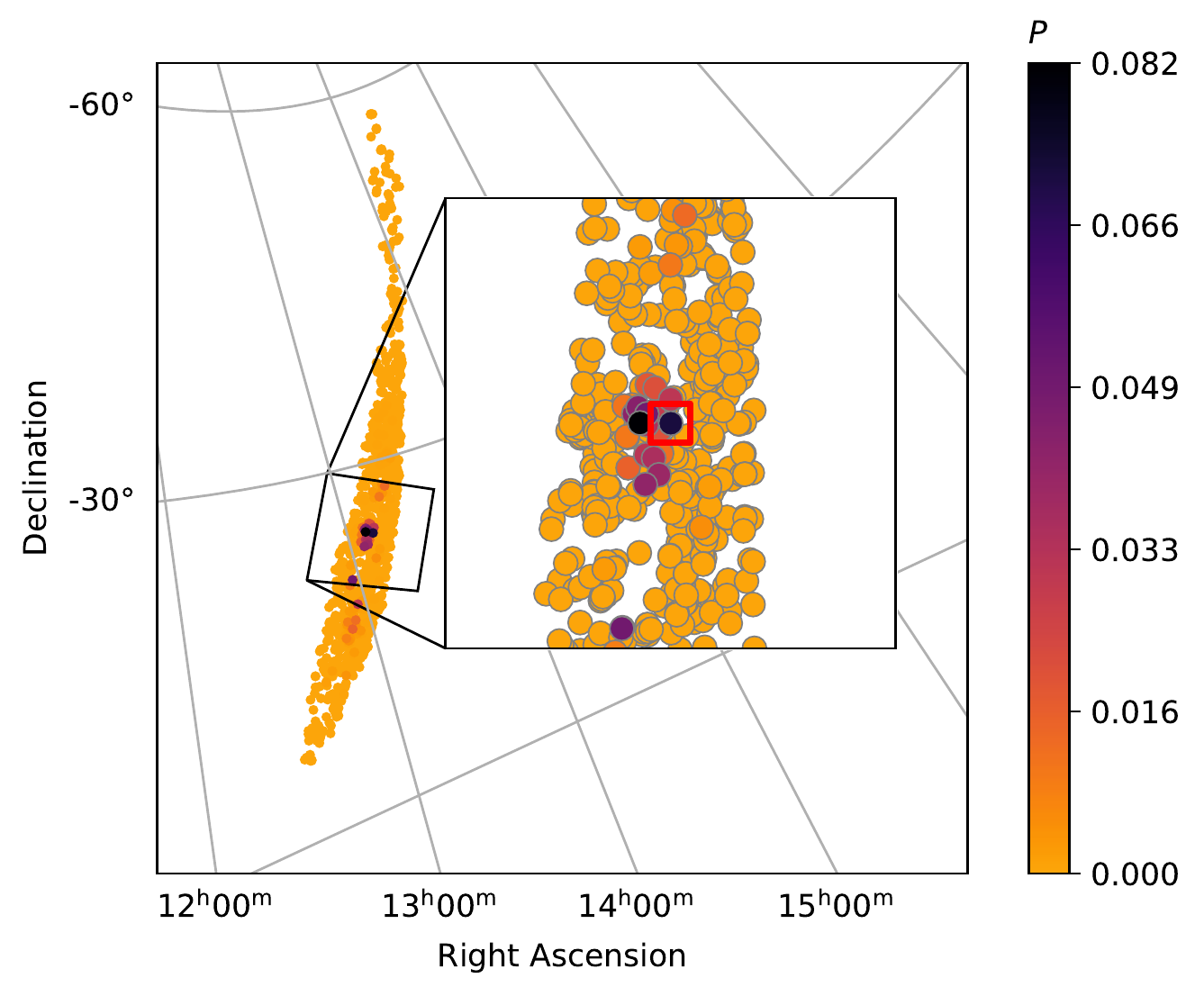}
    \caption{3-D probability combined with merger rate as a function of \mass{} and SFR, using the \citet{Ducoin20} weighting scheme.}
    \end{subfigure}
    \caption{
        \textbf{(a)} 
            The sky distribution of the 2,248 galaxies in the \hec{} that lie inside the 99.9\% confidence region of the 2-D localisation map for the GW170817 event. The inset zooms in the region of the galaxy \ngc{} (red square). The colour indicates the normalised 2-D probability across all the host candidates.
        In \textbf{(b)}, 
            the 3-D probability is used accounting for the distance estimates at each direction in the sky.
        In \textbf{(c)}, 
            the 3-D probability is multiplied by the stellar mass. Similarly, in \textbf{(d)} the 3-D probability is multiplied by the merger rate as computed by the fits in \citet{Artale20a}, and the weighting scheme of \citet{Ducoin20} is used to allow galaxies without stellar population parameters estimates to enter in the prioritisation list.
        Note that in (a) and (b) the contrast of the highest-probability galaxies to the rest is small. The introduction of the \lq{}astrophysical\rq{} terms in (c) and (d), gives prominence to \ngc{}.
        \label{fig:gwloc}
    }
\end{figure*}

We find that \ngc{} is given the highest priority by the schemes involving the $L_{\rm B}$ or \mass{}, and second priority for those also involving the SFRs\footnote{Following \S\ref{txt:limitations}, because GW170817 falls outside the \sdss{} footprint, for this application we supplement the \iras{} photometry with mid-IR photometry from the {\it AllWISE} catalogue.}. Except for the priority lists based only on the 2-D or 3-D position, the lists feature the same top-five galaxies as in the first prioritisation list published after the GW170817 alert \citep{Kasliwal17} based on the {\it CLU} catalogue \citep{Gehrels16,GCN21519}. The same holds for the top-three galaxies reported in (i) \citet{Artale20b} who use \mass{} and SFR estimates from the {\it Mangrove} catalogue \citep{Ducoin20}, and (ii) in \citet{Yang19} who used the $B$-band luminosity from {\it GLADE} \citep{Dalya18} as the \lq{}astrophysical\rq{} term. The top-ranked \hec{} galaxies based on the different prioritisation schemes, agree to a high degree with the results of the same schemes in \citet{Ducoin20} using the {\it Mangrove} catalogue.

While in the case of GW170817 there is no significant difference between the use of $B$-band luminosity, \mass{} or the fits with \mass{} and SFR (\ngc{} was always first, or second with a small difference in the probability), refined prioritisation schemes will be important for the quick identification of EM counterparts of future BNS coalescence signals with poorer localisation or at higher distances. The \hec{}, making readily available a large set of intrinsic properties for the candidate host galaxies, offers versatility in the choice, design, and assessment of different priority schemes.

We note that the practice of initially narrowing down the galaxy sample by deciding on a confidence region based on the 2-D probability, increases the risk of missing the true host. This is indicated by the rank of \ngc{} in the $P_{\rm 2D}$-based priority list (461), and the fact that total 2-D probability of galaxies closer to the centroid (a few degrees from \ngc{}) is as high as 75\%. We suggest using the full galaxy catalogue together with priority schemes involving distance (and astrophysical information where possible). For example, the inclusion of the distance information in $P_{\rm 3D}$, promotes \ngc{} to the 7th position, and shifts the centroid by a few degrees (see panel (b) of \autoref{fig:gwloc}), a consequence of the non-homogeneity of the Universe at the distance of the event (out of the 15 galaxies in \autoref{tab:priorities} for schemes with 3-D positional term, 10 are considered members of the NGC4970 group; \citealt{Kourkchi17}).

Finally, as we show in \autoref{fig:SFRcomp}, infrared estimates of the SFR (as those provided by the \hec{}) can be overestimated up to ${\sim}2\,\rm dex$ with respect to SED estimates, in the case of low-SFR galaxies (such as NGC\,4993). Therefore, grading schemes that account for the expected SFR of the host galaxy may overestimate the probability of a low-SFR galaxy to host the GW event. As an example, the formula given in \citet{Artale20a} for the number of GW events for a BNS event in a galaxy at $z=0.1$ (which is the case most sensitive to the SFR), an overestimation by two orders of magnitude in the SFR, results to an overestimation by a factor of ${\sim}4$ in the BNS rate, and consequently the assigned probability of the galaxy as a host to an event. For GW detections in the \sdss{} footprint, the use of SED-based stellar population parameters (also given in the catalogue), is advised.

\subsubsection{Application in short gamma-ray bursts}

Another manifestation of BNS mergers are short GRBs \citep[sGRBs; e.g.][]{Tanvir13} as it has been shown by the association of GW170817 to GRB170817A \citep{Goldstein17}. The identification of the host galaxies of sGRBs is important for two reasons: 
    (a) connecting their populations with the star-formation history of their host galaxies, and 
    (b) measuring the displacement of the GRBs from their host galaxies.
The former is key for modelling the evolutionary paths of sGRBs and their cosmological evolution \citep[e.g.,][]{Leibler10,Selsing18,Abbot17}. The latter is important for constraining the effect of kicks in the populations of sGRBs \citep[e.g.,][]{Zevin19}. and studying the enrichment of the interstellar medium in r-process elements \citep[e.g.,][]{Andrews19}. The \hec{} can provide the initial information required to quickly associate a GRB with their host galaxy which is also important for prompt follow-up observations.

\subsubsection{Localisation of neutrino and cosmic-ray sources}

In the case of neutrino events, the large error-box of their localisation poses the same challenges as the GW detections \citep[e.g.,][]{Krauss2020}. Therefore, a catalogue of galaxies, which is as complete as possible, can be a valuable resource for the identification of their origin when one could follow a similar approach as the prioritised host list developed for GW events (\S\ref{txt:gwloc}).
Despite of the lack of distance estimates in the case of neutrino events, which can significantly affect the numbers and locations of the host candidates (compare panels (a) and (b) in \autoref{fig:gwloc}), the availability of multi-wavelength data, stellar population parameters, and nuclear activity classifications in the \hec{} is particularly useful for weighting the candidate galaxies according to their astrophysical properties since proposed extragalactic neutrino sources may be star-forming galaxies, AGN etc. \citep[see][and references therein]{Ahlers14,IceCube18}. In addition, future observations with a combination of catalogues such a the \hec{}, may aid in constraining the correlation of the neutrino emission and host galaxy properties.

The same holds for the case of cosmic-ray detections, which also have very large error-circles \citep[e.g.,][]{Aab15}. Recent studies of anisotropy in the arrival direction of high-energy cosmic rays, and the existence of a dipole at high Galactic latitude, indicate that their origin is neither exclusively Galactic, nor cosmological \citep[e.g.,][]{Auger17}. Since the Greisen-Zatsepin-Kuzmin effect limits the propagation of high-energy cosmic rays to ${\lesssim}100\Mpc{}$ \citep[e.g.,][]{Bhattacharjee00}, the \hec{}, as an all-sky galaxy catalogue at this distance range, can be used for the detailed study of their origin \citep[e.g.,][]{Abreu10,He16}.

\subsubsection{Applications in transient astronomy}
\label{txt:transient}

In the following paragraphs we outline potential applications of the \hec{} in various other fields of transient and multi-messenger astrophysics.

Tidal disruption events (TDEs) are typically witnessed as an outburst in X-ray or optical wavelengths resulting from accretion of the material shredded from a star under the effect of the tidal field of a supermassive black-hole (SMBH). Such events are expected to be routinely detected in the {\it eROSITA} all sky X-ray survey, and the {\it LSST} optical survey. 
\hec{} can provide the basis for the quick identification of the host of such an event and its basic properties. In particular, information on the distance, the presence of an AGN, and the velocity dispersion (used to initially estimate the SMBH mass; all available in the \hec{}) are valuable for a quick interpretation of transient events \citep[e.g.,][]{French20}.

A value-added catalogue providing robust distances and stellar population parameters is also useful for the characterisation and study of the populations of transient events observed in on-going or future multi-wavelength surveys. For example, {\it LSST} is expected to provide a host of supernov\ae{} every night. The association of these events with a catalogue like \hec{} will facilitate systematic studies of their populations in the context of their host galaxies (e.g., \mass{}, SFR, and most importantly metallicity; e.g., \citealt{Greggio19}). These pilot studies can be used to effectively plan more focused follow-up observations. The same holds for the identification of hosts of fast radio bursts \citep[e.g.,][]{Marcote20}.

\section{Conclusions and future work}
\label{txt:summary}

We present a new catalogue of galaxies which includes all known galaxies within a distance limit of $D{\lesssim}200\Mpc{}$. We
\begin{enumerate}
\item
    base our sample on the \hyper{} database, incorporating 204733 galaxies with radial velocity ${\lesssim{}}14,000\kms{}$,
\item 
    use all available distance measurements for the sample to get robust redshift-independent distances, which are preferred over recessional velocity based estimates for galaxies in the local Universe, for as many galaxies as possible (10\%),
\item 
    compute redshift-dependent distances for the rest of the galaxies (90\%) which are consistent with the redshift-independent distances (Kernel Regression method), while quantifying their uncertainties due to the unknown peculiar velocity component,
\item 
    incorporate integrated multi-band photometry with special treatment for nearby and/or extended galaxies,
\item 
    derive SFRs, \mass{}, metallicities, and nuclear activity classifications utilising the best available information for each galaxy, 
\item 
    offer five different IR-based SFR indicators, as well as, a homogenised SFR indicator, while providing all the necessary information for user-defined calibrations.
\end{enumerate}

Despite its limitations in terms of the completeness of the catalogue (\S\ref{txt:completeness}), and data coverage (\S\ref{txt:limitations}), the \hec{} is a highly complete sample of known galaxies in the local Universe. Owing to its wealth of information, the \hec{} can be a useful tool for a wide range of applications. By providing positions and size information the catalogue can be used as the basis of future associations of galaxies with additional multi-wavelength surveys. We discuss a wide range of applications, including the prioritisation of host galaxies for follow-up searches for EM counterparts of GW sources, as well as, the initial characterisation of transient sources which will be critical in the era of Big Data of astronomy.

Future versions of the catalogue will expand the distance range beyond the current limit of 200\Mpc{}, and provide a wider coverage in terms of the stellar population parameters. SFR and \mass{} estimates will be improved by: 
(i) including additional multi-wavelength data, 
(ii) adoption of forced-photometry techniques allowing the full exploitation of existing all-sky surveys also for extended objects,
and 
(iii) performing SED analysis. Finally, incorporation of different sources of spectroscopic data will not only extend the coverage of metallicity and nuclear activity classifications, but more importantly will serve as a cross-validation dataset for AGN classifications. This is crucial for many areas of applications (e.g., screening for AGN in X-ray studies of galaxies, identification of candidate sources of high-energy $\gamma$-ray or cosmic rays).

\section*{Acknowledgements}

We are grateful to the anonymous referee for critically and thoroughly reviewing this paper, motivating improvements to the manuscript and the catalogue.

KK thanks D. Makarov for providing support with \hyper{} database, D. Lang for his help with the {\it WISE} forced photometry catalogue, and C. Berry and N. Stergioulas for discussions regarding the application on gravitational wave sources. KK also thanks P. Bonfini, A. Maragkoudakis, P. Sell and S. J. Williams for their helpful comments on the compilation of the catalogue. We also acknowledge early users of the \hec{}: K. Anastasopoulou, M. Colom i Bernadich, F. Haberl, A. Schwope, N. Vulic, and J. Wilms, for providing feedback that led to improvements, and identifying mistakes, limitations as well as possible extensions of the catalogue.

The research leading to these results has received funding from the {\it European Research Council} under the European Union's {\it Seventh Framework Programme} (FP/2007-2013) / {\it ERC} Grant Agreement n.~617001,
and the {\it European Union’s Horizon 2020} research and innovation programme under the {\it Marie Sk\l{}odowska-Curie RISE} action, Grant Agreement n.~691164 ({\it ASTROSTAT}).
J.J.A. acknowledges funding from CIERA and Northwestern University through a Postdoctoral Fellowship.

We acknowledge the usage of the \hyper{} database, the \texttt{TOPCAT} astronomical software (\url{http://www.starlink.ac.uk/topcat/}), and the \texttt{ligo.skymap} package for Python.
This research made use of the cross-match service provided by CDS, Strasbourg. 
This research has made use of the VizieR catalogue access tool, CDS, Strasbourg, France (DOI : 10.26093/cds/vizier). The original description of the VizieR service was published in 2000, A\&AS 143, 23.
This research has made use of the NASA/IPAC Extragalactic Database (NED), which is operated by the Jet Propulsion Laboratory, California Institute of Technology, under contract with the National Aeronautics and Space Administration.
Funding for SDSS-III has been provided by the Alfred P. Sloan Foundation, the Participating Institutions, the National Science Foundation, and the U.S. Department of Energy Office of Science. The SDSS-III web site is \url{http://www.sdss3.org/}.
This publication makes use of data products from the Two Micron All Sky Survey, which is a joint project of the University of Massachusetts and the Infrared Processing and Analysis Center/California Institute of Technology, funded by the National Aeronautics and Space Administration and the National Science Foundation.

\section*{Data availability}

The data underlying this article are available in the \hec{} Portal, at \portal{}.


\bibliographystyle{mnras}
\bibliography{bibliography.bib} 

\begin{thebibliography}{}
\makeatletter
\relax
\def\mn@urlcharsother{\let\do\@makeother \do\$\do\&\do\#\do\^\do\_\do\%\do\~}
\def\mn@doi{\begingroup\mn@urlcharsother \@ifnextchar [ {\mn@doi@}
  {\mn@doi@[]}}
\def\mn@doi@[#1]#2{\def\@tempa{#1}\ifx\@tempa\@empty \href
  {http://dx.doi.org/#2} {doi:#2}\else \href {http://dx.doi.org/#2} {#1}\fi
  \endgroup}
\def\mn@eprint#1#2{\mn@eprint@#1:#2::\@nil}
\def\mn@eprint@arXiv#1{\href {http://arxiv.org/abs/#1} {{\tt arXiv:#1}}}
\def\mn@eprint@dblp#1{\href {http://dblp.uni-trier.de/rec/bibtex/#1.xml}
  {dblp:#1}}
\def\mn@eprint@#1:#2:#3:#4\@nil{\def\@tempa {#1}\def\@tempb {#2}\def\@tempc
  {#3}\ifx \@tempc \@empty \let \@tempc \@tempb \let \@tempb \@tempa \fi \ifx
  \@tempb \@empty \def\@tempb {arXiv}\fi \@ifundefined
  {mn@eprint@\@tempb}{\@tempb:\@tempc}{\expandafter \expandafter \csname
  mn@eprint@\@tempb\endcsname \expandafter{\@tempc}}}

\bibitem[\protect\citeauthoryear{{Abbott} et~al.}{{Abbott}
  et~al.}{2017a}]{LigoVirgo17}
{Abbott} B.~P.,  et~al., 2017a, \mn@doi [\prl]
  {10.1103/PhysRevLett.119.161101}, \href
  {https://ui.adsabs.harvard.edu/abs/2017PhRvL.119p1101A} {119, 161101}

\bibitem[\protect\citeauthoryear{{Abbott} et~al.}{{Abbott}
  et~al.}{2017b}]{AbbottNature}
{Abbott} B.~P.,  et~al., 2017b, \mn@doi [\nat] {10.1038/nature24471}, \href
  {https://ui.adsabs.harvard.edu/abs/2017Natur.551...85A} {551, 85}

\bibitem[\protect\citeauthoryear{{Abbott} et~al.}{{Abbott}
  et~al.}{2017c}]{Abbot17}
{Abbott} B.~P.,  et~al., 2017c, \mn@doi [\apjl] {10.3847/2041-8213/aa93fc},
  \href {https://ui.adsabs.harvard.edu/abs/2017ApJ...850L..40A} {850, L40}

\bibitem[\protect\citeauthoryear{{Abbott} et~al.}{{Abbott}
  et~al.}{2020}]{AbbottLocal}
{Abbott} B.~P.,  et~al., 2020, \mn@doi [Living Reviews in Relativity]
  {10.1007/s41114-020-00026-9}, \href
  {https://ui.adsabs.harvard.edu/abs/2020LRR....23....3A} {23, 3}

\bibitem[\protect\citeauthoryear{{Ackermann} et~al.,}{{Ackermann}
  et~al.}{2012}]{Ackermann12}
{Ackermann} M.,  et~al., 2012, \mn@doi [\apj] {10.1088/0004-637X/755/2/164},
  \href {https://ui.adsabs.harvard.edu/abs/2012ApJ...755..164A} {755, 164}

\bibitem[\protect\citeauthoryear{{Adhikari}, {Fishbach}, {Holz}, {Wechsler}  \&
  {Fang}}{{Adhikari} et~al.}{2020}]{Adhikari20}
{Adhikari} S.,  {Fishbach} M.,  {Holz} D.~E.,  {Wechsler} R.~H.,   {Fang} Z.,
  2020, arXiv e-prints, \href
  {https://ui.adsabs.harvard.edu/abs/2020arXiv200101025A} {p. arXiv:2001.01025}

\bibitem[\protect\citeauthoryear{{Aguado} et~al.,}{{Aguado}
  et~al.}{2019}]{sdss15}
{Aguado} D.~S.,  et~al., 2019, \mn@doi [\apjs] {10.3847/1538-4365/aaf651},
  \href {https://ui.adsabs.harvard.edu/abs/2019ApJS..240...23A} {240, 23}

\bibitem[\protect\citeauthoryear{{Ahlers} \& {Halzen}}{{Ahlers} \&
  {Halzen}}{2014}]{Ahlers14}
{Ahlers} M.,  {Halzen} F.,  2014, \mn@doi [\prd] {10.1103/PhysRevD.90.043005},
  \href {https://ui.adsabs.harvard.edu/abs/2014PhRvD..90d3005A} {90, 043005}

\bibitem[\protect\citeauthoryear{{Andrews} \& {Zezas}}{{Andrews} \&
  {Zezas}}{2019}]{Andrews19}
{Andrews} J.~J.,  {Zezas} A.,  2019, \mn@doi [\mnras] {10.1093/mnras/stz1066},
  \href {https://ui.adsabs.harvard.edu/abs/2019MNRAS.486.3213A} {486, 3213}

\bibitem[\protect\citeauthoryear{{Arcavi} et~al.,}{{Arcavi}
  et~al.}{2017}]{Arcavi17}
{Arcavi} I.,  et~al., 2017, \mn@doi [\apjl] {10.3847/2041-8213/aa910f}, \href
  {https://ui.adsabs.harvard.edu/abs/2017ApJ...848L..33A} {848, L33}

\bibitem[\protect\citeauthoryear{{Artale}, {Mapelli}, {Giacobbo}, {Sabha},
  {Spera}, {Santoliquido}  \& {Bressan}}{{Artale} et~al.}{2019}]{Artale19}
{Artale} M.~C.,  {Mapelli} M.,  {Giacobbo} N.,  {Sabha} N.~B.,  {Spera} M.,
  {Santoliquido} F.,   {Bressan} A.,  2019, \mn@doi [\mnras]
  {10.1093/mnras/stz1382}, \href
  {https://ui.adsabs.harvard.edu/abs/2019MNRAS.487.1675A} {487, 1675}

\bibitem[\protect\citeauthoryear{{Artale}, {Mapelli}, {Bouffanais}, {Giacobbo},
  {Pasquato}  \& {Spera}}{{Artale} et~al.}{2020a}]{Artale20a}
{Artale} M.~C.,  {Mapelli} M.,  {Bouffanais} Y.,  {Giacobbo} N.,  {Pasquato}
  M.,   {Spera} M.,  2020a, \mn@doi [\mnras] {10.1093/mnras/stz3190}, \href
  {https://ui.adsabs.harvard.edu/abs/2020MNRAS.491.3419A} {491, 3419}

\bibitem[\protect\citeauthoryear{{Artale}, {Bouffanais}, {Mapelli}, {Giacobbo},
  {Sabha}, {Santoliquido}, {Pasquato}  \& {Spera}}{{Artale}
  et~al.}{2020b}]{Artale20b}
{Artale} M.~C.,  {Bouffanais} Y.,  {Mapelli} M.,  {Giacobbo} N.,  {Sabha}
  N.~B.,  {Santoliquido} F.,  {Pasquato} M.,   {Spera} M.,  2020b, \mn@doi
  [\mnras] {10.1093/mnras/staa1252}, \href
  {https://ui.adsabs.harvard.edu/abs/2020MNRAS.495.1841A} {495, 1841}

\bibitem[\protect\citeauthoryear{{Basu-Zych} et~al.,}{{Basu-Zych}
  et~al.}{2020}]{Basu20}
{Basu-Zych} A.~R.,  et~al., 2020, \mn@doi [\mnras] {10.1093/mnras/staa2343},
  \href {https://ui.adsabs.harvard.edu/abs/2020MNRAS.498.1651B} {498, 1651}

\bibitem[\protect\citeauthoryear{{Bavera} et~al.,}{{Bavera}
  et~al.}{2020}]{Bavera20}
{Bavera} S.~S.,  et~al., 2020, \mn@doi [\aap] {10.1051/0004-6361/201936204},
  \href {https://ui.adsabs.harvard.edu/abs/2020A&A...635A..97B} {635, A97}

\bibitem[\protect\citeauthoryear{{Bell}, {McIntosh}, {Katz}  \&
  {Weinberg}}{{Bell} et~al.}{2003}]{Bell03}
{Bell} E.~F.,  {McIntosh} D.~H.,  {Katz} N.,   {Weinberg} M.~D.,  2003, \mn@doi
  [\apjs] {10.1086/378847}, \href
  {http://esoads.eso.org/abs/2003ApJS..149..289B} {149, 289}

\bibitem[\protect\citeauthoryear{{Bhattacharjee}}{{Bhattacharjee}}{2000}]{Bhattacharjee00}
{Bhattacharjee} P.,  2000, \mn@doi [\physrep] {10.1016/S0370-1573(99)00101-5},
  \href {https://ui.adsabs.harvard.edu/abs/2000PhR...327..109B} {327, 109}

\bibitem[\protect\citeauthoryear{{Blanton} \& {Roweis}}{{Blanton} \&
  {Roweis}}{2007}]{Blanton07}
{Blanton} M.~R.,  {Roweis} S.,  2007, \mn@doi [\aj] {10.1086/510127}, \href
  {http://esoads.eso.org/abs/2007AJ....133..734B} {133, 734}

\bibitem[\protect\citeauthoryear{{Brinchmann}, {Charlot}, {White}, {Tremonti},
  {Kauffmann}, {Heckman}  \& {Brinkmann}}{{Brinchmann}
  et~al.}{2004}]{Brinchmann04}
{Brinchmann} J.,  {Charlot} S.,  {White} S.~D.~M.,  {Tremonti} C.,  {Kauffmann}
  G.,  {Heckman} T.,   {Brinkmann} J.,  2004, \mn@doi [\mnras]
  {10.1111/j.1365-2966.2004.07881.x}, \href
  {http://adsabs.harvard.edu/abs/2004MNRAS.351.1151B} {351, 1151}

\bibitem[\protect\citeauthoryear{{Bruzual A.} \& {Charlot}}{{Bruzual A.} \&
  {Charlot}}{1993}]{Bruzual93}
{Bruzual A.} G.,  {Charlot} S.,  1993, \mn@doi [\apj] {10.1086/172385}, \href
  {https://ui.adsabs.harvard.edu/abs/1993ApJ...405..538B} {405, 538}

\bibitem[\protect\citeauthoryear{{Buikema} et~al.,}{{Buikema}
  et~al.}{2020}]{Sensitivity20}
{Buikema} A.,  et~al., 2020, \mn@doi [\prd] {10.1103/PhysRevD.102.062003},
  \href {https://ui.adsabs.harvard.edu/abs/2020PhRvD.102f2003B} {102, 062003}

\bibitem[\protect\citeauthoryear{{Chambers} et~al.,}{{Chambers}
  et~al.}{2016}]{Chambers16}
{Chambers} K.~C.,  et~al., 2016, arXiv e-prints, \href
  {https://ui.adsabs.harvard.edu/abs/2016arXiv161205560C} {p. arXiv:1612.05560}

\bibitem[\protect\citeauthoryear{{Chen}, {Fishbach}  \& {Holz}}{{Chen}
  et~al.}{2018}]{Chen18}
{Chen} H.-Y.,  {Fishbach} M.,   {Holz} D.~E.,  2018, \mn@doi [\nat]
  {10.1038/s41586-018-0606-0}, \href
  {https://ui.adsabs.harvard.edu/abs/2018Natur.562..545C} {562, 545}

\bibitem[\protect\citeauthoryear{{Cluver}, {Jarrett}, {Dale}, {Smith}, {August}
   \& {Brown}}{{Cluver} et~al.}{2017}]{Cluver17}
{Cluver} M.~E.,  {Jarrett} T.~H.,  {Dale} D.~A.,  {Smith} J.-D.~T.,  {August}
  T.,   {Brown} M.~J.~I.,  2017, \mn@doi [\apj] {10.3847/1538-4357/aa92c7},
  \href {http://esoads.eso.org/abs/2017ApJ...850...68C} {850, 68}

\bibitem[\protect\citeauthoryear{{Colless} et~al.,}{{Colless}
  et~al.}{2001}]{Colless01}
{Colless} M.,  et~al., 2001, \mn@doi [\mnras]
  {10.1046/j.1365-8711.2001.04902.x}, \href
  {https://ui.adsabs.harvard.edu/abs/2001MNRAS.328.1039C} {328, 1039}

\bibitem[\protect\citeauthoryear{{Cook}, {van Sistine}, {Singer}, {Kasliwal}
  \& {Growth (Global Relay Of Observatories Watching Transients Happen)
  Collaboration}}{{Cook} et~al.}{2017}]{GCN21519}
{Cook} D.~O.,  {van Sistine} A.,  {Singer} L.,  {Kasliwal} M.~M.,   {Growth
  (Global Relay Of Observatories Watching Transients Happen) Collaboration}
  2017, GRB Coordinates Network, \href
  {https://ui.adsabs.harvard.edu/abs/2017GCN.21519....1C} {21519, 1}

\bibitem[\protect\citeauthoryear{{Cook} et~al.,}{{Cook} et~al.}{2019}]{Cook19}
{Cook} D.~O.,  et~al., 2019, \mn@doi [\apj] {10.3847/1538-4357/ab2131}, \href
  {https://ui.adsabs.harvard.edu/abs/2019ApJ...880....7C} {880, 7}

\bibitem[\protect\citeauthoryear{{Cutri} et~al.,}{{Cutri}
  et~al.}{2012}]{Cutri12}
{Cutri} R.~M.,  et~al., 2012, VizieR Online Data Catalog, \href
  {https://ui.adsabs.harvard.edu/abs/2012yCat.2281....0C} {p. II/281}

\bibitem[\protect\citeauthoryear{{Dale} \& {Helou}}{{Dale} \&
  {Helou}}{2002}]{Dale02}
{Dale} D.~A.,  {Helou} G.,  2002, \mn@doi [\apj] {10.1086/341632}, \href
  {http://esoads.eso.org/abs/2002ApJ...576..159D} {576, 159}

\bibitem[\protect\citeauthoryear{{D{\'a}lya}, {B{\'e}csy}, {Raffai}  \& {Glade
  Team}}{{D{\'a}lya} et~al.}{2017}]{GCN21516}
{D{\'a}lya} G.,  {B{\'e}csy} B.,  {Raffai} P.,   {Glade Team} 2017, GRB
  Coordinates Network, \href
  {https://ui.adsabs.harvard.edu/abs/2017GCN.21516....1D} {21516, 1}

\bibitem[\protect\citeauthoryear{{D{\'a}lya} et~al.,}{{D{\'a}lya}
  et~al.}{2018}]{Dalya18}
{D{\'a}lya} G.,  et~al., 2018, \mn@doi [\mnras] {10.1093/mnras/sty1703}, \href
  {http://adsabs.harvard.edu/abs/2018MNRAS.479.2374D} {479, 2374}

\bibitem[\protect\citeauthoryear{{Del Pozzo}, {Berry}, {Ghosh}, {Haines},
  {Singer}  \& {Vecchio}}{{Del Pozzo} et~al.}{2018}]{DelPozzo18}
{Del Pozzo} W.,  {Berry} C.~P.~L.,  {Ghosh} A.,  {Haines} T.~S.~F.,  {Singer}
  L.~P.,   {Vecchio} A.,  2018, \mn@doi [\mnras] {10.1093/mnras/sty1485}, \href
  {https://ui.adsabs.harvard.edu/abs/2018MNRAS.479..601D} {479, 601}

\bibitem[\protect\citeauthoryear{{Deng} et~al.,}{{Deng} et~al.}{2012}]{Deng12}
{Deng} L.-C.,  et~al., 2012, \mn@doi [Research in Astronomy and Astrophysics]
  {10.1088/1674-4527/12/7/003}, \href
  {https://ui.adsabs.harvard.edu/abs/2012RAA....12..735D} {12, 735}

\bibitem[\protect\citeauthoryear{{Ducoin}, {Corre}, {Leroy}  \& {Le
  Floch}}{{Ducoin} et~al.}{2020}]{Ducoin20}
{Ducoin} J.~G.,  {Corre} D.,  {Leroy} N.,   {Le Floch} E.,  2020, \mn@doi
  [\mnras] {10.1093/mnras/staa114}, \href
  {https://ui.adsabs.harvard.edu/abs/2020MNRAS.492.4768D} {492, 4768}

\bibitem[\protect\citeauthoryear{{Evans} et~al.,}{{Evans}
  et~al.}{2010}]{Evans10}
{Evans} I.~N.,  et~al., 2010, \mn@doi [\apjs] {10.1088/0067-0049/189/1/37},
  \href {http://adsabs.harvard.edu/abs/2010ApJS..189...37E} {189, 37}

\bibitem[\protect\citeauthoryear{{Flaugher}}{{Flaugher}}{2005}]{Flaugher05}
{Flaugher} B.,  2005, \mn@doi [International Journal of Modern Physics A]
  {10.1142/S0217751X05025917}, \href
  {https://ui.adsabs.harvard.edu/abs/2005IJMPA..20.3121F} {20, 3121}

\bibitem[\protect\citeauthoryear{{Freedman} et~al.,}{{Freedman}
  et~al.}{2001}]{Freedman01}
{Freedman} W.~L.,  et~al., 2001, \mn@doi [\apj] {10.1086/320638}, \href
  {http://adsabs.harvard.edu/abs/2001ApJ...553...47F} {553, 47}

\bibitem[\protect\citeauthoryear{{French}, {Wevers}, {Law-Smith}, {Graur}  \&
  {Zabludoff}}{{French} et~al.}{2020}]{French20}
{French} K.~D.,  {Wevers} T.,  {Law-Smith} J.,  {Graur} O.,   {Zabludoff}
  A.~I.,  2020, \mn@doi [\ssr] {10.1007/s11214-020-00657-y}, \href
  {https://ui.adsabs.harvard.edu/abs/2020SSRv..216...32F} {216, 32}

\bibitem[\protect\citeauthoryear{{Galliano}, {Galametz}  \& {Jones}}{{Galliano}
  et~al.}{2018}]{Galliano18}
{Galliano} F.,  {Galametz} M.,   {Jones} A.~P.,  2018, \mn@doi [\araa]
  {10.1146/annurev-astro-081817-051900}, \href
  {https://ui.adsabs.harvard.edu/abs/2018ARA&A..56..673G} {56, 673}

\bibitem[\protect\citeauthoryear{{Gardner}, {Sharples}, {Frenk}  \&
  {Carrasco}}{{Gardner} et~al.}{1997}]{Gardner97}
{Gardner} J.~P.,  {Sharples} R.~M.,  {Frenk} C.~S.,   {Carrasco} B.~E.,  1997,
  \mn@doi [\apjl] {10.1086/310630}, \href
  {https://ui.adsabs.harvard.edu/abs/1997ApJ...480L..99G} {480, L99}

\bibitem[\protect\citeauthoryear{{Gehrels}, {Cannizzo}, {Kanner}, {Kasliwal},
  {Nissanke}  \& {Singer}}{{Gehrels} et~al.}{2016}]{Gehrels16}
{Gehrels} N.,  {Cannizzo} J.~K.,  {Kanner} J.,  {Kasliwal} M.~M.,  {Nissanke}
  S.,   {Singer} L.~P.,  2016, \mn@doi [\apj] {10.3847/0004-637X/820/2/136},
  \href {http://esoads.eso.org/abs/2016ApJ...820..136G} {820, 136}

\bibitem[\protect\citeauthoryear{{Goldstein} et~al.,}{{Goldstein}
  et~al.}{2017}]{Goldstein17}
{Goldstein} A.,  et~al., 2017, \mn@doi [\apjl] {10.3847/2041-8213/aa8f41},
  \href {https://ui.adsabs.harvard.edu/abs/2017ApJ...848L..14G} {848, L14}

\bibitem[\protect\citeauthoryear{{G{\'o}rski}, {Hivon}, {Banday}, {Wand elt},
  {Hansen}, {Reinecke}  \& {Bartelmann}}{{G{\'o}rski} et~al.}{2005}]{Gorski05}
{G{\'o}rski} K.~M.,  {Hivon} E.,  {Banday} A.~J.,  {Wand elt} B.~D.,  {Hansen}
  F.~K.,  {Reinecke} M.,   {Bartelmann} M.,  2005, \mn@doi [\apj]
  {10.1086/427976}, \href
  {https://ui.adsabs.harvard.edu/abs/2005ApJ...622..759G} {622, 759}

\bibitem[\protect\citeauthoryear{{Greggio} \& {Cappellaro}}{{Greggio} \&
  {Cappellaro}}{2019}]{Greggio19}
{Greggio} L.,  {Cappellaro} E.,  2019, \mn@doi [\aap]
  {10.1051/0004-6361/201834932}, \href
  {https://ui.adsabs.harvard.edu/abs/2019A&A...625A.113G} {625, A113}

\bibitem[\protect\citeauthoryear{{Hao}, {Kennicutt}, {Johnson}, {Calzetti},
  {Dale}  \& {Moustakas}}{{Hao} et~al.}{2011}]{Hao11}
{Hao} C.-N.,  {Kennicutt} R.~C.,  {Johnson} B.~D.,  {Calzetti} D.,  {Dale}
  D.~A.,   {Moustakas} J.,  2011, \mn@doi [\apj] {10.1088/0004-637X/741/2/124},
  \href {https://ui.adsabs.harvard.edu/abs/2011ApJ...741..124H} {741, 124}

\bibitem[\protect\citeauthoryear{{Hawkins} et~al.,}{{Hawkins}
  et~al.}{2003}]{Hawkins03}
{Hawkins} E.,  et~al., 2003, \mn@doi [\mnras]
  {10.1046/j.1365-2966.2003.07063.x}, \href
  {http://adsabs.harvard.edu/abs/2003MNRAS.346...78H} {346, 78}

\bibitem[\protect\citeauthoryear{{Hayward} et~al.,}{{Hayward}
  et~al.}{2014}]{Hayward14}
{Hayward} C.~C.,  et~al., 2014, \mn@doi [\mnras] {10.1093/mnras/stu1843}, \href
  {https://ui.adsabs.harvard.edu/abs/2014MNRAS.445.1598H} {445, 1598}

\bibitem[\protect\citeauthoryear{{He}, {Kusenko}, {Nagataki}, {Zhang}, {Yang}
  \& {Fan}}{{He} et~al.}{2016}]{He16}
{He} H.-N.,  {Kusenko} A.,  {Nagataki} S.,  {Zhang} B.-B.,  {Yang} R.-Z.,
  {Fan} Y.-Z.,  2016, \mn@doi [\prd] {10.1103/PhysRevD.93.043011}, \href
  {https://ui.adsabs.harvard.edu/abs/2016PhRvD..93d3011H} {93, 043011}

\bibitem[\protect\citeauthoryear{{Helou} \& {Walker}}{{Helou} \&
  {Walker}}{1988}]{Helou88}
{Helou} G.,  {Walker} D.~W.,  1988, in NASA RP-1190, Vol. 7 (1988).

\bibitem[\protect\citeauthoryear{{Helou}, {Madore}, {Schmitz}, {Bicay}, {Wu}
  \& {Bennett}}{{Helou} et~al.}{1991}]{Helou91}
{Helou} G.,  {Madore} B.~F.,  {Schmitz} M.,  {Bicay} M.~D.,  {Wu} X.,
  {Bennett} J.,  1991, in {Albrecht} M.~A.,  {Egret} D.,  eds,  Astrophysics
  and Space Science Library Vol. 171, Databases and On-line Data in Astronomy.
  pp 89--106, \mn@doi{10.1007/978-94-011-3250-3_10}

\bibitem[\protect\citeauthoryear{{IceCube Collaboration}}{{IceCube
  Collaboration}}{2018}]{IceCube18}
{IceCube Collaboration} 2018, \mn@doi [Science] {10.1126/science.aat1378},
  \href {https://ui.adsabs.harvard.edu/abs/2018Sci...361.1378I} {361, eaat1378}

\bibitem[\protect\citeauthoryear{{Ivezi{\'c}} et~al.,}{{Ivezi{\'c}}
  et~al.}{2019}]{Ivezic19}
{Ivezi{\'c}} {\v{Z}}.,  et~al., 2019, \mn@doi [\apj]
  {10.3847/1538-4357/ab042c}, \href
  {https://ui.adsabs.harvard.edu/abs/2019ApJ...873..111I} {873, 111}

\bibitem[\protect\citeauthoryear{{Jarrett}, {Chester}, {Cutri}, {Schneider},
  {Skrutskie}  \& {Huchra}}{{Jarrett} et~al.}{2000}]{Jarrett00}
{Jarrett} T.~H.,  {Chester} T.,  {Cutri} R.,  {Schneider} S.,  {Skrutskie} M.,
   {Huchra} J.~P.,  2000, \mn@doi [\aj] {10.1086/301330}, \href
  {http://esoads.eso.org/abs/2000AJ....119.2498J} {119, 2498}

\bibitem[\protect\citeauthoryear{{Jarrett}, {Chester}, {Cutri}, {Schneider}  \&
  {Huchra}}{{Jarrett} et~al.}{2003}]{Jarrett03}
{Jarrett} T.~H.,  {Chester} T.,  {Cutri} R.,  {Schneider} S.~E.,   {Huchra}
  J.~P.,  2003, \mn@doi [\aj] {10.1086/345794}, \href
  {http://adsabs.harvard.edu/abs/2003AJ....125..525J} {125, 525}

\bibitem[\protect\citeauthoryear{{Kalogera}, {Belczynski}, {Kim},
  {O'Shaughnessy}  \& {Willems}}{{Kalogera} et~al.}{2007}]{Kalogera2007}
{Kalogera} V.,  {Belczynski} K.,  {Kim} C.,  {O'Shaughnessy} R.,   {Willems}
  B.,  2007, \mn@doi [\physrep] {10.1016/j.physrep.2007.02.008}, \href
  {https://ui.adsabs.harvard.edu/abs/2007PhR...442...75K} {442, 75}

\bibitem[\protect\citeauthoryear{{Kanner}, {Huard}, {M{\'a}rka}, {Murphy},
  {Piscionere}, {Reed}  \& {Shawhan}}{{Kanner} et~al.}{2008}]{Kanner08}
{Kanner} J.,  {Huard} T.~L.,  {M{\'a}rka} S.,  {Murphy} D.~C.,  {Piscionere}
  J.,  {Reed} M.,   {Shawhan} P.,  2008, \mn@doi [Classical and Quantum
  Gravity] {10.1088/0264-9381/25/18/184034}, \href
  {https://ui.adsabs.harvard.edu/abs/2008CQGra..25r4034K} {25, 184034}

\bibitem[\protect\citeauthoryear{{Karachentsev} \& {Makarov}}{{Karachentsev} \&
  {Makarov}}{1996}]{Karachentsev96}
{Karachentsev} I.~D.,  {Makarov} D.~A.,  1996, \mn@doi [\aj] {10.1086/117825},
  \href {http://esoads.eso.org/abs/1996AJ....111..794K} {111, 794}

\bibitem[\protect\citeauthoryear{{Karachentsev}, {Karachentseva}  \&
  {Huchtmeier}}{{Karachentsev} et~al.}{2001}]{Karachentsev01}
{Karachentsev} I.~D.,  {Karachentseva} V.~E.,   {Huchtmeier} W.~K.,  2001,
  \mn@doi [\aap] {10.1051/0004-6361:20000262}, \href
  {https://ui.adsabs.harvard.edu/abs/2001A&A...366..428K} {366, 428}

\bibitem[\protect\citeauthoryear{{Karachentsev}, {Karachentseva}  \&
  {Huchtmeier}}{{Karachentsev} et~al.}{2007}]{Karachentsev07}
{Karachentsev} I.~D.,  {Karachentseva} V.~E.,   {Huchtmeier} W.~K.,  2007,
  \mn@doi [Astronomy Letters] {10.1134/S1063773707080026}, \href
  {https://ui.adsabs.harvard.edu/abs/2007AstL...33..512K} {33, 512}

\bibitem[\protect\citeauthoryear{{Karachentsev}, {Makarov}  \&
  {Kaisina}}{{Karachentsev} et~al.}{2013}]{Kara13}
{Karachentsev} I.~D.,  {Makarov} D.~I.,   {Kaisina} E.~I.,  2013, \mn@doi [\aj]
  {10.1088/0004-6256/145/4/101}, \href
  {http://adsabs.harvard.edu/abs/2013AJ....145..101K} {145, 101}

\bibitem[\protect\citeauthoryear{{Kasliwal} et~al.,}{{Kasliwal}
  et~al.}{2017}]{Kasliwal17}
{Kasliwal} M.~M.,  et~al., 2017, \mn@doi [Science] {10.1126/science.aap9455},
  \href {https://ui.adsabs.harvard.edu/abs/2017Sci...358.1559K} {358, 1559}

\bibitem[\protect\citeauthoryear{{Kauffmann} et~al.,}{{Kauffmann}
  et~al.}{2003}]{Kauffmann03}
{Kauffmann} G.,  et~al., 2003, \mn@doi [\mnras]
  {10.1111/j.1365-2966.2003.07154.x}, \href
  {https://ui.adsabs.harvard.edu/abs/2003MNRAS.346.1055K} {346, 1055}

\bibitem[\protect\citeauthoryear{{Kelly}}{{Kelly}}{2007}]{Kelly07}
{Kelly} B.~C.,  2007, \mn@doi [\apj] {10.1086/519947}, \href
  {http://esoads.eso.org/abs/2007ApJ...665.1489K} {665, 1489}

\bibitem[\protect\citeauthoryear{{Kennicutt}}{{Kennicutt}}{1998}]{Kennicutt98}
{Kennicutt} Jr. R.~C.,  1998, \mn@doi [\araa] {10.1146/annurev.astro.36.1.189},
  \href {http://adsabs.harvard.edu/abs/1998ARA%26A..36..189K} {36, 189}

\bibitem[\protect\citeauthoryear{{Kennicutt} \& {Evans}}{{Kennicutt} \&
  {Evans}}{2012}]{Kennicutt12}
{Kennicutt} R.~C.,  {Evans} N.~J.,  2012, \mn@doi [\araa]
  {10.1146/annurev-astro-081811-125610}, \href
  {http://adsabs.harvard.edu/abs/2012ARA%26A..50..531K} {50, 531}

\bibitem[\protect\citeauthoryear{{Kewley} \& {Ellison}}{{Kewley} \&
  {Ellison}}{2008}]{Kewley08}
{Kewley} L.~J.,  {Ellison} S.~L.,  2008, \mn@doi [\apj] {10.1086/587500}, \href
  {http://adsabs.harvard.edu/abs/2008ApJ...681.1183K} {681, 1183}

\bibitem[\protect\citeauthoryear{{Kim}, {Wilkes}, {Kim}, {Green}, {Barkhouse},
  {Lee}, {Silverman}  \& {Tananbaum}}{{Kim} et~al.}{2007}]{Kim07}
{Kim} M.,  {Wilkes} B.~J.,  {Kim} D.-W.,  {Green} P.~J.,  {Barkhouse} W.~A.,
  {Lee} M.~G.,  {Silverman} J.~D.,   {Tananbaum} H.~D.,  2007, \mn@doi [\apj]
  {10.1086/511630}, \href {http://adsabs.harvard.edu/abs/2007ApJ...659...29K}
  {659, 29}

\bibitem[\protect\citeauthoryear{{Kim} et~al.,}{{Kim} et~al.}{2014}]{Kim2014}
{Kim} S.,  et~al., 2014, \mn@doi [\apjs] {10.1088/0067-0049/215/2/22}, \href
  {http://esoads.eso.org/abs/2014ApJS..215...22K} {215, 22}

\bibitem[\protect\citeauthoryear{{Komis}, {Pavlidou}  \& {Zezas}}{{Komis}
  et~al.}{2019}]{Komis19}
{Komis} I.,  {Pavlidou} V.,   {Zezas} A.,  2019, \mn@doi [\mnras]
  {10.1093/mnras/sty3354}, \href
  {https://ui.adsabs.harvard.edu/abs/2019MNRAS.483.4020K} {483, 4020}

\bibitem[\protect\citeauthoryear{{Kopparapu}, {Hanna}, {Kalogera},
  {O'Shaughnessy}, {Gonz{\'a}lez}, {Brady}  \& {Fairhurst}}{{Kopparapu}
  et~al.}{2008}]{Kopparapu08}
{Kopparapu} R.~K.,  {Hanna} C.,  {Kalogera} V.,  {O'Shaughnessy} R.,
  {Gonz{\'a}lez} G.,  {Brady} P.~R.,   {Fairhurst} S.,  2008, \mn@doi [\apj]
  {10.1086/527348}, \href
  {https://ui.adsabs.harvard.edu/abs/2008ApJ...675.1459K} {675, 1459}

\bibitem[\protect\citeauthoryear{{Kourkchi} \& {Tully}}{{Kourkchi} \&
  {Tully}}{2017}]{Kourkchi17}
{Kourkchi} E.,  {Tully} R.~B.,  2017, \mn@doi [\apj]
  {10.3847/1538-4357/aa76db}, \href
  {https://ui.adsabs.harvard.edu/abs/2017ApJ...843...16K} {843, 16}

\bibitem[\protect\citeauthoryear{{Kovlakas}, {Zezas}, {Andrews}, {Basu-Zych},
  {Fragos}, {Hornschemeier}, {Lehmer}  \& {Ptak}}{{Kovlakas}
  et~al.}{2020}]{Kovlakas20}
{Kovlakas} K.,  {Zezas} A.,  {Andrews} J.~J.,  {Basu-Zych} A.,  {Fragos} T.,
  {Hornschemeier} A.,  {Lehmer} B.,   {Ptak} A.,  2020, \mn@doi [\mnras]
  {10.1093/mnras/staa2481}, \href
  {https://ui.adsabs.harvard.edu/abs/2020MNRAS.498.4790K} {498, 4790}

\bibitem[\protect\citeauthoryear{{Krau{\ss}} et~al.,}{{Krau{\ss}}
  et~al.}{2020}]{Krauss2020}
{Krau{\ss}} F.,  et~al., 2020, \mn@doi [\mnras] {10.1093/mnras/staa2148}, \href
  {https://ui.adsabs.harvard.edu/abs/2020MNRAS.497.2553K} {497, 2553}

\bibitem[\protect\citeauthoryear{{Kroupa}}{{Kroupa}}{2001}]{Kroupa01}
{Kroupa} P.,  2001, \mn@doi [\mnras] {10.1046/j.1365-8711.2001.04022.x}, \href
  {https://ui.adsabs.harvard.edu/abs/2001MNRAS.322..231K} {322, 231}

\bibitem[\protect\citeauthoryear{{Lang}}{{Lang}}{2014}]{unwise14}
{Lang} D.,  2014, \mn@doi [\aj] {10.1088/0004-6256/147/5/108}, \href
  {https://ui.adsabs.harvard.edu/abs/2014AJ....147..108L} {147, 108}

\bibitem[\protect\citeauthoryear{{Lang}, {Hogg}  \& {Schlegel}}{{Lang}
  et~al.}{2016}]{Lang16}
{Lang} D.,  {Hogg} D.~W.,   {Schlegel} D.~J.,  2016, \mn@doi [\aj]
  {10.3847/0004-6256/151/2/36}, \href
  {http://adsabs.harvard.edu/abs/2016AJ....151...36L} {151, 36}

\bibitem[\protect\citeauthoryear{{Leibler} \& {Berger}}{{Leibler} \&
  {Berger}}{2010}]{Leibler10}
{Leibler} C.~N.,  {Berger} E.,  2010, \mn@doi [\apj]
  {10.1088/0004-637X/725/1/1202}, \href
  {https://ui.adsabs.harvard.edu/abs/2010ApJ...725.1202L} {725, 1202}

\bibitem[\protect\citeauthoryear{{Leitherer} \& {Heckman}}{{Leitherer} \&
  {Heckman}}{1995}]{Leitherer95}
{Leitherer} C.,  {Heckman} T.~M.,  1995, \mn@doi [\apjs] {10.1086/192112},
  \href {https://ui.adsabs.harvard.edu/abs/1995ApJS...96....9L} {96, 9}

\bibitem[\protect\citeauthoryear{{Leitherer} et~al.,}{{Leitherer}
  et~al.}{1999}]{Leitherer99}
{Leitherer} C.,  et~al., 1999, \mn@doi [\apjs] {10.1086/313233}, \href
  {https://ui.adsabs.harvard.edu/abs/1999ApJS..123....3L} {123, 3}

\bibitem[\protect\citeauthoryear{{Madau} \& {Dickinson}}{{Madau} \&
  {Dickinson}}{2014}]{Madau14}
{Madau} P.,  {Dickinson} M.,  2014, \mn@doi [\araa]
  {10.1146/annurev-astro-081811-125615}, \href
  {https://ui.adsabs.harvard.edu/abs/2014ARA&A..52..415M} {52, 415}

\bibitem[\protect\citeauthoryear{{Makarov}, {Prugniel}, {Terekhova}, {Courtois}
   \& {Vauglin}}{{Makarov} et~al.}{2014}]{Makarov14}
{Makarov} D.,  {Prugniel} P.,  {Terekhova} N.,  {Courtois} H.,   {Vauglin} I.,
  2014, \mn@doi [\aap] {10.1051/0004-6361/201423496}, \href
  {http://adsabs.harvard.edu/abs/2014A%26A...570A..13M} {570, A13}

\bibitem[\protect\citeauthoryear{{Mapelli}, {Giacobbo}, {Ripamonti}  \&
  {Spera}}{{Mapelli} et~al.}{2017}]{Mapelli17}
{Mapelli} M.,  {Giacobbo} N.,  {Ripamonti} E.,   {Spera} M.,  2017, \mn@doi
  [\mnras] {10.1093/mnras/stx2123}, \href
  {https://ui.adsabs.harvard.edu/abs/2017MNRAS.472.2422M} {472, 2422}

\bibitem[\protect\citeauthoryear{{Mapelli}, {Giacobbo}, {Toffano}, {Ripamonti},
  {Bressan}, {Spera}  \& {Branchesi}}{{Mapelli} et~al.}{2018}]{Mapelli18}
{Mapelli} M.,  {Giacobbo} N.,  {Toffano} M.,  {Ripamonti} E.,  {Bressan} A.,
  {Spera} M.,   {Branchesi} M.,  2018, \mn@doi [\mnras]
  {10.1093/mnras/sty2663}, \href
  {https://ui.adsabs.harvard.edu/abs/2018MNRAS.481.5324M} {481, 5324}

\bibitem[\protect\citeauthoryear{{Marcote} et~al.,}{{Marcote}
  et~al.}{2020}]{Marcote20}
{Marcote} B.,  et~al., 2020, \mn@doi [\nat] {10.1038/s41586-019-1866-z}, \href
  {https://ui.adsabs.harvard.edu/abs/2020Natur.577..190M} {577, 190}

\bibitem[\protect\citeauthoryear{{Moshir} et~al.}{{Moshir}
  et~al.}{1990}]{Moshir90}
{Moshir} M.,  et~al., 1990, IRAS Faint Source Catalogue, \href
  {https://ui.adsabs.harvard.edu/abs/1990IRASF.C......0M} {p.~0}

\bibitem[\protect\citeauthoryear{Nadaraya}{Nadaraya}{1964}]{Nadaraya64}
Nadaraya E.~A.,  1964, \mn@doi [Theory of Probability \& Its Applications]
  {10.1137/1109020}, 9, 141

\bibitem[\protect\citeauthoryear{{Neijssel} et~al.,}{{Neijssel}
  et~al.}{2019}]{Neijssel19}
{Neijssel} C.~J.,  et~al., 2019, \mn@doi [\mnras] {10.1093/mnras/stz2840},
  \href {https://ui.adsabs.harvard.edu/abs/2019MNRAS.490.3740N} {490, 3740}

\bibitem[\protect\citeauthoryear{{Nilson}}{{Nilson}}{1973}]{Nilson73}
{Nilson} P.,  1973, Uppsala Astron. Obs. Ann., \href
  {http://adsabs.harvard.edu/abs/1973ugcg.book.....N} {6}

\bibitem[\protect\citeauthoryear{{Nissanke}, {Kasliwal}  \&
  {Georgieva}}{{Nissanke} et~al.}{2013}]{Nissanke13}
{Nissanke} S.,  {Kasliwal} M.,   {Georgieva} A.,  2013, \mn@doi [\apj]
  {10.1088/0004-637X/767/2/124}, \href
  {https://ui.adsabs.harvard.edu/abs/2013ApJ...767..124N} {767, 124}

\bibitem[\protect\citeauthoryear{{Nuttall} \& {Sutton}}{{Nuttall} \&
  {Sutton}}{2010}]{Nuttall10}
{Nuttall} L.~K.,  {Sutton} P.~J.,  2010, \mn@doi [\prd]
  {10.1103/PhysRevD.82.102002}, \href
  {https://ui.adsabs.harvard.edu/abs/2010PhRvD..82j2002N} {82, 102002}

\bibitem[\protect\citeauthoryear{{O'Shaughnessy}, {Bellovary}, {Brooks},
  {Shen}, {Governato}  \& {Christensen}}{{O'Shaughnessy}
  et~al.}{2017}]{OShaughnessy17}
{O'Shaughnessy} R.,  {Bellovary} J.~M.,  {Brooks} A.,  {Shen} S.,  {Governato}
  F.,   {Christensen} C.~R.,  2017, \mn@doi [\mnras] {10.1093/mnras/stw2550},
  \href {https://ui.adsabs.harvard.edu/abs/2017MNRAS.464.2831O} {464, 2831}

\bibitem[\protect\citeauthoryear{{Parkash}, {Brown}, {Jarrett}  \&
  {Bonne}}{{Parkash} et~al.}{2018}]{Parkash18}
{Parkash} V.,  {Brown} M. J.~I.,  {Jarrett} T.~H.,   {Bonne} N.~J.,  2018,
  \mn@doi [\apj] {10.3847/1538-4357/aad3b9}, \href
  {https://ui.adsabs.harvard.edu/abs/2018ApJ...864...40P} {864, 40}

\bibitem[\protect\citeauthoryear{{Pettini} \& {Pagel}}{{Pettini} \&
  {Pagel}}{2004}]{PP04}
{Pettini} M.,  {Pagel} B. E.~J.,  2004, \mn@doi [\mnras]
  {10.1111/j.1365-2966.2004.07591.x}, \href
  {https://ui.adsabs.harvard.edu/abs/2004MNRAS.348L..59P} {348, L59}

\bibitem[\protect\citeauthoryear{{Phinney}}{{Phinney}}{1991}]{Phinney91}
{Phinney} E.~S.,  1991, \mn@doi [\apjl] {10.1086/186163}, \href
  {https://ui.adsabs.harvard.edu/abs/1991ApJ...380L..17P} {380, L17}

\bibitem[\protect\citeauthoryear{{Pierre Auger Collaboration}}{{Pierre Auger
  Collaboration}}{2010}]{Abreu10}
{Pierre Auger Collaboration} 2010, \mn@doi [Astroparticle Physics]
  {10.1016/j.astropartphys.2010.08.010}, \href
  {https://ui.adsabs.harvard.edu/abs/2010APh....34..314A} {34, 314}

\bibitem[\protect\citeauthoryear{{Pierre Auger Collaboration}}{{Pierre Auger
  Collaboration}}{2015}]{Aab15}
{Pierre Auger Collaboration} 2015, \mn@doi [\apj] {10.1088/0004-637X/804/1/15},
  \href {https://ui.adsabs.harvard.edu/abs/2015ApJ...804...15A} {804, 15}

\bibitem[\protect\citeauthoryear{{Pierre Auger Collaboration}}{{Pierre Auger
  Collaboration}}{2017}]{Auger17}
{Pierre Auger Collaboration} 2017, \mn@doi [Science] {10.1126/science.aan4338},
  \href {https://ui.adsabs.harvard.edu/abs/2017Sci...357.1266P} {357, 1266}

\bibitem[\protect\citeauthoryear{{Pietrzy{\'n}ski} et~al.,}{{Pietrzy{\'n}ski}
  et~al.}{2013}]{LMCmodulus}
{Pietrzy{\'n}ski} G.,  et~al., 2013, \mn@doi [\nat] {10.1038/nature11878},
  \href {https://ui.adsabs.harvard.edu/abs/2013Natur.495...76P} {495, 76}

\bibitem[\protect\citeauthoryear{{Planck Collaboration}}{{Planck
  Collaboration}}{2016}]{PLANCK15}
{Planck Collaboration} 2016, \mn@doi [\aap] {10.1051/0004-6361/201525830},
  \href {http://esoads.eso.org/abs/2016A%26A...594A..13P} {594, A13}

\bibitem[\protect\citeauthoryear{{Rowan-Robinson}}{{Rowan-Robinson}}{1999}]{Rowan99}
{Rowan-Robinson} M.,  1999, \apss, \href
  {https://ui.adsabs.harvard.edu/abs/1999Ap&SS.266..291R} {266, 291}

\bibitem[\protect\citeauthoryear{{Salim} et~al.,}{{Salim}
  et~al.}{2016}]{Salim16}
{Salim} S.,  et~al., 2016, \mn@doi [\apjs] {10.3847/0067-0049/227/1/2}, \href
  {http://adsabs.harvard.edu/abs/2016ApJS..227....2S} {227, 2}

\bibitem[\protect\citeauthoryear{{Salim}, {Boquien}  \& {Lee}}{{Salim}
  et~al.}{2018}]{Salim18}
{Salim} S.,  {Boquien} M.,   {Lee} J.~C.,  2018, \mn@doi [\apj]
  {10.3847/1538-4357/aabf3c}, \href
  {https://ui.adsabs.harvard.edu/abs/2018ApJ...859...11S} {859, 11}

\bibitem[\protect\citeauthoryear{{Salmon}, {Hanlon}, {Jeffrey}  \&
  {Martin-Carrillo}}{{Salmon} et~al.}{2020}]{Salmon20}
{Salmon} L.,  {Hanlon} L.,  {Jeffrey} R.~M.,   {Martin-Carrillo} A.,  2020,
  \mn@doi [\aap] {10.1051/0004-6361/201936573}, \href
  {https://ui.adsabs.harvard.edu/abs/2020A&A...634A..32S} {634, A32}

\bibitem[\protect\citeauthoryear{{Salpeter}}{{Salpeter}}{1955}]{Salpeter55}
{Salpeter} E.~E.,  1955, \mn@doi [\apj] {10.1086/145971}, \href
  {https://ui.adsabs.harvard.edu/abs/1955ApJ...121..161S} {121, 161}

\bibitem[\protect\citeauthoryear{{Sanders}, {Mazzarella}, {Kim}, {Surace}  \&
  {Soifer}}{{Sanders} et~al.}{2003}]{Sanders03}
{Sanders} D.~B.,  {Mazzarella} J.~M.,  {Kim} D.-C.,  {Surace} J.~A.,   {Soifer}
  B.~T.,  2003, \mn@doi [\aj] {10.1086/376841}, \href
  {http://adsabs.harvard.edu/abs/2003AJ....126.1607S} {126, 1607}

\bibitem[\protect\citeauthoryear{{Saxton}, {Read}, {Esquej}, {Freyberg},
  {Altieri}  \& {Bermejo}}{{Saxton} et~al.}{2008}]{Saxton08}
{Saxton} R.~D.,  {Read} A.~M.,  {Esquej} P.,  {Freyberg} M.~J.,  {Altieri} B.,
   {Bermejo} D.,  2008, \mn@doi [\aap] {10.1051/0004-6361:20079193}, \href
  {https://ui.adsabs.harvard.edu/abs/2008A&A...480..611S} {480, 611}

\bibitem[\protect\citeauthoryear{{Schutz}}{{Schutz}}{1986}]{Schutz86}
{Schutz} B.~F.,  1986, \mn@doi [\nat] {10.1038/323310a0}, \href
  {https://ui.adsabs.harvard.edu/abs/1986Natur.323..310S} {323, 310}

\bibitem[\protect\citeauthoryear{{Selsing} et~al.,}{{Selsing}
  et~al.}{2018}]{Selsing18}
{Selsing} J.,  et~al., 2018, \mn@doi [\aap] {10.1051/0004-6361/201731475},
  \href {https://ui.adsabs.harvard.edu/abs/2018A&A...616A..48S} {616, A48}

\bibitem[\protect\citeauthoryear{{She}, {Ho}  \& {Feng}}{{She}
  et~al.}{2017}]{She17}
{She} R.,  {Ho} L.~C.,   {Feng} H.,  2017, \mn@doi [\apj]
  {10.3847/1538-4357/835/2/223}, \href
  {http://adsabs.harvard.edu/abs/2017ApJ...835..223S} {835, 223}

\bibitem[\protect\citeauthoryear{{Singer} et~al.,}{{Singer}
  et~al.}{2016}]{Singer16}
{Singer} L.~P.,  et~al., 2016, \mn@doi [\apjl] {10.3847/2041-8205/829/1/L15},
  \href {https://ui.adsabs.harvard.edu/abs/2016ApJ...829L..15S} {829, L15}

\bibitem[\protect\citeauthoryear{{Skrutskie} et~al.,}{{Skrutskie}
  et~al.}{2006}]{Skrutskie06}
{Skrutskie} M.~F.,  et~al., 2006, \mn@doi [\aj] {10.1086/498708}, \href
  {http://adsabs.harvard.edu/abs/2006AJ....131.1163S} {131, 1163}

\bibitem[\protect\citeauthoryear{{Springob}, {Haynes}, {Giovanelli}  \&
  {Kent}}{{Springob} et~al.}{2005}]{Springob04}
{Springob} C.~M.,  {Haynes} M.~P.,  {Giovanelli} R.,   {Kent} B.~R.,  2005,
  \mn@doi [\apjs] {10.1086/431550}, \href
  {https://ui.adsabs.harvard.edu/abs/2005ApJS..160..149S} {160, 149}

\bibitem[\protect\citeauthoryear{{Stampoulis}, {van Dyk}, {Kashyap}  \&
  {Zezas}}{{Stampoulis} et~al.}{2019}]{Stampoulis19}
{Stampoulis} V.,  {van Dyk} D.~A.,  {Kashyap} V.~L.,   {Zezas} A.,  2019,
  \mn@doi [\mnras] {10.1093/mnras/stz330}, \href
  {https://ui.adsabs.harvard.edu/abs/2019MNRAS.485.1085S} {485, 1085}

\bibitem[\protect\citeauthoryear{{Steer} et~al.,}{{Steer}
  et~al.}{2017}]{Steer17}
{Steer} I.,  et~al., 2017, \mn@doi [\aj] {10.3847/1538-3881/153/1/37}, \href
  {http://adsabs.harvard.edu/abs/2017AJ....153...37S} {153, 37}

\bibitem[\protect\citeauthoryear{{Tanvir}, {Levan}, {Fruchter}, {Hjorth},
  {Hounsell}, {Wiersema}  \& {Tunnicliffe}}{{Tanvir} et~al.}{2013}]{Tanvir13}
{Tanvir} N.~R.,  {Levan} A.~J.,  {Fruchter} A.~S.,  {Hjorth} J.,  {Hounsell}
  R.~A.,  {Wiersema} K.,   {Tunnicliffe} R.~L.,  2013, \mn@doi [\nat]
  {10.1038/nature12505}, \href
  {https://ui.adsabs.harvard.edu/abs/2013Natur.500..547T} {500, 547}

\bibitem[\protect\citeauthoryear{{Terry}, {Paturel}  \& {Ekholm}}{{Terry}
  et~al.}{2002}]{Terry02}
{Terry} J.~N.,  {Paturel} G.,   {Ekholm} T.,  2002, \mn@doi [\aap]
  {10.1051/0004-6361:20021018}, \href
  {http://esoads.eso.org/abs/2002A%26A...393...57T} {393, 57}

\bibitem[\protect\citeauthoryear{{Toffano}, {Mapelli}, {Giacobbo}, {Artale}  \&
  {Ghirlanda}}{{Toffano} et~al.}{2019}]{Toffano19}
{Toffano} M.,  {Mapelli} M.,  {Giacobbo} N.,  {Artale} M.~C.,   {Ghirlanda} G.,
   2019, \mn@doi [\mnras] {10.1093/mnras/stz2415}, \href
  {https://ui.adsabs.harvard.edu/abs/2019MNRAS.489.4622T} {489, 4622}

\bibitem[\protect\citeauthoryear{{Tremonti} et~al.,}{{Tremonti}
  et~al.}{2004}]{Tremonti04}
{Tremonti} C.~A.,  et~al., 2004, \mn@doi [\apj] {10.1086/423264}, \href
  {http://adsabs.harvard.edu/abs/2004ApJ...613..898T} {613, 898}

\bibitem[\protect\citeauthoryear{{Tully}, {Courtois}  \& {Sorce}}{{Tully}
  et~al.}{2016}]{Tully16}
{Tully} R.~B.,  {Courtois} H.~M.,   {Sorce} J.~G.,  2016, \mn@doi [\aj]
  {10.3847/0004-6256/152/2/50}, \href
  {https://ui.adsabs.harvard.edu/abs/2016AJ....152...50T} {152, 50}

\bibitem[\protect\citeauthoryear{{Varela} et~al.,}{{Varela}
  et~al.}{2009}]{Varela09}
{Varela} J.,  et~al., 2009, VizieR Online Data Catalog, \href
  {https://ui.adsabs.harvard.edu/abs/2009yCat..34970667V} {pp J/A+A/497/667}

\bibitem[\protect\citeauthoryear{{Verley} et~al.,}{{Verley}
  et~al.}{2007}]{Verley07}
{Verley} S.,  et~al., 2007, \mn@doi [\aap] {10.1051/0004-6361:20077307}, \href
  {https://ui.adsabs.harvard.edu/abs/2007A&A...470..505V} {470, 505}

\bibitem[\protect\citeauthoryear{{Wang}, {Rowan-Robinson}, {Norberg}, {Heinis}
  \& {Han}}{{Wang} et~al.}{2014}]{Wang14}
{Wang} L.,  {Rowan-Robinson} M.,  {Norberg} P.,  {Heinis} S.,   {Han} J.,
  2014, \mn@doi [\mnras] {10.1093/mnras/stu915}, \href
  {http://adsabs.harvard.edu/abs/2014MNRAS.442.2739W} {442, 2739}

\bibitem[\protect\citeauthoryear{{Webb} et~al.,}{{Webb} et~al.}{2020}]{Webb20}
{Webb} N.~A.,  et~al., 2020, \mn@doi [\aap] {10.1051/0004-6361/201937353},
  \href {https://ui.adsabs.harvard.edu/abs/2020A&A...641A.136W} {641, A136}

\bibitem[\protect\citeauthoryear{{Wenger} et~al.,}{{Wenger}
  et~al.}{2000}]{Wenger00}
{Wenger} M.,  et~al., 2000, \mn@doi [\aaps] {10.1051/aas:2000332}, \href
  {https://ui.adsabs.harvard.edu/abs/2000A&AS..143....9W} {143, 9}

\bibitem[\protect\citeauthoryear{{White}, {Daw}  \& {Dhillon}}{{White}
  et~al.}{2011}]{White11}
{White} D.~J.,  {Daw} E.~J.,   {Dhillon} V.~S.,  2011, \mn@doi [Classical and
  Quantum Gravity] {10.1088/0264-9381/28/8/085016}, \href
  {https://ui.adsabs.harvard.edu/abs/2011CQGra..28h5016W} {28, 085016}

\bibitem[\protect\citeauthoryear{{Wolf} et~al.,}{{Wolf} et~al.}{2018}]{Wolf18}
{Wolf} C.,  et~al., 2018, \mn@doi [\pasa] {10.1017/pasa.2018.5}, \href
  {https://ui.adsabs.harvard.edu/abs/2018PASA...35...10W} {35, e010}

\bibitem[\protect\citeauthoryear{{Wyatt}, {Tohuvavohu}, {Arcavi}, {Lundquist},
  {Howell}  \& {Sand}}{{Wyatt} et~al.}{2020}]{Wyatt20}
{Wyatt} S.~D.,  {Tohuvavohu} A.,  {Arcavi} I.,  {Lundquist} M.~J.,  {Howell}
  D.~A.,   {Sand} D.~J.,  2020, \mn@doi [\apj] {10.3847/1538-4357/ab855e},
  \href {https://ui.adsabs.harvard.edu/abs/2020ApJ...894..127W} {894, 127}

\bibitem[\protect\citeauthoryear{{Yang} et~al.,}{{Yang} et~al.}{2019}]{Yang19}
{Yang} S.,  et~al., 2019, \mn@doi [\apj] {10.3847/1538-4357/ab0e06}, \href
  {https://ui.adsabs.harvard.edu/abs/2019ApJ...875...59Y} {875, 59}

\bibitem[\protect\citeauthoryear{{Zevin}, {Kelley}, {Nugent}, {Fong}, {Berry}
  \& {Kalogera}}{{Zevin} et~al.}{2019}]{Zevin19}
{Zevin} M.,  {Kelley} L.~Z.,  {Nugent} A.,  {Fong} W.-f.,  {Berry} C. P.~L.,
  {Kalogera} V.,  2019, arXiv e-prints, \href
  {https://ui.adsabs.harvard.edu/abs/2019arXiv191003598Z} {p. arXiv:1910.03598}

\bibitem[\protect\citeauthoryear{{de Vaucouleurs}}{{de
  Vaucouleurs}}{1991}]{RC91}
{de Vaucouleurs} G.,  1991, Science, \href
  {http://adsabs.harvard.edu/abs/1991Sci...254..592D} {254, 592}

\bibitem[\protect\citeauthoryear{{de Vaucouleurs}, {de Vaucouleurs}  \&
  {Corwin}}{{de Vaucouleurs} et~al.}{1976}]{RC2}
{de Vaucouleurs} G.,  {de Vaucouleurs} A.,   {Corwin} J.~R.,  1976, in Second
  reference catalogue of bright galaxies, Vol. 1976, p. Austin: University of
  Texas Press..

\makeatother
\end{thebibliography}


\appendix

\section{Computation of the Virgo-infall corrected radial velocities}
\label{app:vvir}

Starting from the heliocentric velocity, $v_{hc}$, of a galaxy at galactic coordinates $\left(l, b\right)$, we adopt the correction of \citet{Karachentsev96} for solar motion in the Local Standard of Rest (LSR) and the Milky Way's motion with respect to the Local Group (LG) centroid:
\begin{equation}
    v_{\rm lg} = v_{\rm hc} + V_a \left[ \cos b \cos b_a \cos\left(l - l_a\right) + \sin b \sin b_a \right],
\end{equation}
where $V_a{=}\left(316{\pm}5\right)\kms$ is the velocity of the Sun towards the LG centroid at galactic coordinates
$l_a{=}\left(93{\pm}2\right)\degr$ and $b_a{=}\left(-4{\pm}2\right)\degr$. Then, we correct for the Local Group's infall to the Virgo cluster following \citet{Terry02}:
\begin{equation}
    \vvir = v_{\rm lg} + V_{\rm lg-infall} \cos{\Theta},
\end{equation}
where $V_{\rm lg-infall}{=}\left(208{\pm}9\right)\kms$ is the infall velocity of LG to Virgo cluster, and $\Theta$ is the great-circle distance between the galaxy's supergalactic coordinates and LG's apex $\left(102\fdg88, -2\fdg34\right)$.

\section{Cross-matching procedures and obtained data}

The following paragraphs provide additional details regarding some of the cross-matching procedures described in \S\ref{txt:multi}.

\subsection{\hyper{} vs. \ned{}}
\label{app:ned}

The cross-correlation of the \hyper{} and the \ned{} is an essential step to 
(i) obtain missing radial velocities, 
(ii) use the associations to match \hyper{} objects to $z$-independent distance measurements in \nedd{}, and 
(iii) provide quick links to \ned{} entries for the galaxies.
This step of the pipeline is executed before applying the recession velocity cut, since \ned{} complements our sample with radial velocities. Therefore, 884,766 objects are searched, i.e. galaxies with heliocentric velocity ${<}14,500\kms{}$ (ensuring that no object with $\vvir{<}14,000\kms$ is excluded), and objects without radial velocity information in \hyper{}.
We use the Python \code{astroquery} package to associate the \hyper{} galaxies to \ned{} objects on the basis of their designation: for each object in \hyper{}, we perform two searches: based on their PGC ID (e.g., PGC000002) and principal designation (e.g., UGC12889).
We perform a series of checks to identify cases where:
\begin{enumerate}
\item 
    the two searches (principal designation and PGC number) return different \ned{} objects (1,024).
\item 
    different \hyper{} objects are associated to the same \ned{} object (510),
\item 
    positions or radial velocities disagree (1,232),
\item 
    the \hyper{} object has a large astrometric error and size, and has been associated to a \ned{} object by chance (usually Zone of Avoidance objects; 33,101),
\item 
    there are typographic errors in galaxy pairs (e.g., A in the place of B in \ned{}; 202)
\end{enumerate}
The above situations are resolved automatically (e.g., positional disagreement larger than 1~arcmin), or after manual inspection. In total, 137,586 galaxies in the \hec{} (67\%) are associated to \ned{} objects.

\subsection{Supplementary size information}
\label{app:diameters}

\hyper{} provides the size of the galaxies based on the $D_{25}$ isophote in the $B$-band. However, for 39,251 objects (19\%) this information is not available. Using the associations of \hyper{} to \ned{} objects, we find that for the majority of these objects, the diameters can be obtained from \twomass{} and \sdss{}. In addition, using the CDS XMatch service (\url{http://cdsxmatch.u-strasbg.fr/}), we find eight other catalogues that can provide diameters for the majority of the rest of these objects. The catalogues used to draw this information are listed in \autoref{tbl:dsources}.

The supplementary size information is incorporated by rescaling the semi-major axis from the external catalogue, $a_{\rm ext}$, using as reference the \hyper{} semi-major axis, $a_{\rm hyp}$. To do so, we:
\begin{enumerate}
\item 
    associate all \hyper{} objects to the external catalogue,
\item 
    use the associated galaxies for which both $a_{\rm hyp}$ and $a_{\rm ext}$ are defined to compute the scaling factor $c{=}{\langle}a_{\rm hyp}/a_{\rm ext}{\rangle}$, and
\item 
    fill in the $a_{\rm hyp}$ for the galaxies in \hec{} without semi-major axis from \hyper{}: $a_{\rm hyp}{=}c{\times}a_{\rm ext}$.
\end{enumerate}

The priority of the external catalogues was based on the number of common objects in the external catalogue and the \hyper{}, the proximity of the band to the $B$-band which is available in \hyper{}, and the scatter in the scaling relation (samples with smaller scatter are considered as more reliable). More details can be found in \autoref{tbl:dsources} where the external catalogues are listed in the order of their priority. 

\begin{table*}
\centering
\caption{
    The various sources of semi-major axis information incorporated in the \hec{}. Where available, axis ratios and position angles are also obtained. The semi-major axes are rescaled to match the $D_{25}$ isophotal one in \hyper{}. The columns are:
    (1) the name of the source; 
    (2) The flag in the column \code{dsource} in the provided catalogue; 
    (3) the number of objects in the \hec{} for which the sizes where obtained from the source;
    (4) Number of common objects in both catalogues;
    (5) the scaling factor $C$ in dex, used to homogenise the sizes $r_1$ from the given source to $R_1$ (adopted semi-major axis) as in $\log R_1 = \log r_1 + C$; 
    (6) the scatter (in dex) between the size in \hyper{} and the source for the common objects; 
    (7) notes;
    (8) reference of the source.
    \label{tbl:dsources}
}
\begin{tabular}{lcrrrrp{0.35\linewidth}l}
\hline
    Source & Flag & $N$ & $N_{\rm tot}$ & Scale & Scatter & Notes & Reference \\
    (1) & (2) & (3) & (4) & (5) & (6) & (7) & (8) \\
\hline
    \hyper{} & H & 165,482 & & & & 
            This is the reference sample. Adopted as they are. & \citet{Makarov14} \\
    \sdss{} & S & 12,214 & 124,055 &  0.208 & 0.188 & 
            Petrosian radius in the $g$-band from Data Release 15. The $g$-band was selected because of the small scatter in the scaling factor, as expected due to its proximity to the $B$-band. & \citet{sdss15} \\
    \twomass{} & 2 & 12,918 & 143546 & 0.236 & 0.118 &  
            Super-coadd 3-sigma isophotal semi-major axis radius (\code{sup\_r\_3sig}). The $J$-band 21 mag/arcsec$^2$ isophotal semi-major axis presents a slightly small scatter of $0.115$ but it was not available for all objects. & \citet{Jarrett00} \\
    {\it 2dFGS} & 6 & 6,327 & 21,404 & 1.940 & 0.065 & 
            Areas, eccentricities and orientations in $B$-band. Computed the corresponding semi-major
            and semi-minor axes. The scaling factor converts from pixels to arcmin. & \citet{Colless01} \\
    {\it WINGS} & W & 740 & 2229 & -0.056 & 0.104 & $B$-band isophotal ellipses. & \citet{Varela09} \\
    {\it SkyMapper} & Y & 1,814 & 76,488 & 0.355 & 0.203 &
             Data Release 1.1. Mean $r$-band isophotal diameters. & \citet{Wolf18} \\
    {\it AMIGA-CIG} & A & 65 & 5708 & -0.255 & 0.137 & $R$-band isophotal major axis.
             & \citet{Verley07} \\
    {\it UNGC} & K & 60 & 658 & -0.068 & 0.170 & $B$-band Holmberg isophotal semi-major axis. & \citet{Kara13} \\
    {\it VIII/77} & V & 28 & 8472 & -0.069 & 0.121 & Semi-major axis taken from UGC and ESO, or estimated from POSS-I. & \citet{Springob04} \\
    {\it KKH2001} & 1 & 26 & 101 & & & 
            No correction applied ($B$-band isophotes). & \citet{Karachentsev01} \\
    {\it KKH2007} & 7 & 9 & 90 & & & 
            No correction applied ($B$-band isophotes). & \citet{Karachentsev07} \\
    {\it NED} & N & 212 & 130586 & & & 
            No correction applied. Miscellaneous diameters based on $B$-band, mainly from {\it ESO-LV} & \url{http://ned.ipac.caltech.edu} \\
\hline
\end{tabular}
\end{table*}

When available, semi-minor axes and position angles are also taken from the external catalogues (the axis ratio in the \hec{} is the same as the one reported by the external catalogue). In total, we complete the size information for 34,413 galaxies, leaving 4,837 (2.4\%) galaxies without such information in the \hec{}. Because of the different wavebands and methods used by the external catalogues, the application of a scaling factor is over-simplistic and may have introduced biases. Users of the catalogue are suggested to use the corresponding flag, \code{dsource}, to either filter out these galaxies, or study any biases.

\subsection{\rbgs{}}
\label{app:rbgs}

We cross-match objects in the \hec{} and \rbgs{} on the basis of their $D_{25}$ elliptical regions. 589 galaxies out of the 629 objects in \rbgs{} are associated to \hec{} galaxies. The remaining 40 objects are not cross-linked for the following reasons.
19 associations are rejected because they are galaxy pairs that are resolved in the \hec{} but unresolved in \rbgs{}: 
NGC\,3395/6, NGC\,4038/9, NGC\,4568/7, ESO\,255-IG007, NGC\,4922, ESO\,343-IG013, NGC\,7592, NGC\,3994/5, NGC\,5394/5, NGC\,6670A/B, ESO\,60-IG016, NGC\,7752/3, IC\,2810, IC\,0563/4, IC\,4518A/B, NGC\,5257/8, UGC\,12914/5, NGC\,6052, and AM1633-682.
In addition, 21 \rbgs{} objects are not found in the \hec{} because: (i) their radial velocity exceeds our recession velocity limit $\vvir{}{=}14,000\kms{}$ (18 galaxies), (ii) their object type in \hyper{} is unknown (ESO\,221-IG010 and ESO 350-IG038), or (iii) is identified as a star (IRAS\,F05170+0535).

\subsection{\rifz{}}
\label{app:rifz}

Before the cross-matching of \rifz{} and \hec{}, we corrected an object designation in \rifz{} which was appearing twice in the catalogue (column \code{FSCNAME}): two instances of F14012+5434, one of which was corrected to F01339+1532, after manual inspection using the provided coordinates.
We associate the \hec{} objects (without associations to \rbgs{}) to \rifz{} objects, if the $D_{25}$ elliptical region of the former and the 6\asec{} circle (i.e., the resolution of \iras{}) of the latter overlap. We find thousands of multiple matches. In order to resolve the multiple matches, we apply a four-step procedure:
\begin{enumerate}
\item 
    Since \rifz{} provides better positional accuracy than \iras{} (through associations to other surveys such as \twomass{}), we use a matching radius of 3\asec{} to cross-link the \hec{} and \rifz{}. 18,147 matches are accepted, after resolving manually six multiple matches on the basis of radial velocities and offsets of the matched sources.
\item 
    For the objects in \hec{} and \rifz{} that remain unmatched after step (i), we use a 6\asec{} match radius for both catalogues. We find 550 matches, after resolving manually six multiple matches with the same criteria as in (i).
\item 
    The unmatched objects (after (i) and (ii)), are cross-linked using the $D_{25}$ region in the \hec{} and the 6\asec{} circle around the position in the \rifz{}. Multiple matches are resolved with the requirement that radial velocities match (${<}100\kms{}$ difference). 407 matches are found, leaving only 168 unmatched objects in the \hec{}, and 175 in \rifz{}.
\item 
    The matches of the steps (i)-(iii) are joined and inspected for ambiguous matches, i.e., galaxy pairs may be resolved in the \hec{} but not in the \rifz{}. We reject 22 such associations.
\end{enumerate}
The above steps provide 19,082 unique associations between \hec{} and \rifz{} objects.

\subsection{\twomass{}}
\label{app:twomass}

We sequentially cross-match the \hec{} with the three catalogues providing \twomass{} data: \lga{}, \masx{} and \masp{}. This order ensures that the associated photometric data reflect the full extent of the galaxies.

Out of the 665 objects in \lga{}, we exclude 35 because they are not galaxies (see \url{https://irsa.ipac.caltech.edu/data/LGA/overview.html}). Out of the remaining 620 galaxies, 609 are cross-matched to \hec{} objects. The unassociated objects were either exceeding the radial velocity criterion (7 objects), or \hyper{} did not classify them as galaxies (3 objects), or belonged to the galaxy pair Arp\,244 that is resolved in the \hec{} (1 object).

Then, we cross-match the \hec{} and the \masx{} using a 3\asec{} match radius. From this procedure, we exclude the \hec{} objects that are already associated to the \lga{} galaxies. We also exclude the following extended galaxies that are resolved in the \twomass{}, and would produce thousands of chance coincidence matches: Draco Dwarf, Leo B, Sextans Dwarf Spheroidal, the Magellanic Clouds and Carina Dwarf Spheroidal. In total, we find 117,713 matches.

Finally, considering objects not associated to either the \lga{} or the \masx{}, we cross-match \hec{} and \masp{} and find 25,224 matches.

\section{Empirical formul\AE{} for the distances of the \hec{} galaxies}
\label{txt:app:empirical}

\begin{figure}
    \includegraphics[width=\columnwidth]{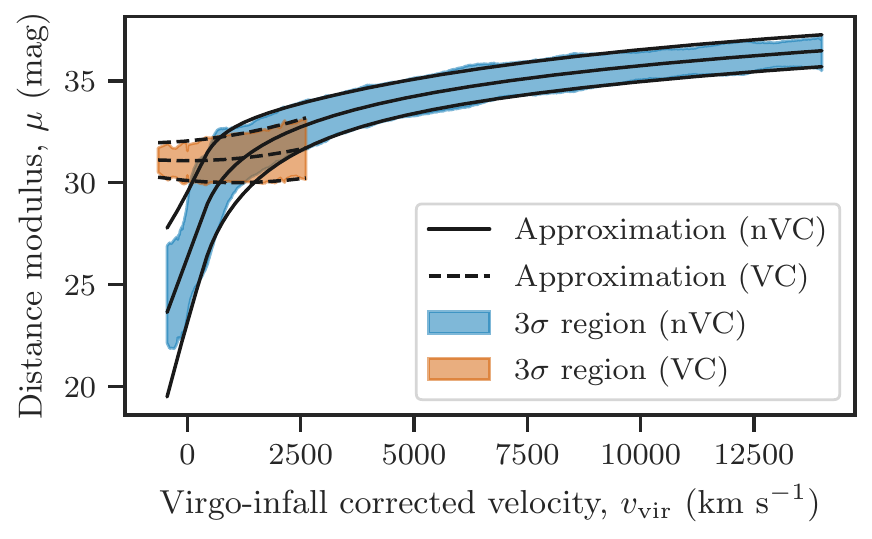}
    \caption{
        The local $99.7\%$ intervals of the distance modulus of the two models, orange for the Virgo cluster and blue for the rest. The black lines correspond to the mean and same interval (dashed for VC model) computed using the approximation formul\ae{} (Appendix~\ref{txt:app:empirical}).
        }
    \label{fig:approx}
\end{figure}

The intrinsic distance modulus $\mu_{\text{int}}$ and its uncertainty $\epsilon_\text{int}$ of a galaxy with Virgo-infall corrected radial velocity $\vvir$, inferred by the Kernel Regression explained in \S\ref{txt:localaverage}, can be approximated by the following formul\ae{} for nVC galaxies:
\begin{align}
    \mu_{\text{int}}
        &\approx \begin{cases}
                    26.34 + 0.006057 u, & u \leq 358.5 \\
                    15.74 + 5 \log_{10}{u}, & u > 358.5
                 \end{cases}, \\
    \epsilon_{\text{int}} &
        \approx 0.2611 + 0.8016 \exp\left(-\frac{u}{1341}\right),
\end{align}
where $u{\equiv}\vvir{/}\kms{}$. The above relations are valid for the range $u{\in}\left[-481.7, 14,033.0\right]$. Similarly, for VC galaxies:
\begin{align}
    \mu_{\text{int,VC}}       &  \approx 31.08 + 9.177\times 10^{-8} u^2, \\
    \epsilon_{\text{int,VC}}  &  \approx 0.3235 + 6.464\times 10^{-5} u,
\end{align}
valid in the range $u{\in}\left[-792.5, 2764\right]$. These approximating formul\ae{} for the distance modulus $\mu_{\text{int}}$, and the $3\epsilon_{\text{int}}$ region are plotted in \autoref{fig:approx}, on top of the corresponding quantities computed using the regression models.

\section{Description of columns in the provided catalogue}
\label{app:columns}

The columns of the \hec{} are described in \autoref{tbl:columns}.

\newcommand{\col}[1]{\texttt{\detokenize{#1}}}

\begin{table*}
    \centering
    \caption{
        Description of the columns in the machine-readable catalogue. In cases of adopted values from external catalogues, the middle column reports the source: H=\hyper{}, N=\ned{}, I=\iras{}, F=\wise{} forced photometry, M=\twomass{}, S=\sdss{}, G=\gsw{}. Unflagged columns were computed by us.
        \label{tbl:columns}
    }
\begin{tabular}{p{0.22\textwidth}p{0.03\textwidth}p{0.66\textwidth}} \hline \small
Column & Flag & Description \\\hline
\col{pgc}, \col{objname} & H
    & {\it Principal Catalogue of Galaxies} number, and object name in the \hyper{}. \\
\col{id_ned}, \col{id_nedd} & N  
    & Name in \ned{} and \nedd{} respectively. \\
\col{id_iras} & I
    & Name in \rbgs{}, or in \rifz{} if in the form Fxxxxx+xxxx. \\
\col{id_2mass} & M
    & ID in \lga{}, \masx{}, or \masp{} (see \col{flag_2mass}). \\
\col{sdss_photid}, \col{sdss_specid} & S
    & \sdss{} photometric and spectroscopic IDs (consistent with DR8 and later releases). \\\hline
\col{ra}, \col{dec} & H
    & Decimal J2000.0 equatorial coordinates (deg). \\
\col{f_astrom} & H
    & Astrometric precision flag. -1 for ${\sim}0.1\asec{}$; 0 for ${\sim}1\asec{}$; 1 for ${\sim}10\asec{}$; and so on. \\
\col{r1}, \col{r2}, \col{pa} & H
    & $D_{25}$ semi-major and semi-minor axes (arcmin), and North-to-Northeast position angle (deg). \\
\col{rsource}, \col{rflag} &
    & Source (see \autoref{tbl:dsources}) and flag of the size information: 0=missing, 1=all size information defined, 2=either \col{r2} or \col{pa} were missing and they were set equal to \col{r1} and $0.0$ respectively (circular isophote).\\
\hline
\col{t}, \col{e_t} & H
    & Numerical Hubble-type and its uncertainty. See \citet{RC2}. \\
\col{incl} & H
    & Inclination (deg). \\
\col{v}, \col{e_v} & HN
    & Heliocentric radial velocity, and its uncertainty ($\rm km\,s^{-1}$). \\
\col{v_vir}, \col{e_v_vir} & &
    Virgo-infall corrected radial velocity and its uncertainty ($\rm km\,s^{-1}$). \\\hline
\col{ndist} & 
    & Number of distance measurements in \nedd{} used for the computation of \col{d}. \\
\col{edist} & 
    & If True, the \nedd{} distance measurements had uncertainties. \\
\col{d}, \col{e_d} & 
    & Distance, and its uncertainty (Mpc). \\
\col{d_lo68}, \col{d_hi68}, \col{d_lo95}, \col{d_hi95} &
    & 68\% and 95\% confidence intervals of the distance. \\
\col{dmethod} & 
    & Method for the estimation of the distance: N=from \nedd{}, Z=regressor, Zv=VC-regressor, C(v)=distance from \nedd{} but uncertainty from (VC-)regressor. \\
\hline
\col{ut}, \col{bt}, \col{vt}, \col{it} & H
    & Total U, B, V, and I apparent magnitudes (mag). \\
\col{e_ut}, \col{e_bt}, \col{e_vt}, \col{e_it} & H
    & Uncertainties on \col{ut}, \col{bt}, \col{vt}, \col{it} (mag).\\
\col{ag}, \col{ai} & H
    & Galactic and intrinsic absorption in $B$-band. \\
\col{s12}, \col{s25}, \col{s60}, \col{s100} & I
    & \iras{} fluxes at 12, 25, 60, and 100\microns{} respectively (Jy). \\
\col{q12}, \col{q25}, \col{q60}, \col{q100} & I
    & Quality flags for \col{s12}, \col{s25}, \col{s60}, \col{s100}: 0=not in \iras{}, 1=upper limit, 2=moderate quality, 3=high quality in {\it FSC} or 4=flux from {\it RBGS}. \\
\col{wf1}, \col{wf2}, \col{wf3}, \col{wf4} & F
    & 3.3, 4.6, 12 and 22\microns{} fluxes in the \wise{} forced photometry catalogue (mag). \\
\col{e_wf1}, \col{e_wf2}, \col{e_wf3}, \col{e_wf4} & F
    & Uncertainties on \col{wf1}, \col{wf2}, \col{wf3}, \col{wf4} (mag). \\
\col{wfpoint}, \col{wftreat} & F
    & \lq{}True\rq{} if point source, and \lq{}True\rq{} if treated as such, respectively, in the \wise{} forced photometry catalogue. \\
\col{j}, \col{h}, \col{k} & M
    & $J$, $H$, and $K_s$-band apparent magnitudes in \twomass{} (mag). \\
\col{e_j}, \col{e_h}, \col{e_k} & M
    & Uncertainties on \col{j}, \col{h}, \col{k} (mag). \\
\col{flag_2mass} &
    & Source of the \twomass{} ID and JHK magnitudes: 0=none, 1={\it LGA}, 2={\it XSC}, 3={\it PSC}. \\
\col{u}, \col{g}, \col{r}, \col{i}, \col{z} & S
    & $u$, $g$, $r$, $i$, and $z$-band apparent magnitudes in \sdss{} (mag). \\
\col{e_u}, \col{e_g}, \col{e_r}, \col{e_i}, \col{e_z} & S
    & Uncertainties on \col{u}, \col{g}, \col{r}, \col{i}, \col{z} (mag). \\\hline
\col{logL_TIR} &
    & Decimal logarithm of the TIR luminosity ($L_\odot{=}3.83\times10^{33}\,\mathrm{erg\,s^{-1}}$). \\
\col{logL_FIR} &   
    & Decimal logarithm of the FIR luminosity ($L_\odot$). \\
\col{logL_60u} &
    & Decimal logarithm of the 60\microns{}-band luminosity ($L_\odot$). \\
\col{logL_12u} & 
    & Decimal logarithm of the 12\microns{}-band luminosity ($L_\odot$). \\
\col{logL_22u} & 
    & Decimal logarithm of the 22\microns{}-band luminosity ($L_\odot$). \\
\col{logL_K} & 
    & Decimal logarithm of the $K_s$-band luminosity ($L_\odot$). \\
\col{ML_ratio} &
    & Mass-to-light ratio (\S\ref{txt:stellarmass}). \\
\col{logSFR_TIR} & 
    & Decimal logarithm of the TIR-based SFR estimate (\sfrunit{}).\\
\col{logSFR_FIR} & 
    & Decimal logarithm of the FIR-based SFR estimate (\sfrunit{}).\\
\col{logSFR_60u} &
    & Decimal logarithm of the 60\microns{}-based SFR estimate (\sfrunit{}).\\
\col{logSFR_12u} &
    & Decimal logarithm of the W3-based SFR estimate (\sfrunit{}).\\
\col{logSFR_22u} &
    & Decimal logarithm of the W4-based SFR estimate (\sfrunit{}).\\
\col{logSFR_HEC} &
    & Homogenised $\log {\rm SFR}$ (\sfrunit{}). Rescaling of SFR indicators is performed only here (\S\ref{txt:sfr}). \\
\col{SFR_HEC_flag} & 
    & Flag indicating photometry source and SFR indicator used for \col{logSFR_HEC} (\autoref{tbl:sfrindicators}). \\
\col{logM_HEC} &
    & Decimal logarithm of the \mass{} (\massunit{}). \\
\col{logSFR_GSW} & G
    & Decimal logarithm of the SFR in \gsw{} (\sfrunit{}). \\
\col{logM_GSW} & G
    & Decimal logarithm of the \mass{} in \gsw{} (\massunit{}). \\
\col{min_snr} &
    & Minimum signal-to-noise ratio of the emission lines used for the activity classification (\col{class_sp}). \\
\col{metal}, \col{flag_metal} &
    & Metallcity [$12+\log{\left(O/H\right)}$] and its quality flag (\S\ref{txt:metallicity}). \\
\col{class_sp} &
    & Nuclear activity classification (\S\ref{txt:activity}): 0=star forming, 1=Seyfert, 2=LINER, 3=composite, -1=unknown. \\
\col{agn_s17} & E
    & AGN classification in \citet{She17}: Y=AGN, N=non-AGN, ?=unknown. \\
\col{agn_hec} &
    & Combination of SDSS and \citet{She17} classifications (\S\ref{txt:activity}): Y=AGN, N=non-AGN, ?=unknown.\\
\hline
\end{tabular}
\end{table*}


\bsp	
\label{lastpage}
\end{document}